\documentclass[12pt,fleqn]{iopart}
\usepackage{iopams}
\usepackage[abbr,round]{harvard}
\usepackage{graphicx}
\usepackage{amssymb}
\usepackage{bm}
\expandafter\let\csname equation*\endcsname\relax
\expandafter\let\csname endequation*\endcsname\relax
\usepackage{amsmath}

\def\ga{\mathrel{\mathchoice {\vcenter{\offinterlineskip\halign{\hfil
$\displaystyle##$\hfil\cr>\cr\sim\cr}}}
{\vcenter{\offinterlineskip\halign{\hfil$\textstyle##$\hfil\cr
>\cr\sim\cr}}}
{\vcenter{\offinterlineskip\halign{\hfil$\scriptstyle##$\hfil\cr
>\cr\sim\cr}}}
{\vcenter{\offinterlineskip\halign{\hfil$\scriptscriptstyle##$\hfil\cr
>\cr\sim\cr}}}}}

\def\la{\mathrel{\mathchoice {\vcenter{\offinterlineskip\halign{\hfil
$\displaystyle##$\hfil\cr<\cr\sim\cr}}}
{\vcenter{\offinterlineskip\halign{\hfil$\textstyle##$\hfil\cr  
<\cr\sim\cr}}}
{\vcenter{\offinterlineskip\halign{\hfil$\scriptstyle##$\hfil\cr
<\cr\sim\cr}}}
{\vcenter{\offinterlineskip\halign{\hfil$\scriptscriptstyle##$\hfil\cr
<\cr\sim\cr}}}}}

\newcommand{\pof}{Phys. Fluids}   
\newcommand{\pre}{Phys. Rev. E}  
\newcommand{\prl}{Phys. Rev. Lett.}  
\newcommand{\jfm}{J. Fluid Mech.}

\newcommand{\gafd}{Geophys.\ Astrophys.\ Fluid\ Dynam.}  
\newcommand{\ssr}{Space Sci. Rev.}

\renewcommand{\vec}[1]{\mbox{\boldmath$#1$}}

\begin{document}

\title[Triadic resonances in a precessing cylinder]{Triadic resonances
  in non-linear simulations of a fluid flow in 
  a precessing cylinder}

\author[A. Giesecke et al.]{Andr{\'e} Giesecke\dag, Thomas
  Albrecht\ddag, Thomas Gundrum\dag, Johann  Herault\dag, Frank
  Stefani\dag} 
\address{\dag Helmholtz-Zentrum Dresden-Rossendorf\\ 
Institute of Fluid Dynamics, Department Magnetohydrodynamics\\ 
Bautzner Landstrasse 400, D-01314 Dresden, Germany}
\address{\ddag Department of Mechanical and Aerospace Engineering\\ 
Monash University, Victoria 3800, Australia}
\ead{a.giesecke@hzdr.de}

\date{\today}

\begin{abstract}

We present results from three-dimensional non-linear hydrodynamic
simulations of a precession driven flow in cylindrical geometry. The
simulations are motivated by a dynamo experiment currently under
development at Helmholtz-Zentrum Dresden-Rossendorf (HZDR) in which
the possibility of generating a magnetohydrodynamic dynamo will be
investigated in a cylinder filled with liquid sodium and
simultaneously rotating around two axes.

In this study, we focus on the emergence of non-axisymmetric
time-dependent flow structures in terms of inertial waves which -- in
cylindrical geometry -- form so-called Kelvin modes. For a precession
ratio (Poincar{\'e} number)
${\rm{Po}}=\Omega_{\rm{p}}/\Omega_{\rm{c}}=0.014$ considered by us,
the amplitude of the forced Kelvin mode reaches up to one fourth of
the rotation velocity of the cylindrical container confirming that
precession provides a rather efficient flow driving mechanism even at
moderate values of ${\rm{Po}}$.

More relevant for dynamo action might be free Kelvin modes
with higher azimuthal wave number. These free Kelvin modes are
triggered by non-linear interactions and may constitute a triadic
resonance with the fundamental forced mode when the height of the
container matches their axial wave lengths. Our simulations reveal
triadic resonances at aspect ratios close to those predicted by the
linear theory except around the primary resonance of the forced
mode. In that regime we still identify various free Kelvin modes,
however, all of them exhibit a retrograde drift around the symmetry
axis of the cylinder and none of them can be assigned to a triadic
resonance. The amplitudes of the free Kelvin modes always remain below
the forced mode but may reach up to 6\% of the of the container's
angular velocity. The properties of the free Kelvin modes, namely their
amplitude and their frequency, will be used in future simulations of
the magnetic induction equation to investigate their ability to
provide for dynamo action.
 
\end{abstract}


\submitto{\NJP}

\maketitle


\section{Introduction}

Instabilities and waves in rotating fluids are important in numerous
technical applications. In most cases it is important to avoid these
phenomena in order to ensure the stability of fast spinning liquid
filled bodies like gyroscopes \cite{1974pfeiffer}, spacecrafts
\cite{1997bao} or projectiles \cite{1982miller}. Waves in rotating
fluids are also of general interest because of their fundamental
character in oceanographic and atmospheric flows \cite{1939rossby}
like, e.g., zonal flows in the atmospheres of Saturn or Jupiter that
may develop via a modulation instability of meridional Rossby waves
\cite{2010colm}. Moreover, the identification of inertial waves in the
Earth's fluid outer core \cite{1987Natur.325..421A} allows conclusions
on the core dynamics and the associated dynamo process that is
responsible for the generation of the Earth's magnetic
field. Nowadays, it is believed that in general planetary dynamos are
driven by rotating thermal or compositional convection
\cite{2010GGG....11.6016K,2010GeoJI.183..150B,2010SSRv..152..449B}. On
the other hand, other types of flows cannot be ruled out, and, for
example, a superposition of random inertial waves in a rotating
conducting fluid is capable of transferring energy in a magnetic field
as well \cite{FLM:383564}. However, in the model of
\citeasnoun{FLM:383564}, field and flow must decay in the long term
because of the lack of energy sources required for a steady
driving. This problem can be overcome by more complex models in which,
for example, the equatorial heat flux in the Earth's outer core
provides a persistent excitation mechanism for inertial waves with
sufficient helicity for the generation of planetary magnetic fields
\cite{2014GeoJI.198.1832D}. Inertial waves can be excited by
precessional forcing as well, and it has long been discussed whether
precession of the Earth can provide the necessary power to drive the
geodynamo
\cite{1963JFM....17....1S,1968JFM....33..739B,1968Sci...160..259M}.

In a next-generation dynamo experiment currently under development at
Helmholtz-Zentrum Dresden-Rossendorf (HZDR) a precession driven flow
of liquid sodium will be used to validate its suitability to drive a
dynamo \cite{leorat1,leorat2,stefani,2014arXiv1410.8373S}. Since the
precessional forcing provides a natural driving mechanism, it
represents a qualitatively new approach compared to previous dynamo
experiments, where an optimized flow was driven either by impellers
\cite{2000PhRvL..84.4365G,2007PhRvL..98d4502M} or by a system of
electromagnetic pumps \cite{2001PhFl...13..561S}.

A small-scale water experiment is currently running at HZDR in order
to examine the hydrodynamic properties of a precession driven flow in
a cylindrical container with strong forcing and a large precession
angle. Measurements of pressure fluctuations and of axial velocity
profiles show three distinct flow regimes in dependence of the
precession ratio: a laminar state (with only the forced mode) followed
by a non-linear regime with various modes superimposed on the
fundamental mode, eventually leading to a permanent chaotic state when
a critical precession ratio is exceeded \cite{johann}. Presumably, the
ability of the flow to drive a dynamo will be different in the three
regimes and although the precessional dynamo experiment at HZDR will
allow magnetic Reynolds numbers of the order of ${\rm{Rm}}\approx 700$
(based on the rotation velocity and the radius of the cylindrical container) it is a
priori not obvious whether a magnetic field will be self-generated at
all. There are promising indications for dynamo action driven by
precession both from liquid sodium experiments by
\citeasnoun{1971JFM....45..111G} which -- despite of the rather small
precession ratio -- achieved an amplification of an applied field by a
factor of three, as well as from simulations in spheres
\cite{2005PhFl...17c4104T}, spheroids \cite{2009GApFD.103..467W},
ellipsoids \cite{hullermann}, cubes \cite{krauze} and cylinders
\cite{2011PhRvE..84a6317N}. However, kinematic dynamo simulations
demonstrate that the primary forced mode alone, which consists of a
single, non-axisymmetric mode with azimuthal wave number $m=1$ and
axial wave number $k=1$, cannot drive a dynamo at a reasonable
${\rm{Rm}}$ \cite{2014arXiv1411.1195G}. This coincides with
\citeasnoun{2011JFM...679...32H} who showed that a linear wave packet
of inertial waves cannot drive a mean-field dynamo. However, this does
not generally preclude the possibility of quasi-laminar
\citeaffixed{1989RSPSA.425..407D}{in the sense of} or small scale
dynamo action driven by inertial waves or another precession induced
instability with sufficient intricate topology.

Experimental studies in a weakly precessing cylinder filled with a
non-conducting fluid showed a large diversity of flow phenomena with
complex structures frequently resulting in a vigorous turbulent state
\cite{1970JFM....41..865G,1992JFM...243..261M,1994JFM...265..345M,1995JFM...303..233K,1996JFM...315..151M}. 
Other experiments revealed large scale structures like a system of 
intermittent, cyclonic vortices that may provide a strong source of
helicity and hence be beneficial for a dynamo
\cite{2012ExFl...53.1693M}, or free Kelvin modes with an azimuthal
wave number $m=5$ and $m=6$ propagating around the symmetry axis of a
weakly precessing cylinder \cite{2008PhFl...20h1701L,FLM:7951619}.

Free Kelvin modes are the natural eigenmodes in a rotating cylinder
\cite{Kelvin,greenspan,1970JFM....40..603M} and are promising
candidates for driving quasi-laminar dynamo action in case of
precessional forcing because at least a subclass of these modes has a
structure similar to the columnar convection cells that are
responsible for dynamo action in convection driven models of the
geodynamo \cite{2003JFM...492..363L}. Free Kelvin modes may emerge
from non-linear interactions of a forced Kelvin mode with itself or as
a parametric instability that involves the forced mode and two free
Kelvin modes
\cite{1999JFM...382..283K,2008PhFl...20h1701L,FLM:7951619}. The latter  
case requires appropriate combinations of wave numbers, frequencies
and a geometry such that all three modes become resonant
simultaneously which usually is named a triadic resonance. Triadic
resonances have been observed in simulations of precession driven flow
in a spheroid \cite{2003JFM...492..363L} and, experimentally, in
weakly forced precession in a cylinder \cite{2008PhFl...20h1701L} and
a cylindrical annulus \cite{2014PhFl...26d6604L}.

In the present study we conduct numerical simulations of a precessing
flow far below the transition to the chaotic state observed in water
experiments, but with sufficient forcing to excite free
Kelvin modes with azimuthal wave numbers $m>1$. We focus on the impact
of the aspect ratio which -- at least in the linear approximation --
determines whether an inertial wave becomes resonant. We
identify different free Kelvin modes by means of their spatial
structure, and we compute their frequencies from the azimuthal drift
motion in order to conclude whether the free Kelvin modes constitute a
triadic resonance. The results are used to further constrain the modes
that may be observed in the water experiment at HZDR\footnote{Pressure
  measurements in the water experiment at HZDR show a periodic signal
  with two frequencies close to the values expected from the
  dispersion relation for free Kelvin modes (see below). However, the
  measurements do not 
  yet allow a unique identification of azimuthal, axial or radial wave
  numbers.} and will be applied in future simulations of the magnetic
induction equation in order to proof whether free Kelvin modes are
suitable to drive a dynamo.

\section{Theoretical background}

The flow of a fluid with kinematic viscosity $\nu$ in the frame of an
enclosed precessing cylinder is described by the Navier-Stokes
equation including terms for the Coriolis force and the Poincar{\'e}
force \citeaffixed{tilgner98}{see e.g.}: 
\begin{equation}
\frac{\partial \vec{u}}{\partial t}+\vec{u}\cdot\nabla\vec{u}
+2(\vec{\Omega}_{\rm{p}}+\vec{\Omega}_{\rm{c}})\times\vec{u}=-\nabla P+\nu\nabla^2\vec{u}
-(\vec{\Omega}_{\rm{p}}\times \vec{\Omega}_{\rm{c}})\times\vec{r}.\label{eq::navier_stokes1} 
\end{equation}
\begin{figure}[t!]
\begin{center}
\includegraphics[width=7cm]{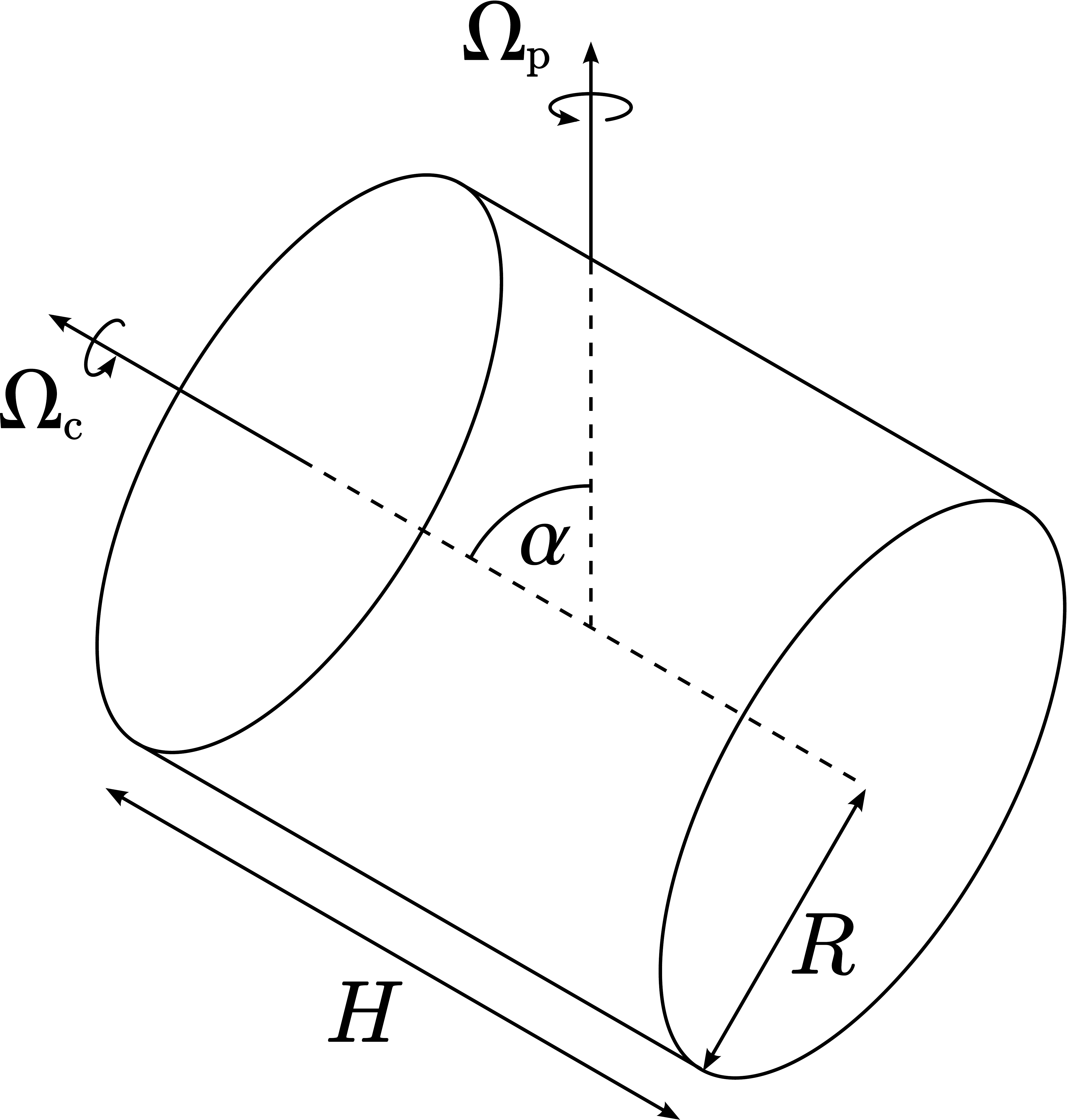}
\caption{Sketch of the problem setup. The cylinder rotates with
  $\Omega_{\rm{c}}$ and the rotation axis precesses with
  $\Omega_{\rm{p}}$. In the simulations we varied the aspect ratio
  $\Gamma=H/R$ and fixed the angle between rotation axis and precession
  axis at $\alpha=\pi/2$.\label{fig::sketch}}
\end{center}
\end{figure}
Here, $\vec{u}$ denotes the velocity field, which additionally obeys
the incompressibility condition $\nabla\cdot\vec{u}=0$, $\vec{r}$ is
the position vector, $P$ is the reduced pressure (including the
centrifugal terms), $\vec{\Omega}_{\rm{c}}$ is the angular frequency
of the cylindrical container and $\vec{\Omega}_{\rm{p}}$ denotes 
the precession (see figure~\ref{fig::sketch}). The precession axis is
time dependent in the cylinder frame, and the unit vector
${\vec{\widehat{\Omega}}}_{\rm{p}}$ that denotes the
orientation of the precession axis is given by
\begin{equation}
{\vec{\widehat{\Omega}}}_{\rm{p}}=\sin \alpha \cos \Omega_{\rm{c}} t
\widehat{\vec{x}}-\sin\alpha\sin \Omega_{\rm{c}} t
\widehat{\vec{y}} +\cos\alpha\widehat{\vec{z}}
\end{equation}
with $\alpha$ the angle between rotation axis and
precession axis. We consider a cylinder with aspect ratio
$\Gamma=H/R$, where $H$ and $R$ are height and radius so that $r\in
[0;R]$ and $z\in[0;\Gamma R]$. In the following we non-dimensionalize
all quantities using $R$ as length scale and $\Omega_{\rm{c}}^{-1}$ as
time scale.  

In the inviscid limit a linear solution that fulfills the 
boundary conditions $\hat{\vec{z}}\cdot\vec{u}=0$ at the top and the
bottom is given by a Kelvin mode \cite{Kelvin,greenspan}:
%
%
\begin{equation}
\vec{U}_{k,m,n}(r,z,\varphi,t)=
\vec{u}_{k,m,n}(r,z)e^{i(\Omega_{k,m,n}t+m\varphi)}+c.c.\label{eq::kelvinmode} 
\end{equation}
with a frequency $\Omega_{k,m,n}$ that depends on the azimuthal wave
number $m$, the axial wave number $k$ and a third number $n$ that
counts the roots of the dispersion relation
\begin{equation}
\Omega_j\lambda_j J_{m-1}(\lambda_j) +
m\left({2}-\Omega_j\right)J_m(\lambda_j)=0\label{eq::dispersion} 
\mbox{ with } 
\Omega_{j}=\pm 2 \sqrt{\left(1 +
\left(\frac{\lambda_{j}}{\Gamma k\pi}\right)^2\right)^{-1}}, 
\end{equation}
in which $J_m$ denotes the Bessel function of order $m$ and the triple
index $(k, m, n)$ is replaced by $j$. Positive
(negative) frequencies correspond to retrograde (prograde)
propagation. Both signs correspond to different radial wave numbers
(different solutions of the dispersion relation) and hence to a
different spatial structure of the corresponding Kelvin mode.

The dispersion relation ensures the fulfillment of the radial boundary
conditions $\hat{\vec{r}}\cdot\vec{u}=0$ at the sidewalls and although
$\lambda_j$ is not an integer it plays a role similar to a radial wave
number with its position in the sequence of zeros corresponding to the
number of half-cycles in the radial direction. The spatial structure of
a Kelvin mode in a meridional plane $\vec{u}_j(r,z)$ is given by
\begin{eqnarray}
u^r_{j}(r,z) & =&
\frac{-i}{(4-\Omega_{j}^2)}
\left[{\Omega_{j}\lambda_{j}}J_{m-1}(\lambda_{j}r)
+\frac{m(2-\Omega_{j})}{r}J_{m}(\lambda_{j}r)\right]
\cos\left(\frac{\pi k z}{\Gamma}\right),
\nonumber
\\
u^{\varphi}_{j}(r,z) & = &
\frac{1}{(4-\Omega_{j}^2)}
\left[{2\lambda_{j}}J_{m-1}(\lambda_{j}r)
-\frac{m(2-\Omega_{j})}{r}J_{m}(\lambda_{j}r)\right]
\cos\left(\frac{\pi k z}{\Gamma}\right),
\label{eq::kelvinmodes}
\\
u^z_{j}(r,z) & =  & -i 
\frac{k \pi}{\Omega_j}J_{m}(\lambda_{j}r)
\sin\left(\frac{\pi k z}{\Gamma}\right).\nonumber 
\end{eqnarray}
with $\lambda_j$ and $\Omega_j$ taken from the solutions of 
the dispersion relation~(\ref{eq::dispersion}). 

In the inviscid linear approximation the total flow of the forced mode with
$m=1$ is given by
\begin{equation}
\vec{u}_{m=1}(r,z,\varphi)=
\sum_k\sum_nA_{1kn}\vec{u}_{1kn}(r,z)
e^{i\varphi+\Omega_{1kn} t}
\end{equation}
where we replaced the index $j$ by the individual contributions and
$A_{1kn}$ represents the amplitude of the forced mode which can be
computed by a projection of the applied forcing onto a particular mode
\cite{1994JFM...265..345M}. For a precession ratio (or Poincar{\'e} number)
${\rm{Po}}=\Omega_{\rm{p}}/\Omega_{\rm{c}}$ and a precession angle
$\alpha$ the corresponding calculation yields
\cite{2012JFM...709..610L}
\begin{equation}
A_{j}=\frac{-{\rm{Po}}\sin\alpha\Gamma^3
(1-(-1)^k)(2+\Omega_{j})^2(2-\Omega_{j})}{\lambda_{j}^2[(k\pi/\Gamma)^2
+\frac{1}{2}(2-\Omega_{j})](1-\Omega_{j})J_1(\lambda_{j})}.\label{eq::amp_invisc}    
\end{equation}

If the wave length that corresponds to an axial wave number of a mode
matches exactly the height of the cylinder, this mode becomes resonant
with an eigenfrequency $\Omega_j=1$ and the expression for the
inviscid amplitude~(\ref{eq::amp_invisc}) diverges. For each $\Gamma$
there exist, in principle, an infinite number of resonant modes. From
the dispersion relation~(\ref{eq::dispersion}) we find that the
primary forced mode, i.e. the mode with $m=1$, the axial wave number
$k=1$ and a radial wave number corresponding to the first root,
becomes resonant at $\Gamma=1.9898174$ which is rather close to the so
called spherical cylinder with height equals diameter
($\Gamma=2$). Further resonances for increasing radial wave number
occur at $\Gamma=0.9560735$ (for $n=2$) and at $\Gamma=0.6206981$ (for
$n=3$). The computation of the amplitude at resonance requires a
consideration of viscous effects in the bulk and in terms of boundary
layers with associated Ekman layer suction. Then the amplitude can
be calculated by matching the Ekman layer suction to the
precessional force \cite{1970JFM....41..865G}. A more general approach
for the computation of the amplitudes of the response created by the
precessional forcing is given by \citeasnoun{2012JFM...709..610L} who
derived an asymptotic solution including viscous effects that is valid
at and away from resonance.

The precessional forcing only excites modes with $m=1$ and $k$
odd. Higher azimuthal modes or modes with even axial wave number must
be triggered by non-linear interactions, e.g., in terms of triadic
resonances that involve this forced Kelvin mode and two free Kelvin
modes $(m_a, k_a)$ and $(m_b, k_b)$ or by the non-linear interaction
of the forced Kelvin mode with itself \cite{2008JFM...599..405M}. In
the following we assume that the resonant case is most promising in
view of the dynamo problem due to the expected larger amplitudes of
the velocity field.

The free Kelvin modes are solutions of the linearized Navier-Stokes
equation with a Coriolis term as a restoring force but without a
precessional driving on the right hand side:
\begin{equation}
\frac{\partial \vec{u}}{\partial t}
+2(\vec{\Omega}_{\rm{p}}+\vec{\Omega}_{\rm{c}})\times\vec{u}=-\nabla
P.\label{eq::navier_stokes_lin} 
\end{equation}
The principle of the interaction of free and forced modes becomes
evident when we consider an exact triad with a forced mode
$\vec{U}_{\!\rm{f}}$, with $m_{\rm{f}}=1$ and $k_{\rm{f}}=1$ and two
free Kelvin modes $\vec{U}_{\! a}$ with $m_{\rm{a}}$ and $k_{\rm{a}}$
and $\vec{U}_{\! b}$ with $m_{\rm{b}}$ and $k_{\rm{b}}$. We ignore
viscosity and further assume a constant amplitude $A_{\rm{f}}$ of the
forced mode so that we can write the total flow as
\begin{equation}
\vec{U}=A_{\rm{f}}\vec{U}_{\rm{f}}+a(t)\vec{U}_{{a}}+b(t)\vec{U}_{{b}},\label{eq::triadenansatz} 
\end{equation} 
where $a(t)$ and $b(t)$ denote the amplitude modulations of the free
mode (without the harmonic oscillations $\Omega_a$ and $\Omega_b$)
which are given by \citeasnoun{1999JFM...382..283K}
\begin{eqnarray}
\frac{{\rm{d}}{a}}{{\rm{d}}t} & = &
\left<\vec{U}_a,\vec{U}_b\times(\nabla\times\vec{U}_{\rm{f}}^{*})
+\vec{U}_{\rm{f}}^{*}\times(\nabla\times\vec{U}_b)\right>b(t),\nonumber\\ 
&&\label{eq::triadenamp}\\[-0.3cm]
\frac{{\rm{d}}{b}}{{\rm{d}}t} & = & 
\left<\vec{U}_b,\vec{U}_a\times(\nabla\times\vec{U}_{\rm{f}}
+\vec{U}_{\rm{f}}\times(\nabla\times\vec{U}_a)\right>a(t).\nonumber
\end{eqnarray}
The square brackets $\left<\cdot,\cdot\right>$ denote a scalar product
defined as $\left<\vec{U}_a,\vec{U}_b\right> = \int
\vec{U}_a^{*}\cdot\,\vec{U}_b {\rm{d}} V$, and a normalization of the
free modes is assumed such that
$\left<\vec{U}_i,\vec{U}_j\right>=\delta_{ij}$. In order to achieve
non-trivial solutions of~(\ref{eq::triadenamp}), and to constitute a
triadic resonance, the wave numbers and frequencies of the free modes
$a$ and $b$ must fulfill the conditions
\begin{equation}
\begin{array}{ccccccc}
\delta m & = & m_{{b}}-m_{{a}} & = & m_{\rm{f}} & = & 1,\\
\delta k & = & |k_{{b}}-k_{{a}}| & = & k_{\rm{f}} & = & 1,\\
\delta\Omega & = & \Omega_{{b}}-\Omega_{{a}} & = & \Omega_{\rm{c}} &
= & 1,\label{eq::triades} 
\end{array}
\end{equation}
where $m_{\rm{f}}=1, k_{\rm{f}}=1$ and $\Omega_{\rm{c}}=1$ denote the
properties of the forced mode. In contrast to sine and
cosine-functions, there are no corresponding addition theorems for
Bessel functions so that no further restrictions are imposed on the
interaction of different radial wave numbers.

\begin{figure}[b!]
\begin{center}
\vspace*{-1.5cm}
\includegraphics[width=13cm]{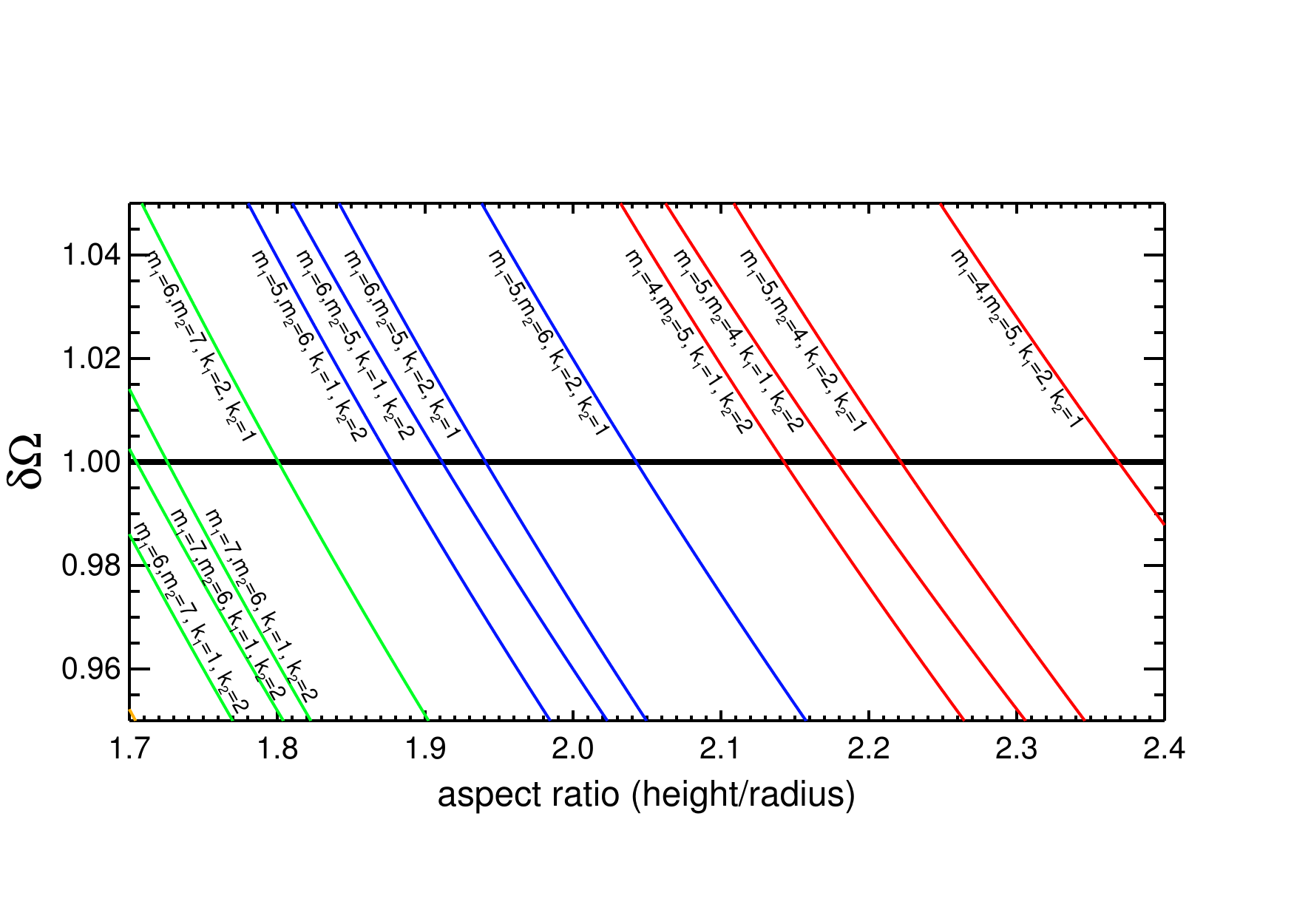}
\vspace*{-1cm}
\caption{Frequency difference $\delta \Omega$ in dependence of the
  aspect ratio $\Gamma$ for two free Kelvin modes which fulfill
  $\delta m=1$ and $\delta k=1$.  The crossings of the curves with the
  horizontal black line at $\delta\Omega=1$ mark the aspect ratio at
  which triadic resonances can be expected.}\label{fig::dispersion}
\end{center}
\end{figure}

In the following, we solve the dispersion
relation~(\ref{eq::dispersion}) for $k=1,2$ and $m=1,\dots, 7$. We
assume that higher radial modes are damped and only consider the first
root for each combination of $m$ and $k$. Figure~\ref{fig::dispersion}
shows the frequency difference $\delta \Omega$ of two free Kelvin
modes with $\delta m = 1$ and $\delta k = 1$ versus the aspect ratio
in the vicinity of the critical value $\delta\Omega=1$. The crossings
of the curves with the horizontal line at $\delta\Omega=1$ mark aspect
ratios for which an exact triadic resonance may occur.
These aspect ratios at which $\delta\Omega=1$ are listed in
table~\ref{tab::dispersion} together with the corresponding frequencies
of the resonant free Kelvin modes.

\begin{table}[h!]
\begin{center}
\begin{tabular}{ccccccc}
$m_a$ & $k_a$ & 
$m_b$ & $k_b$ & 
$\Omega_a$ & $\Omega_b$ & $ \Gamma $\\ 
\hline
6 & 1 & 7 & 2 & 0.3825 & -0.6175 & 1.67420\\
7 & 1 & 6 & 2 & 0.3356 & -0.6644 & 1.70478\\
7 & 2 & 6 & 1 & 0.6494 & -0.3506 & 1.72568\\
6 & 2 & 7 & 1 & 0.6956 & -0.3045 & 1.80074\\
\hline
5 & 1 & 6 & 2 & 0.3888 & -0.6113 & 1.87742\\
6 & 1 & 5 & 2 & 0.3352 & -0.6648 & 1.91146\\
6 & 2 & 5 & 1 & 0.6489 & -0.3511 & 1.94082\\
5 & 2 & 6 & 1 & 0.7013 & -0.2987 & 2.04266\\
\hline
4 & 1 & 5 & 2 & 0.3974 & -0.6026 & 2.14270\\
5 & 1 & 4 & 2 & 0.3350 & -0.6649 & 2.17832\\
5 & 2 & 4 & 1 & 0.6485 & -0.3515 & 2.22216\\
4 & 2 & 5 & 1 & 0.7092 & -0.2907 & 2.36888\\
\hline
\end{tabular}
\caption{Possible triadic resonances with $\delta m=1$, $\delta k=1$ and
  $\delta \Omega = 1$ and corresponding aspect ratios. Negative
  frequencies present  pro-grade modes and positive frequencies
  denote retrograde modes. Only the first radial wave number has been
  considered for each value of $m$ and $k$.  
}\label{tab::dispersion}
\end{center}
\end{table}
A triad is always formed by free Kelvin modes consisting of one
prograde and one retrograde mode \cite{1993PhFl....5..677W}. Note that
the frequencies of free Kelvin modes that constitute different 
triads often are close (if we restrict to triads with $k_a=1$ and
$k_b=2$, see table~\ref{tab::dispersion}) making it difficult to 
reliably identify individual Kelvin modes in realistic setups with
potential frequency shifts due to viscous and/or non-linear effects.

\section{Numerical model}

We conduct numerical simulations in a cylindrical geometry with the
code SEMTEX that applies a spectral element Fourier approach for the
numerical solution of the Navier-Stokes equation
\cite{2004JCoPh.197..759B}. In order to simplify the implementation of
the precessional forcing we switch to the precessional frame in which
the precession is stationary and the cylinder rotates with a frequency
$\Omega_{\rm{c}}$. Thus only a term for the Coriolis force appears in
the Navier-Stokes equation which now reads
\begin{equation}
\frac{\partial \vec{u}}{\partial t}+\vec{u}\cdot\nabla\vec{u}
+2\vec{\Omega}_{\rm{p}}\times\vec{u}=-\nabla P+\nu\nabla^2\vec{u}. 
\end{equation}
Due to the rotation of the cylinder in the precessing frame the
boundary conditions for the azimuthal flow change to
$u_{\varphi}= r\Omega_{\rm{c}}$ (at endcaps and sidewall) whereas
no-slip conditions are still applied
for $u_z$ and $u_r$.

The problem is described by four parameters, the Reynolds number
defined with the angular velocity of the container
${\rm{Re}}=\Omega_{\rm{c}}R^2/\nu$, the precession ratio (or
Poincar{\'e} number)
${\rm{Po}}=\displaystyle{\Omega_{\rm{p}}}/{\Omega_{\rm{c}}}$, the  
aspect ratio $\Gamma=H/R$ and the precession angle $\alpha$ with
$\cos\alpha={\vec{\Omega}_{\rm{p}}\cdot\vec{\Omega}_{\rm{c}}}/{|\vec{\Omega}_{\rm{p}}||\vec{\Omega}_{\rm{c}}|}$. 
In the present study, we keep the precession ratio fixed at
${\rm{Po}}=0.0141$ and the precession angle is set to $\alpha=\pi/2$.
All simulations are performed at ${\rm{Re}}=6500$ except for one run
at ${\rm{Re}}=10000$ to briefly examine the impact of increasing
${\rm{Re}}$. The aspect ratio is varied in the range $\Gamma\in
[1.7;2.4]$ in order to find the container geometry for which
individual modes or triads become resonant.

\section{Results}

\subsection{Pattern of the total flow} 

\begin{figure}[h!]
\begin{center}
\includegraphics[width=15.5cm]{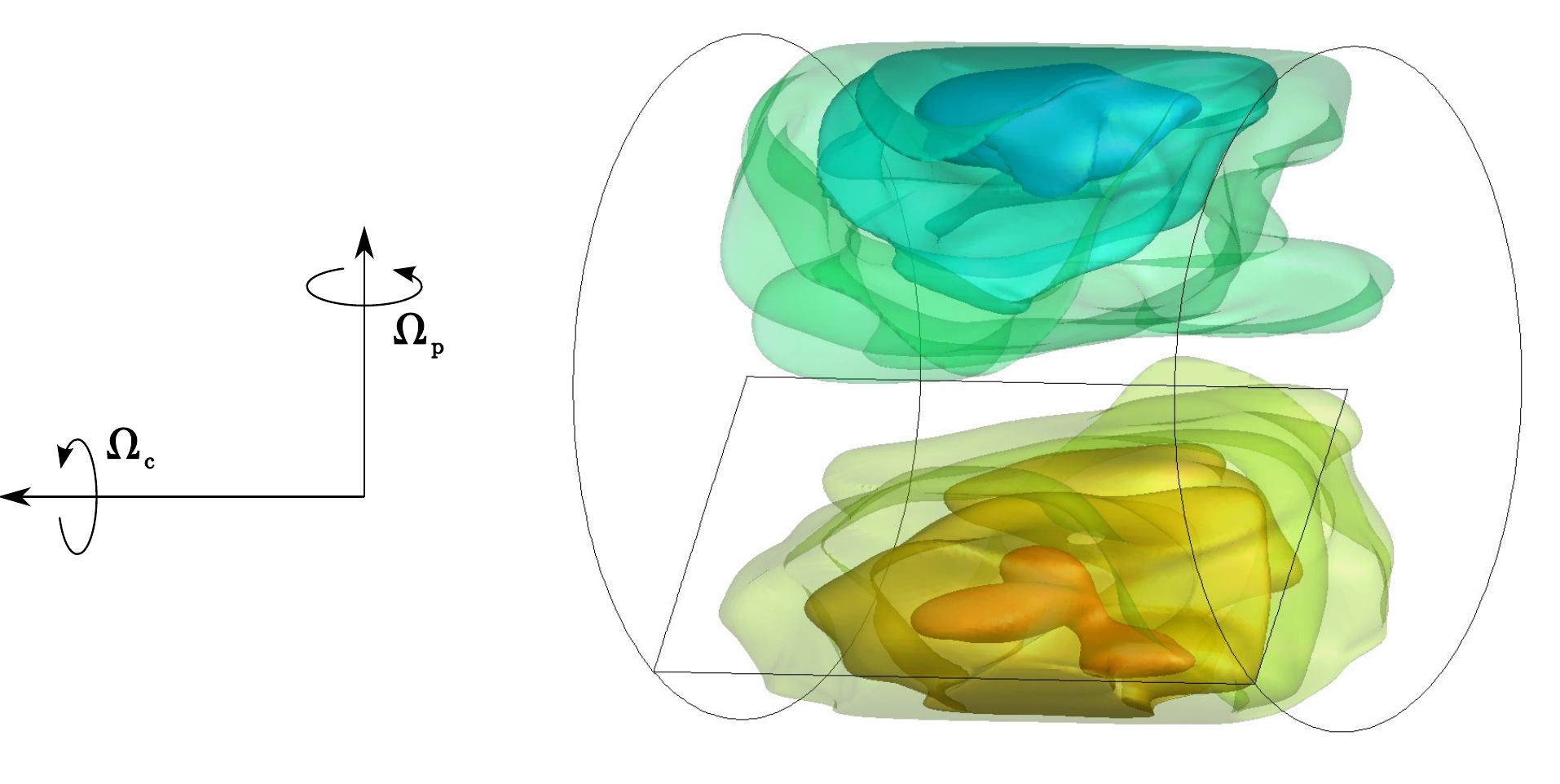}
\caption{Volume rendering of the axial velocity $u_z$ at ${\rm{Re}}=6500$
  and $\Gamma=2$.  The nested isosurfaces show $u_z$ at
  30\% (60\%, 90\%) of its maximum value. An animation of the
  flow's behavior in the statistically stationary state is available at
  {\tt{https://www.hzdr.de/db/VideoDl?pOid=45098}}.}\label{fig::re6500_vz_tot}  
\end{center}
\end{figure}

The simulations are started from an initial state with pure solid-body
rotation $\vec{u}=r\widehat{\vec{e}}_{\varphi}$. After switching on
the Coriolis force at $t=0$ the fluid shows a direct response in form
of a large scale mode with $m=1$. In dependence of the aspect ratio
this is only a transient state until after a finite time (which may
take up to 1000 rotation periods or more) a quasi-steady state is
reached. The typical flow pattern for $\Gamma=2$ and ${\rm{Re}}=6500$
is presented in figure~\ref{fig::re6500_vz_tot} where the nested
isosurfaces show a snapshot of the axial velocity $u_z$ at 30\% (60\%,
90\%) of its maximum value.
The flow is clearly dominated by a velocity mode with $m=1$ and
$k=1$. This is the primary Kelvin mode driven by the precessional
forcing. Superimposed on the $m=1$ mode
we see some non-axisymmetric (time-dependent)
disturbances that will be analyzed below.

\subsection{Kinetic energy}
\subsubsection{Forced mode}

\begin{figure}[b!]
\begin{center}
\includegraphics[width=10.0cm]{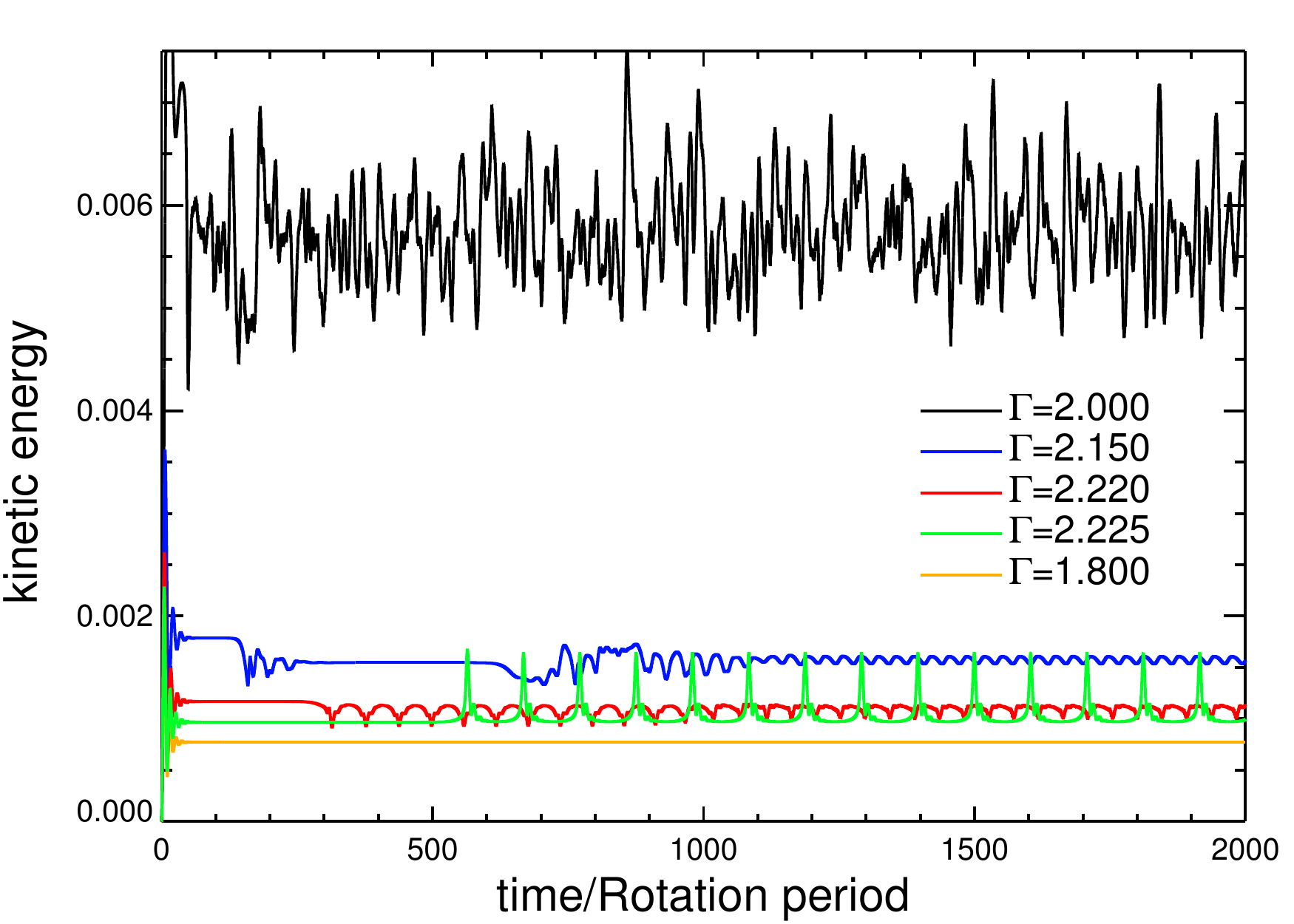}
\caption{Kinetic energy of the $m=1$ mode versus time for various
  $\Gamma$. Note the qualitative diversity for 
  different aspect ratios.}\label{fig::energy_m1_vs_time}
\end{center}
\end{figure}

In the following we discuss the behavior of the flow by means of
azimuthal Fourier modes $\vec{\widetilde{u}}_m(r,z)$ with the total
flow given by
$\vec{u}(r,z,\varphi)=
{2\pi}/({N_{\varphi}}+1)
\sum_{m=0}^{N_{\varphi}}\widetilde{\vec{u}}_m(r,z)e^{im\varphi}$.  
Figure~\ref{fig::energy_m1_vs_time} shows the history of the kinetic
energy of the $m=1$ mode for different aspect ratios.
A final state arises either after a short transitional period
($\Gamma=2.000$) or after one or more bifurcations (see e.g. the blue
curve for $\Gamma=2.150$). The final state can be chaotic
($\Gamma=2.000$, black curve), oscillatory ($\Gamma=2.15$, blue),
quasi-periodic with collapses ($\Gamma=2.2$, red), quasi-periodic with
bursts ($\Gamma=2.225$, green) or stationary ($\Gamma=1.8$,
yellow). We also find transient periods with constant energy
(e.g. $\Gamma=2.15$ for $t=300...600$), and it may take up to 1200
revolutions of the cylinder ($\Gamma=2.15$, blue curve) till a final
quasi-stationary state is reached. This saturation time also depends
on the Reynolds number and decreases with increasing ${\rm{Re}}$.

\begin{figure}[h!]
\begin{center}
\vspace*{-5.0cm}
\includegraphics[width=17.25cm]{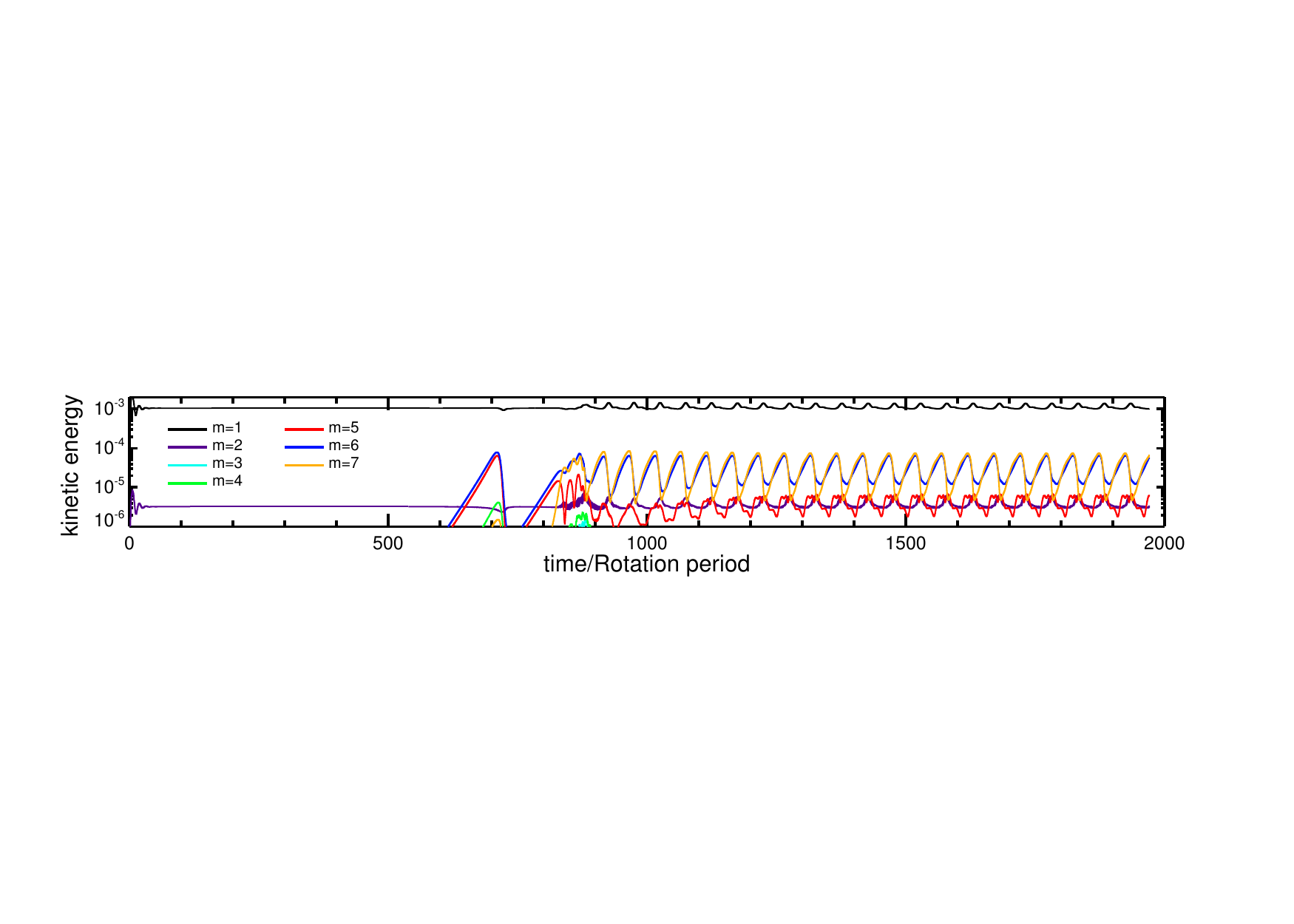}\\[-9.6cm]
\includegraphics[width=17.25cm]{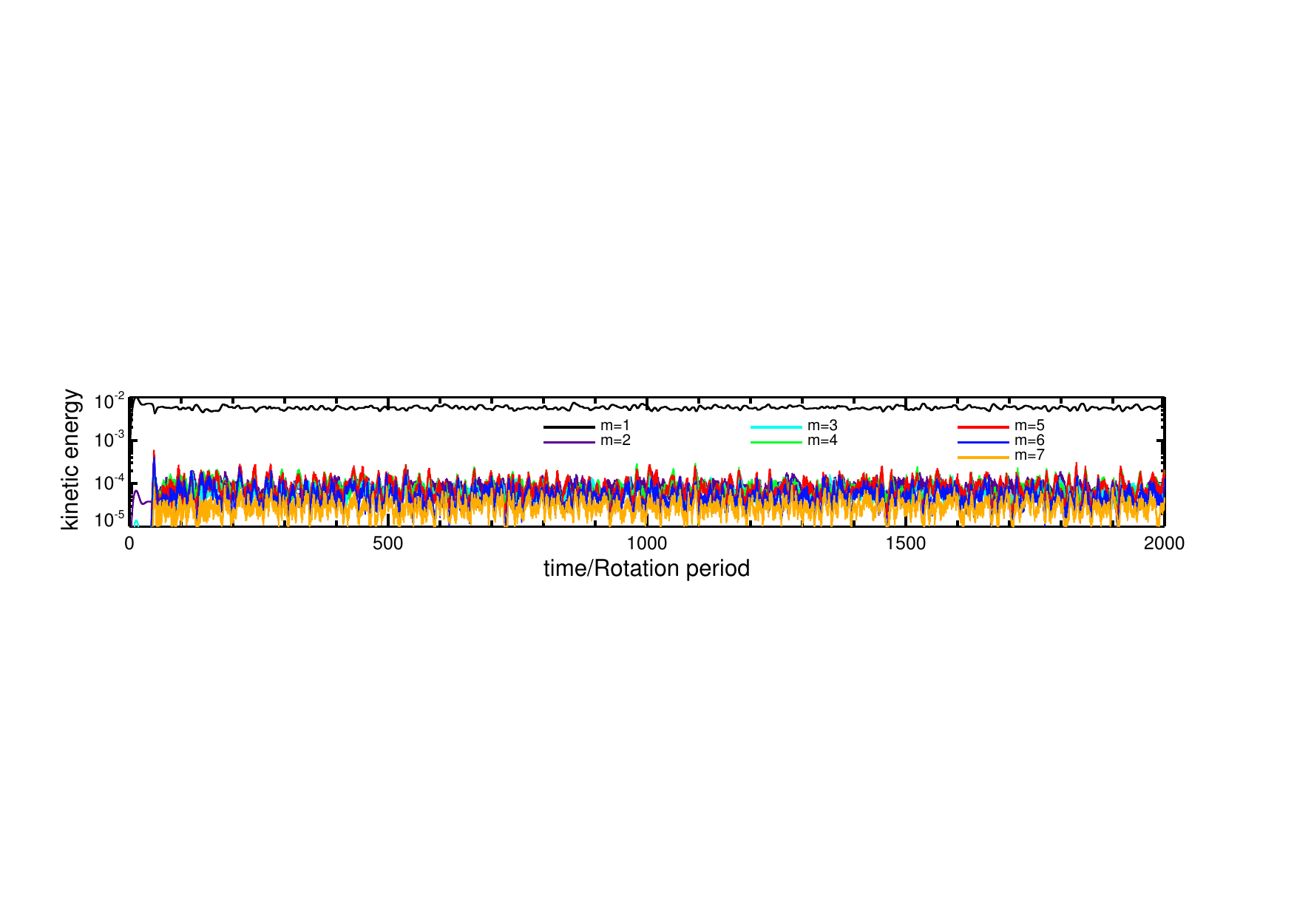}\\[-9.6cm]
\includegraphics[width=17.25cm]{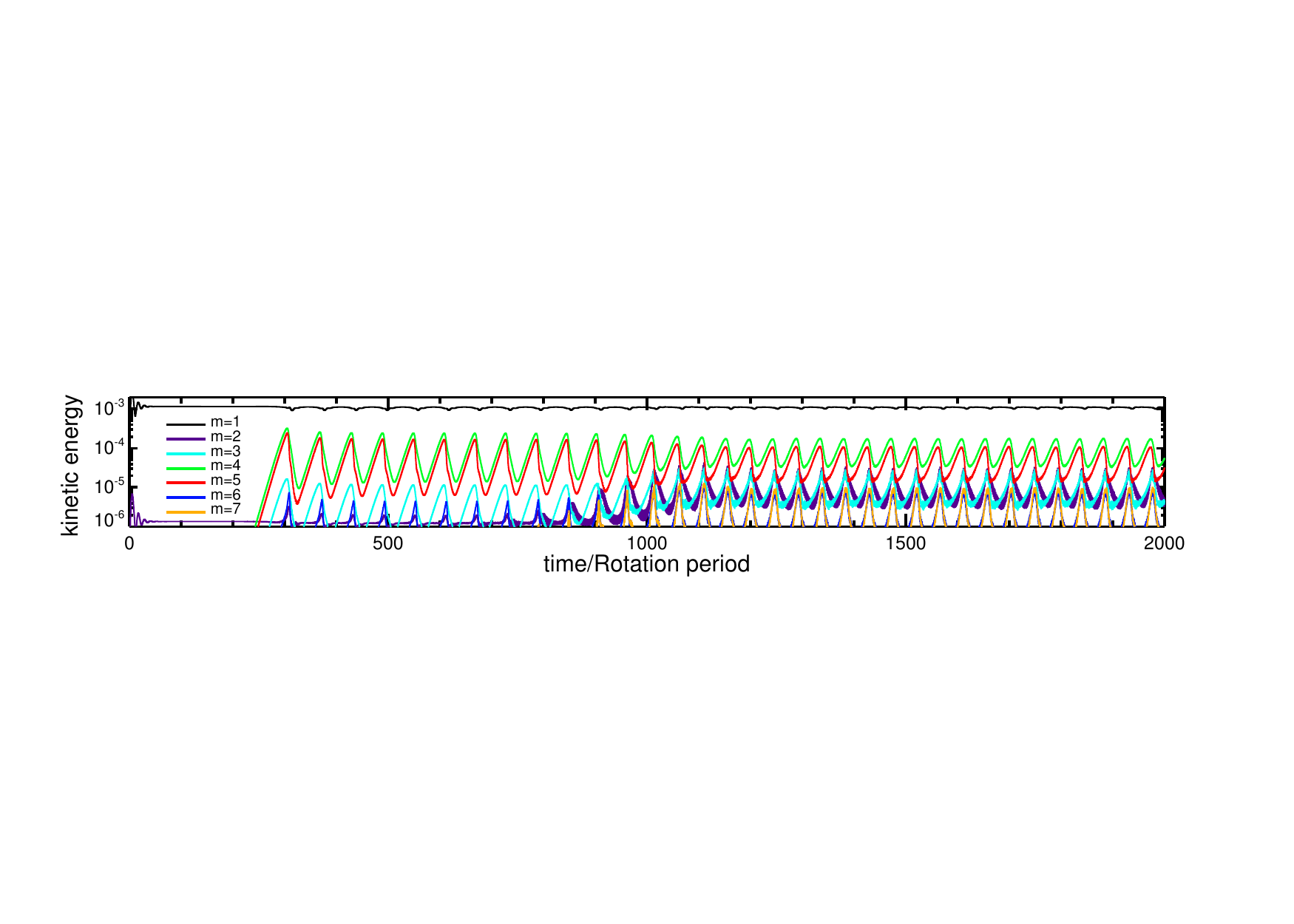}
\vspace*{-5.5cm}
\caption{Temporal behavior of the modes with $m=1,\dots ,7$. From top to
  bottom: $\Gamma=1.825, 2.000, 2.200$. Note the different vertical scale for
  $\Gamma=2.000$ (central panel).}\label{fig::ener_vs_time_mm}
\end{center}
\end{figure}

The bifurcations shown by the $m=1$ mode go along with the emergence
of non-axisymmetric modes with $m\geq 2$
(figure~\ref{fig::ener_vs_time_mm}). These modes become unstable only
after quite a long time, e.g., for $\Gamma=1.825$ it takes up to $600$
rotation periods of the cylinder till modes with $m \ge 2$ start to
grow, and it may take up to twice that time till a quasi-steady state
is reached.  The non-axisymmetric modes with $m \geq 2$ essentially
proceed similar to the behavior of the $m=1$ mode, i.e., we see a
chaotic behavior when the $m=1$ mode behaves chaotic ($\Gamma=2.0$,
figure~\ref{fig::ener_vs_time_mm}, central panel) and we see periodic
growth and decay when the $m=1$ mode is periodic ($\Gamma=1.825$, top
panel in figure \ref{fig::ener_vs_time_mm} and $\Gamma=2.200$, bottom
panel in figure~\ref{fig::ener_vs_time_mm}). In the periodic regime
the amplitude of the energy oscillations of the modes with $m \geq 2$
is rather large so that the kinetic energy may vary in time by more
than one order of magnitude.

In the following we compare analytic expressions for the energy of the
forced mode with the time-averaged energy taken from the
quasi-stationary period of the final state in our simulations. This
comparison is only partly justified because the analytical expressions
are based on a number of serious simplifications, such as the
linearization of the Navier-Stokes equation, the limit
${\rm{Po}}\rightarrow 0$, and the neglect of the time dependent part
of the Coriolis force which is justified only for small precession
angles $\alpha$. Furthermore, the final state in our simulations is
instable and additional caution is advised when comparing simulations
with a linear time independent solution. Nevertheless, such an
analysis is helpful to identify the regimes in which, e.g., a maximum
response of the flow can be expected, or to localize the aspect ratios at
which triadic resonances are possible.

\begin{figure}[t!]
\begin{center}
\includegraphics[width=11cm]{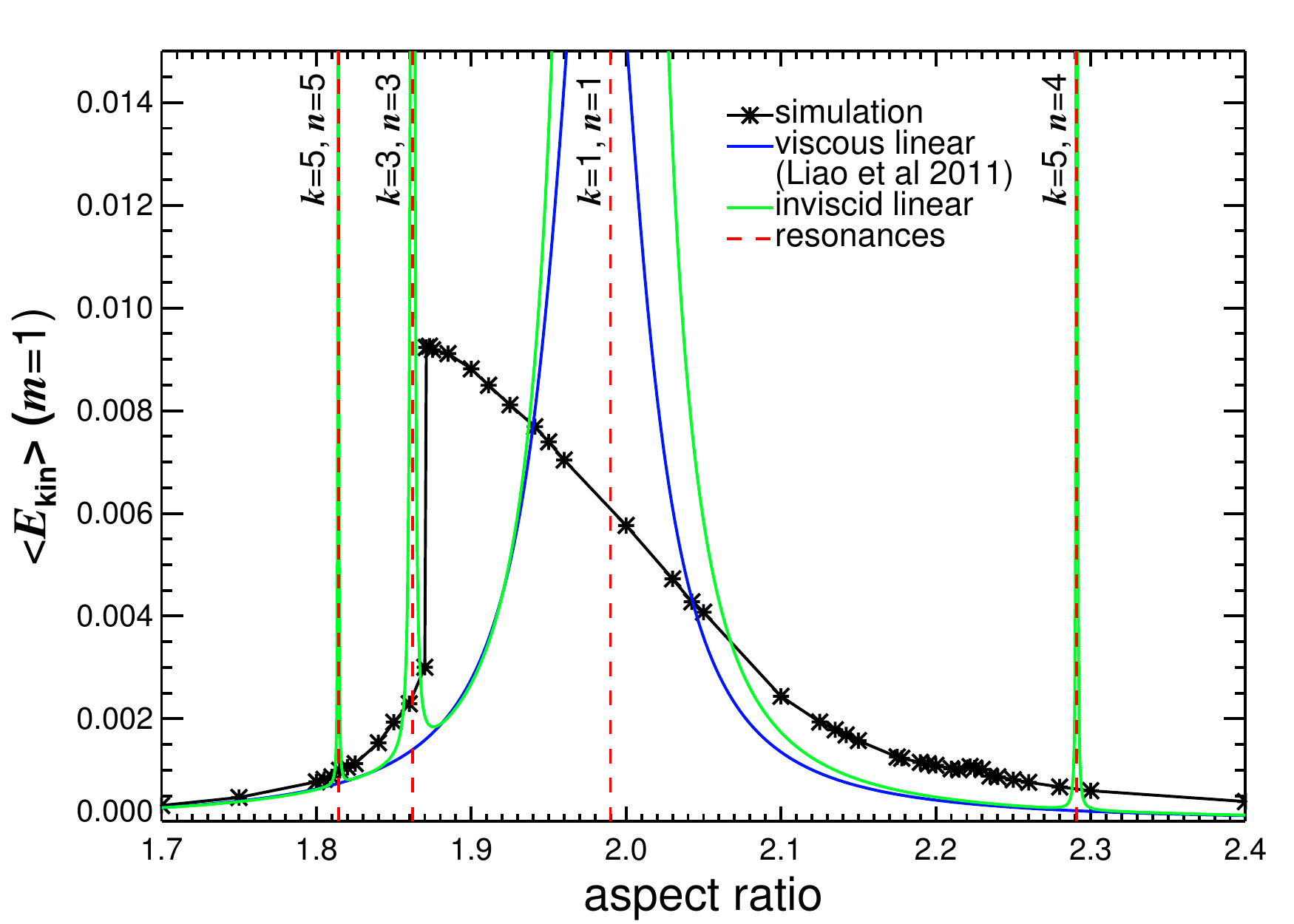}
\caption{Time-averaged kinetic energy of the fundamental mode ($m=1$)
  versus aspect ratio. 
  ${\rm{Re}}=6500, {\rm{Po}}=0.014,
  \alpha=\pi/2$. The dashed red lines denote the resonances with larger
  $k$ and/or $n$. For the computation of the energy,
  equation~(\ref{eq::energym1_inviscid}) has been truncated at $k=6$
  and $n=6$ (with prograde and retrograde modes respectively). Note
  that the inviscid solution (green curve) diverges at the resonances
  whereas the viscous solution has a finite maximum of
  $E_{\rm{kin}}^{\rm{max}}\approx 0.021$ at $\Gamma\approx 1.9814$.
}\label{fig::energy_m1_vs_ar} 
\end{center}
\end{figure}

Analytically, the energy of the forced Kelvin mode is given by \cite{2012JFM...709..610L} 
\begin{equation}
E_{m=1}=\Gamma^2\sum\limits_{j}\frac{(k\pi/\Gamma)^2
+\frac{1}{2}(2-\Omega_{j})}{\Omega_{j}^2(4-\Omega_{j}^2)}|A_{j}|^2
J_1^2(\lambda_{j})\label{eq::energym1_inviscid}  
\end{equation}
where the index $j$ now denotes a
combination of the axial wave number $k$ and the radial wave number
$n$, and $A_{j}$ is the inviscid linear amplitude given
by~(\ref{eq::amp_invisc}) with the azimuthal wave number fixed at
$m=1$. $A_j$ can also be calculated by the method
of~\citeasnoun{2012JFM...709..610L} which includes viscosity and is
valid at and off the resonance. However, in that case
equation~(\ref{eq::energym1_inviscid}) is accurate only up to an error
of $O({\rm{Re}}^{-\frac{1}{2}})$ since
equation~(\ref{eq::energym1_inviscid}) does not consider the energy
from the flow in the boundary
layers. Figure~\ref{fig::energy_m1_vs_ar} shows the time-averaged
energy taken from the simulations (black curve) compared with the
analytic solutions (green curve, inviscid solution and blue curve,
viscous solution).  For the calculation of the analytical solutions
equation~(\ref{eq::energym1_inviscid}) is truncated at $n=6$ and
$k=6$ which is sufficient to reach convergence (off-resonance in the
inviscid case) and avoids the occurrence of too many resonances 
in the inviscid case (green curve) which anyway are suppressed in the more
realistic viscous or non-linear computations. 
Comparing the numerical solutions (black curve) with the linear
solutions we see significant deviations around the primary resonance
($k=1, n=1$, $\Gamma_{\!\!\rm{res}}\approx 1.9898$) while outside of
this regime the agreement of all curves remains good (say for $\Gamma
\la 1.85$ and for $\Gamma \ga 2.05$). Two key features characterize
the behavior of the mean energy in our simulations. On the one hand we
see a clear shift of the maximum to smaller aspect ratios and the
maximum energy arises rather close to the aspect ratio expected for
the resonance with $k=3$ and $n=3$. However, this resonance is
suppressed by viscous effects and we do not see strong contributions
with $k=3$ and/or $n=3$ in our simulations (see also
section~\ref{sec::451}). Therefore the correlation is either
coincidental or we have to assume an indirect impact of the
($k=3,n=3$) resonance. The second scenario is supported by the abrupt
transition from the absolute maximum to a rather low energy state
right below $\Gamma\approx 1.871$. This jump is also connected to a
modification of the character of the $m=1$ mode which changes its
behavior below $\Gamma\approx 1.86$ from the chaotic regime to the
periodic regime. Abrupt transitions of the energy of higher azimuthal
modes can also be seen at the other (theoretical) resonances of the
forced mode ($k=5, n=5$ and, a little further away $k=5, n=4$, see
figure~\ref{fig::energy_m1_vs_ar}) which will be revisited in the next
paragraph. Beside the shift of the maximum to smaller aspect ratios
the energy obtained in the simulations remains considerably smaller
compared to the linear viscous solution. The maximum of the energy
based on the amplitude of \citeasnoun{2012JFM...709..610L} is
$E_{\rm{kin}}^{\rm{max}}\approx 0.021$ (at $\Gamma\approx 1.9814$)
which is roughly two times larger than the maximum kinetic energy
obtained in our simulations ($E_{\rm{kin}}^{\rm{max}}\approx 0.0093$
at $\Gamma=1.871$). The differences are not surprising because the
prerequisites for the computation of the amplitudes $A_j$ are not well
met in our simulations. In particular, we observe a significant impact
of non-linear interactions in the simulations in terms of higher
non-axisymmetric modes and the forming of an azimuthal shear flow that
both draw energy from the forced $m=1$ mode.

\subsubsection{Higher azimuthal wave numbers}

\begin{figure}[h!]
\begin{center}
\includegraphics[width=11cm]{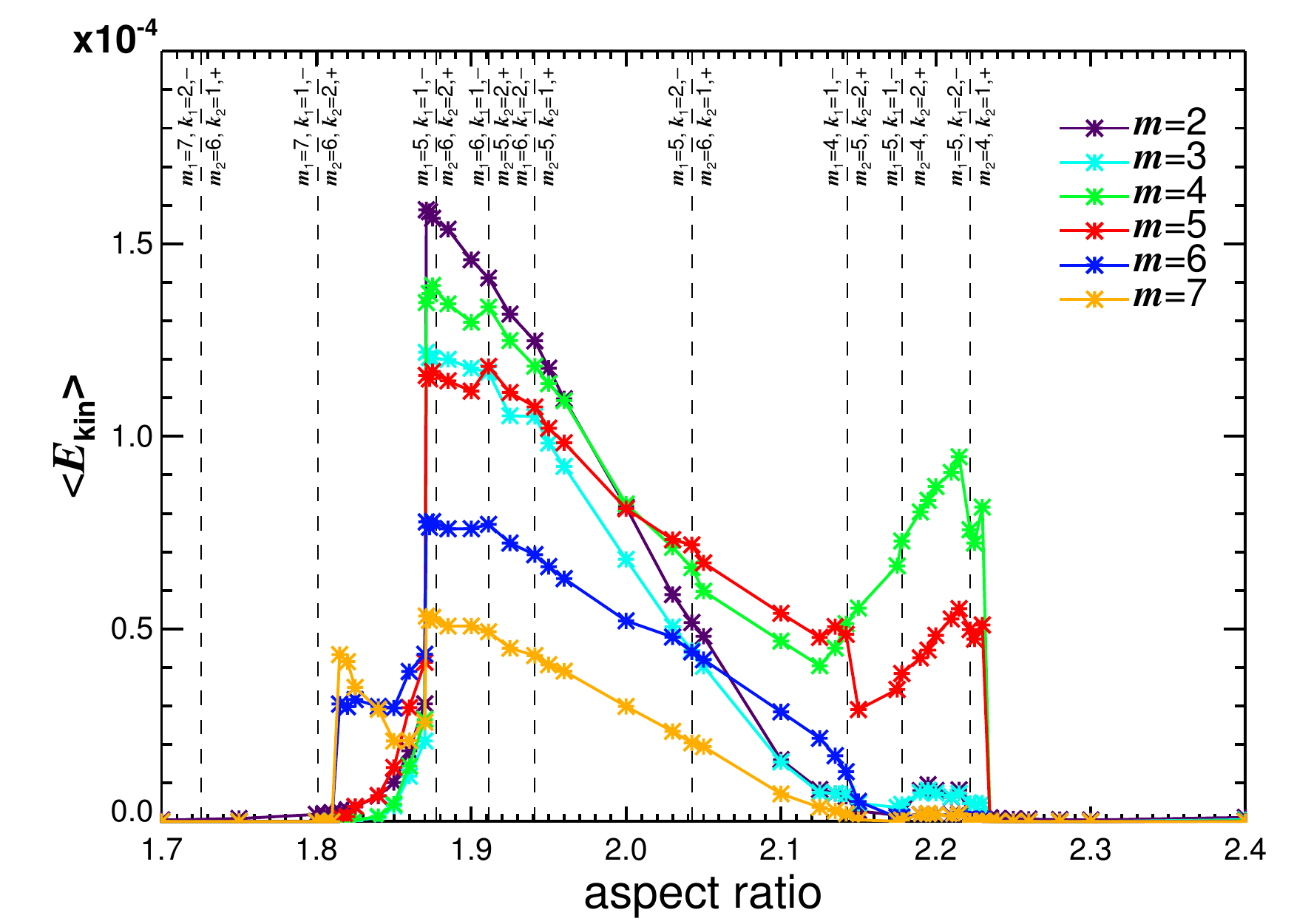}
\caption{Time-averaged kinetic energy of azimuthal
  modes with $m\geq 2$. ${\rm{Re}}=6500, {\rm{Po}}=0.014,
  \alpha=\pi/2$. The vertical lines indicate the aspect ratios
  for which triadic resonances could be expected according to the 
  dispersion relation~(\ref{eq::dispersion}). The labels denote the
  corresponding azimuthal and 
  axial wave numbers with $-$ ($+$) marking retrograde (prograde)
  modes.}\label{fig::energy_mm_vs_ar}  
\end{center}
\end{figure}

We only find higher non-axisymmetric modes with a noteworthy amount of
energy between $\Gamma\approx 1.81$ and $\Gamma\approx 2.24$. This
corresponds approximately to the regime in which the energy of the
$m=1$ mode deviates from the linear prediction (see previous
paragraph). For the sake of brevity we will call this regime the
non-linear regime, assuming that the higher modes are triggered by
non-linear effects. The kinetic energy of the modes with $m\geq 2$ is
roughly 2 orders of magnitude smaller than the energy of the forced
mode but essential characteristics such as the maximum at
$\Gamma=1.871$ and the abrupt drop of kinetic energy below this
threshold are also reflected in the behavior of the higher
non-axisymmetric modes (figure~\ref{fig::energy_mm_vs_ar}).
The sudden drop of the energy for $\Gamma\leq 1.871$ pertains to all
modes which indicates that this breakdown might be a global flow
property that is not connected to an individual mode.

A similar steep decrease of energy can also be found on the two outer
edges of the non-linear regime. It is striking that all the abrupt
transitions are close to theoretical resonances of the forced mode 
($k=3, n=3 \Rightarrow \Gamma_{\rm{res}} \approx 1.8621, 
k=5, n=5 \Rightarrow \Gamma_{\rm{res}} \approx 1.8142$, 
and with some greater distance 
$k=5, n=4 \Rightarrow \Gamma_{\rm{res}} \approx 2.2911$ , see
figure~\ref{fig::energy_m1_vs_ar}). These resonances are, however, 
suppressed and cannot be directly observed in the simulations. So
far, we cannot conclude whether the correlation of the energy jumps
and the occurrence of (theoretical) resonances is a coincidence, or
whether they may be some systematic relationship.

Away from the primary resonance of the forced mode we find two further
regimes with local maxima of the higher modes. For $1.80 \la \Gamma
\la 1.85$ we see a narrow window with the maximum around
$\Gamma\approx 1.825$ in which the $m=6$ and $m=7$ mode dominate (blue
and yellow curve in figure~\ref{fig::energy_mm_vs_ar}). Likewise, but
more explicit, we find the $m=4$ and $m=5$ mode becoming dominant for
$2.05 \la \Gamma\la 2.25$ with a maximum around $\Gamma\approx 2.2$
(green and red curves in figure~\ref{fig::energy_mm_vs_ar}).

The local maxima around $\Gamma \approx 1.825$ and $\Gamma \approx
2.2$ for Fourier modes with $\delta m =1$ indicate the presence of
corresponding triadic resonances and roughly agree with predictions
from the dispersion relation (see table~\ref{tab::dispersion} and
figure~\ref{fig::dispersion}). From the linear predictions we would
also expect triads with $m=5$ and $m=6$ in the intermediate regime
(for $1.85 < \Gamma < 2.05$) but obviously there are other flow
contributions with at least comparable energy which may disguise the
energetic signature of the involved free Kelvin modes (e.g., we see
relatively high energies of the $m=4$ and the $m=5$ modes, whereas in
comparison the $m=6$ mode is clearly suppressed).

\subsection{Spatial Structure of the Fourier modes}

\begin{figure}[b!]
\includegraphics[width=15.50cm]{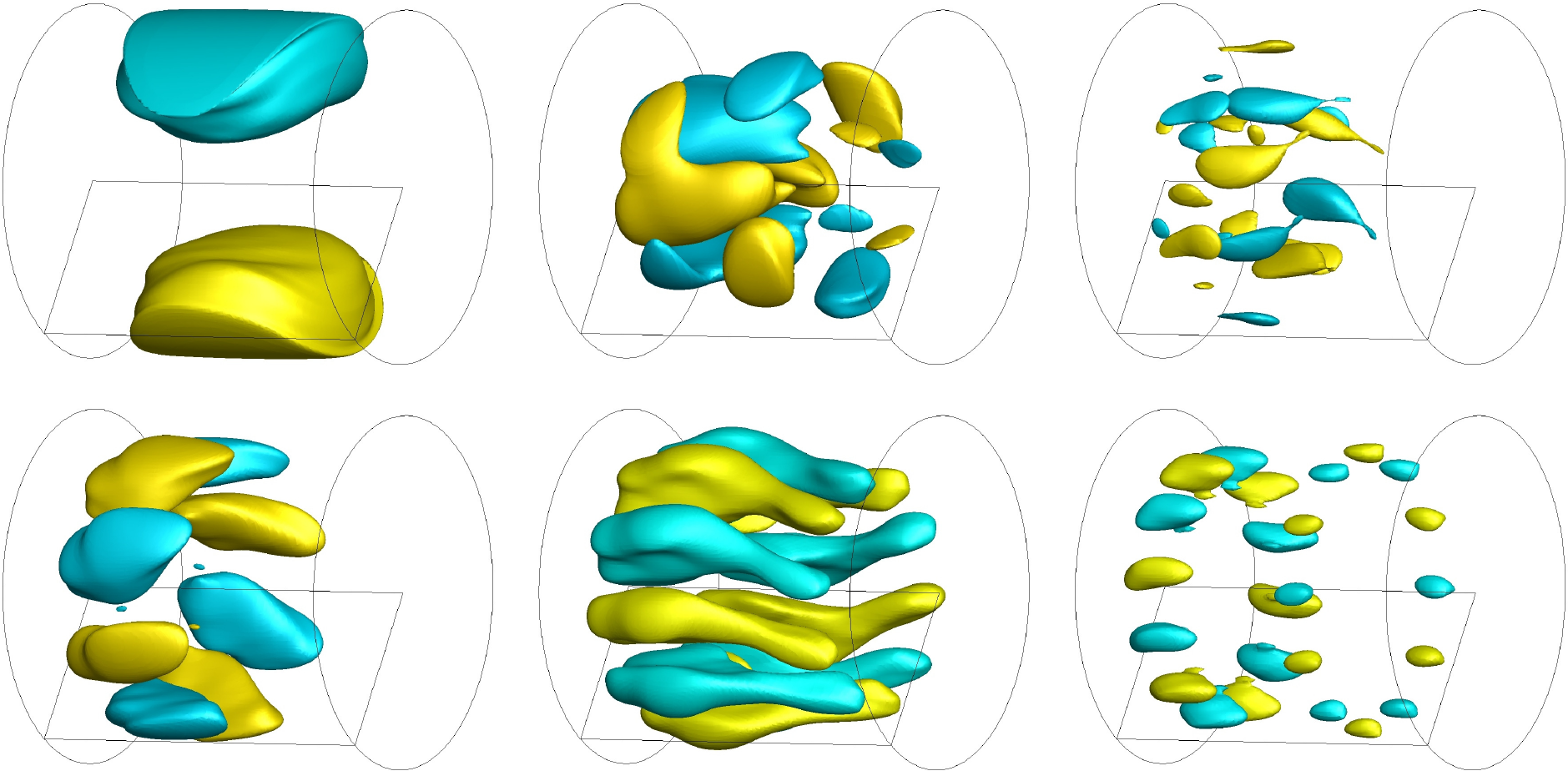}
\caption{Snapshot of the axial velocity field $u_z$ filtered to
  $m=1\dots 6$ (upper left to lower right) for
  $\Gamma=2$ (isosurface at 50\% of the respective maximum
  value). 
  An animation of the flow
  pattern is available at {\tt{https://www.hzdr.de/db/VideoDl?pOid=45104}}.
  The temporal behavior of the modes $m=2$ and $m=3$ is
  complex and chaotic. The modes $m=4, 5$ and $6$ show a more regular
  behavior essentially resulting from the superimposition of free
  Kelvin modes with 
  $k=1$ and $k=2$. 
  ${\rm{Re}}=6500, {\rm{Po}}=0.014, \alpha=\pi/2$.
}\label{fig::flow_structure01} 
\end{figure}
\begin{figure}[b!]
\includegraphics[width=5.10cm]{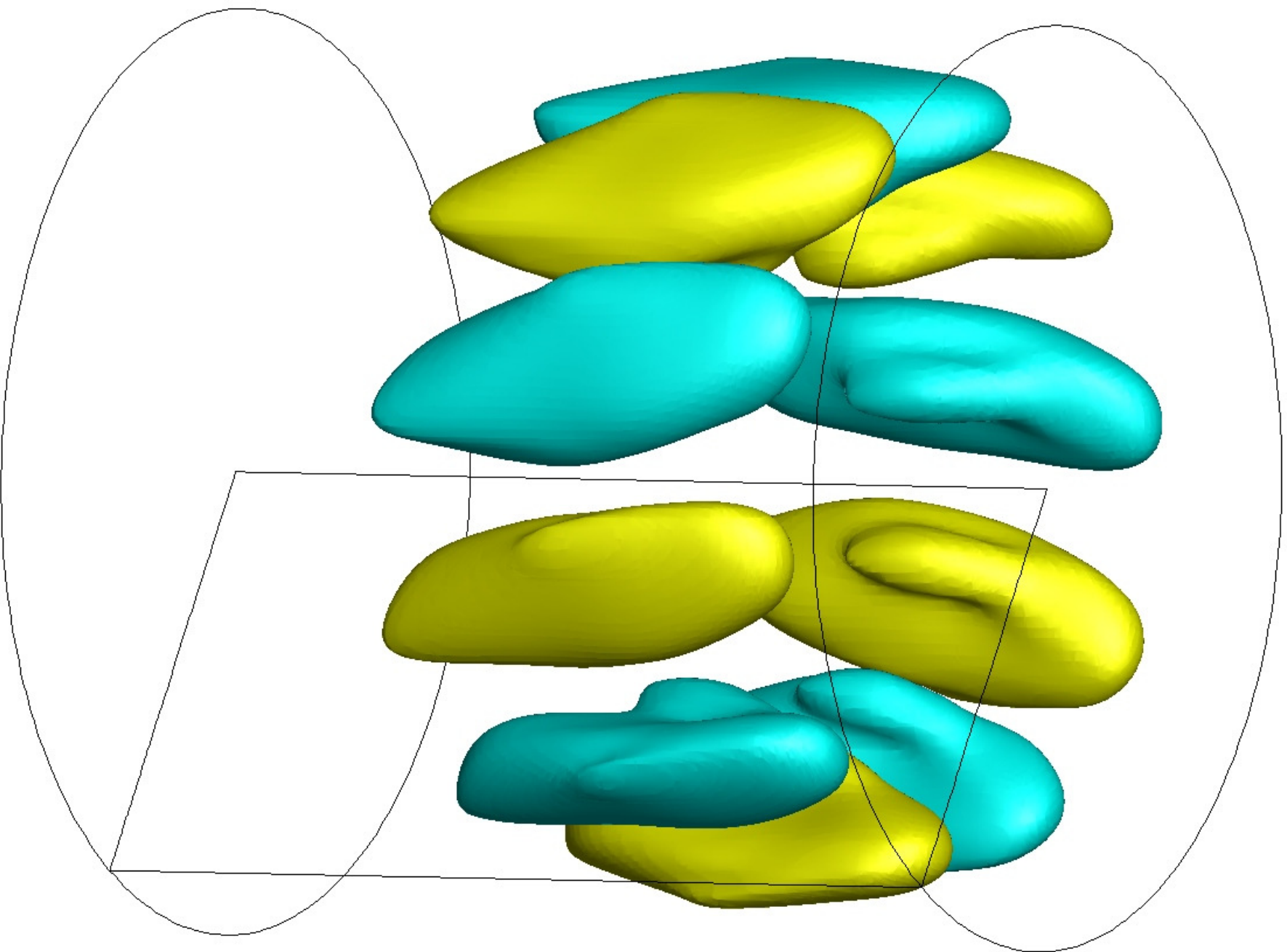}
\includegraphics[width=5.10cm]{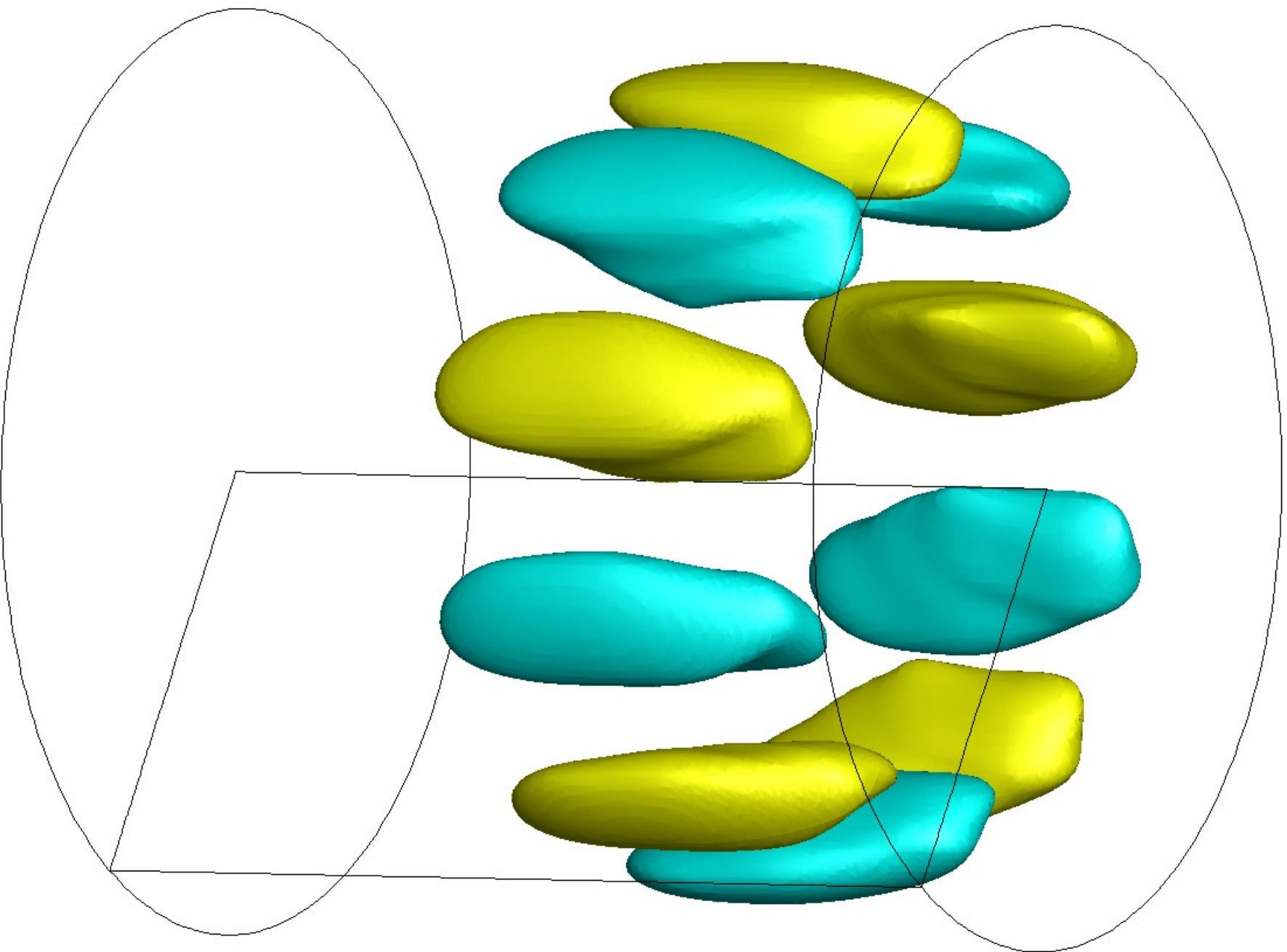}
\includegraphics[width=5.10cm]{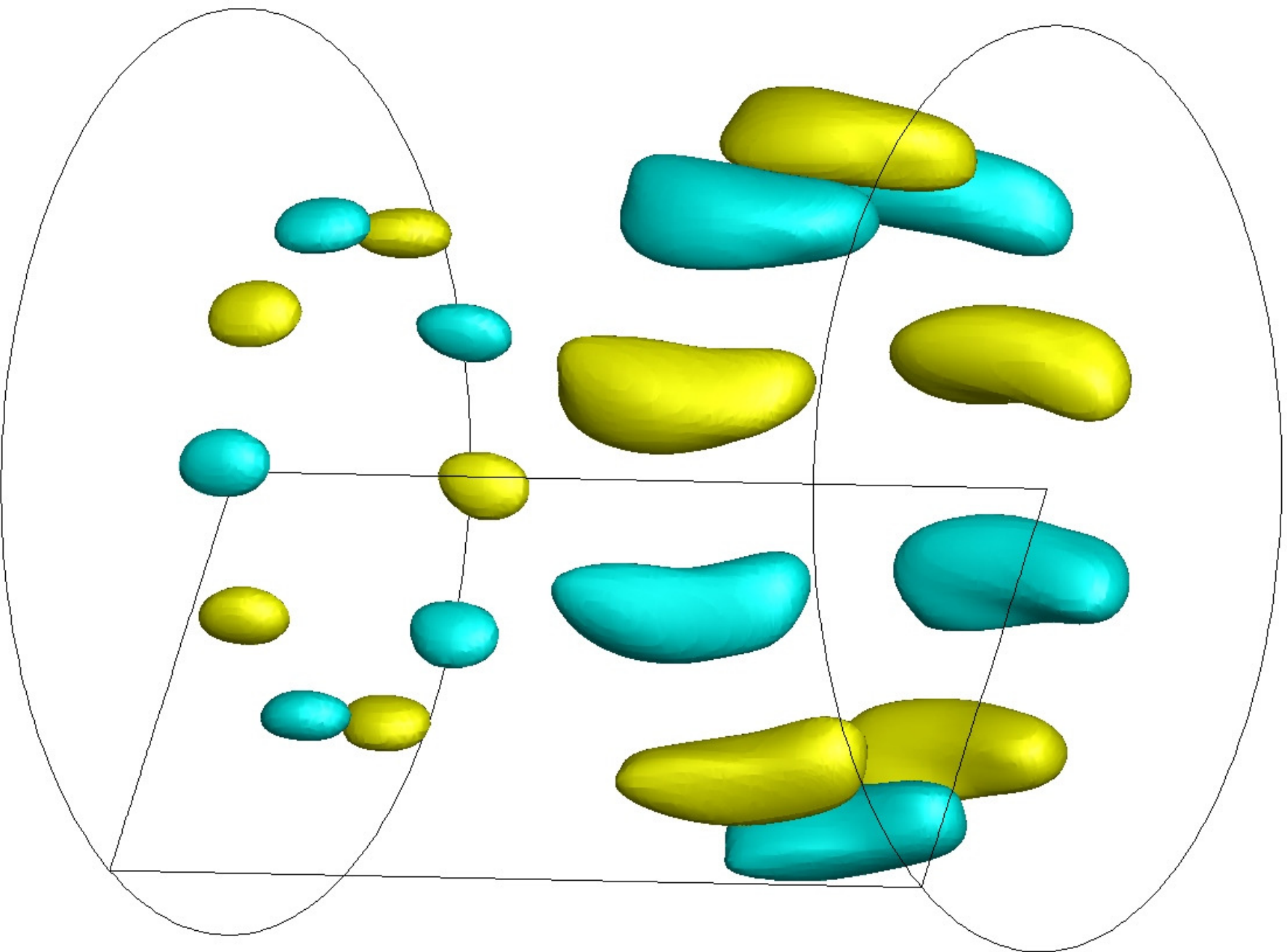}\\
\includegraphics[width=5.10cm]{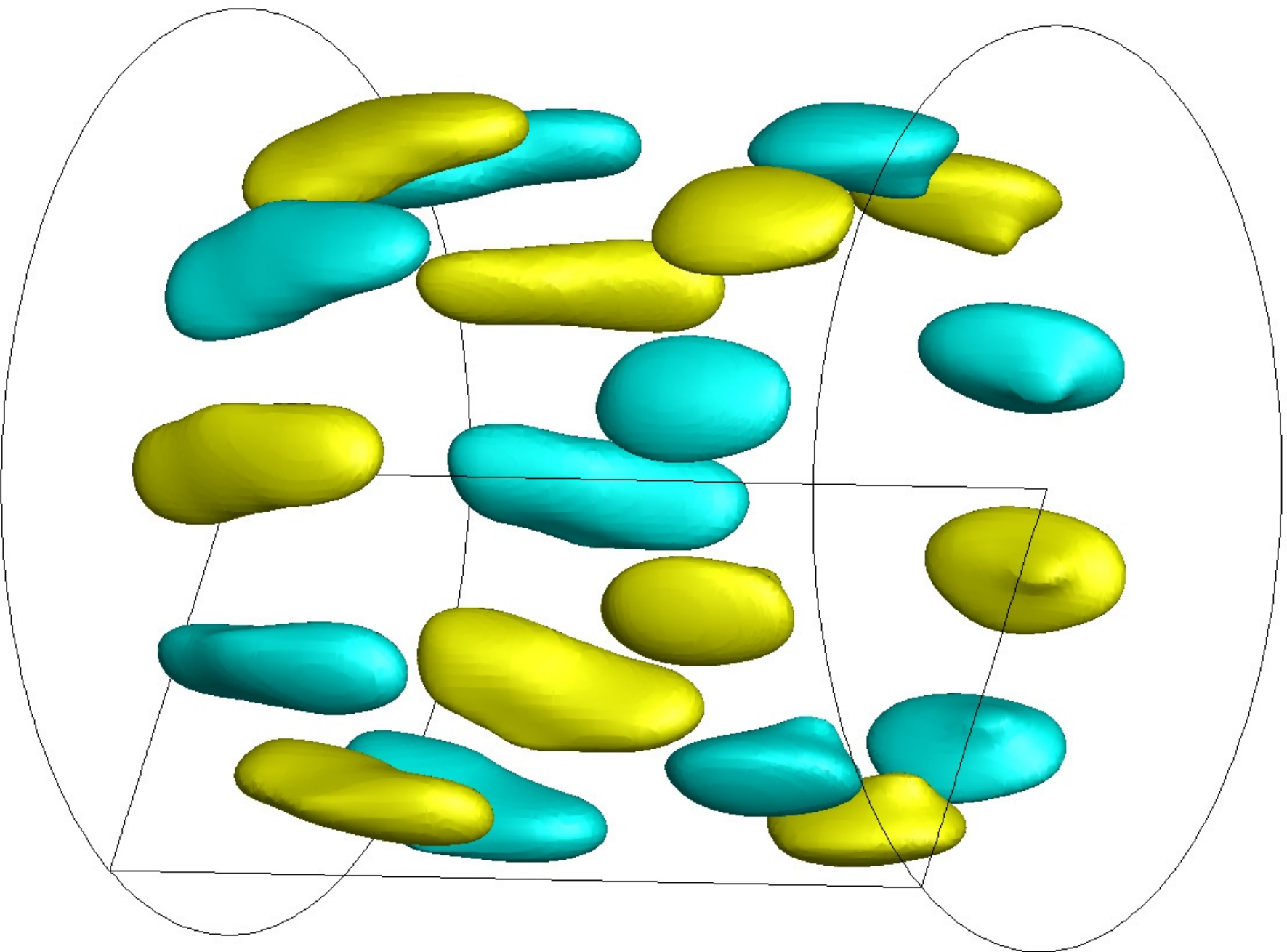}
\includegraphics[width=5.10cm]{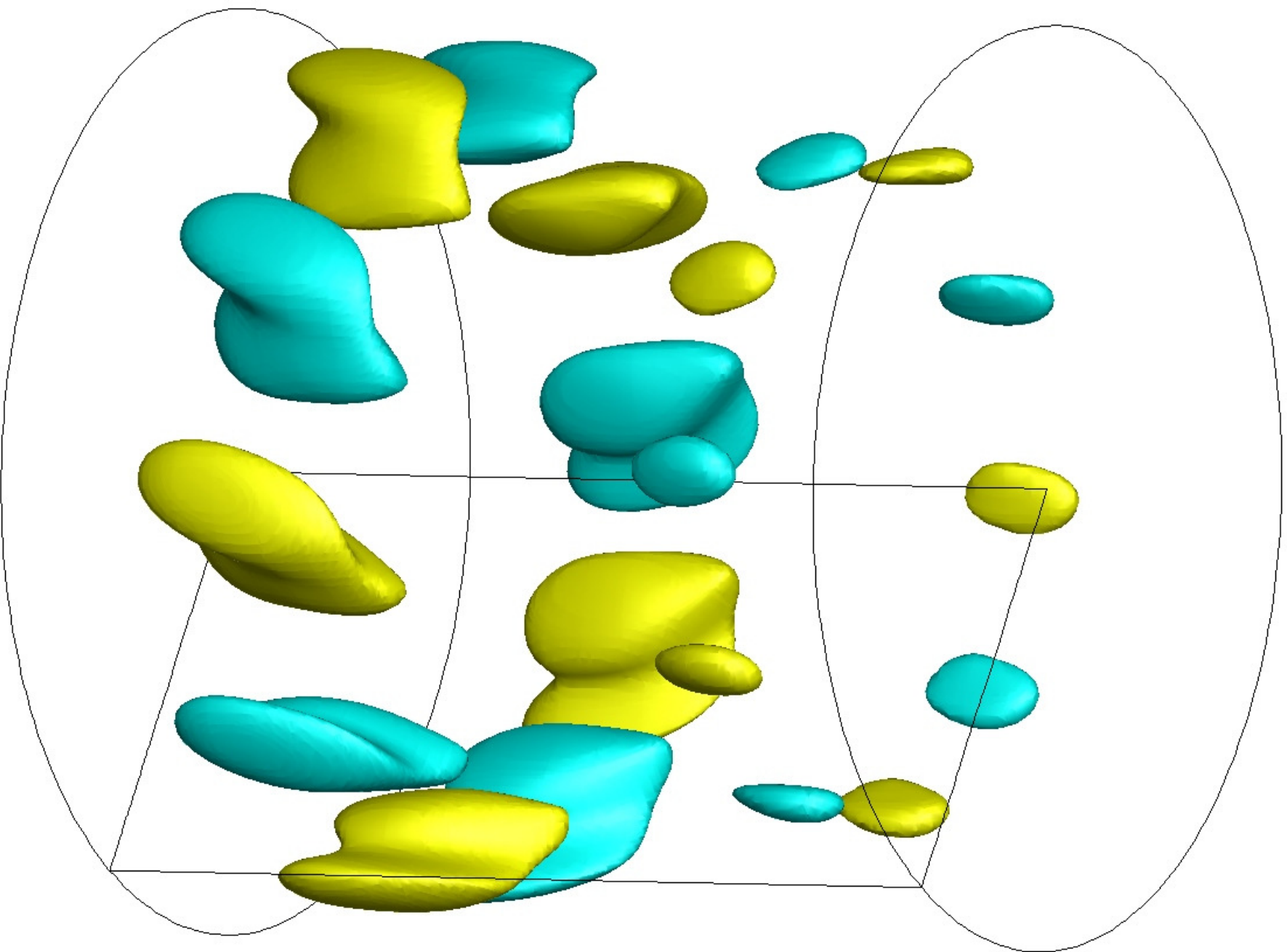}
\includegraphics[width=5.10cm]{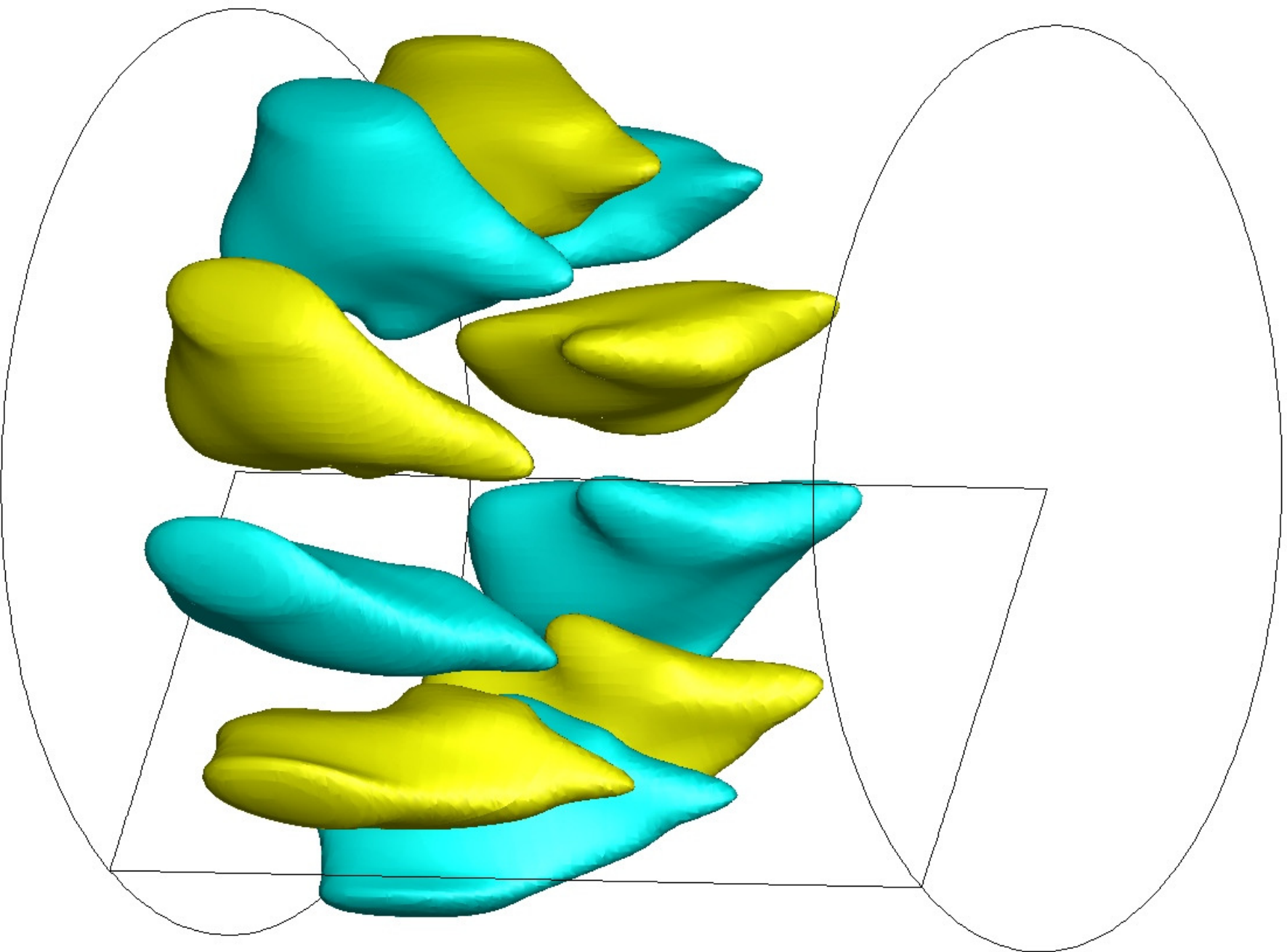}
\caption{Time series of the $m=5$ mode that visualizes the axial
  symmetry breaking. The isosurfaces show 
  $\widetilde{u}_{m=5, z}(r,\varphi,z)$ at 
  50\% of its maximum value at various characteristic time steps (form upper left to
  lower right: t=2617.5418, 2617.8489, 2618.0845, 2618.1958,
  2618.8738, 2618.9599 (time in units of the rotation time); note that the time gaps are not equidistant).  
}\label{fig::timeseries_m5} 
\end{figure}

 
A qualitative impression of the flow structure is provided by
isosurfaces of the individual Fourier modes of the axial velocity
shown in figure~\ref{fig::flow_structure01} for $\Gamma=2$
(for an impression of the temporal behavior see the snapshots of
the $m=5$ mode in figure~\ref{fig::timeseries_m5} and the movie at
{\tt{https://www.hzdr.de/db/VideoDl?pOid=45104}}).  
The $m=1$ mode is obviously dominated by a $k=1$ contribution and
shows comparatively small variations in time. Regarding the higher
non-axisymmetric modes we see a complex pattern for the $m=2$ and
$m=3$ mode with irregular temporal and spatial fluctuations and a more
regular behavior of the modes $m=4, 5$ and $6$ with typical signatures
of $k=1$ and $k=2$. It is striking that the $m=6$ mode, which should
become resonant around $\Gamma\approx 2$, remains weak compared to the
$m=4$ or the $m=5$ mode.

The amplitude of the individual contributions with $m=4,5,6$ varies in
time and the resulting flow perpetually changes its axial
structure. The azimuthal mode resulting from the superposition of $k =
1$ and $k = 2$ contributions has a distinct and time dependent
asymmetry with respect to the equatorial plane of the cylinder, so
that the associated flow is concentrated alternately in both halves of
the cylindrical container (see time series of the $m=5$ mode in
figure~\ref{fig::timeseries_m5}).
The typical timescale for this process is of the order of the rotation
period $2\pi/\Omega_{\rm{c}}$ but we can not derive a reliable
periodicity from our simulations.

\begin{figure}[h!]
\begin{center}
\includegraphics[width=15.5cm]{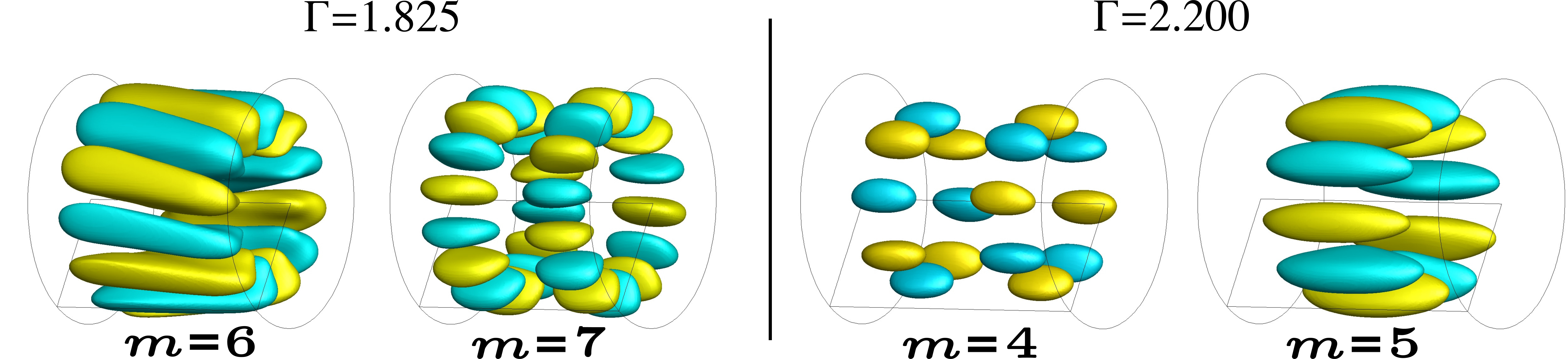}
\caption{Snapshots of the structure of the axial velocity field for
  different Fourier modes. Left:
  $\Gamma=1.825$, dominated by $m=6, k=1$ and $m=7, k=2$. Right:  
  $\Gamma=2.200$, dominated by $m=4, k=2$ and $m=5, k=1$. In both cases
  these modes represent the dominant contributions of the
  non-axisymmetric modes with $m\geq 2$ (see
  figure~\ref{fig::kelvin_amp_vs_time} below and animation at
  {\tt{https://www.hzdr.de/db/VideoDl?pOid=45105}}).}\label{fig::flow_structure02} 
\end{center}
\end{figure} 
Away from the primary resonance of the forced mode the structure of
higher azimuthal modes is much more regular and -- except for periodic
variations of amplitude -- essentially remains constant over time.
Figure~\ref{fig::flow_structure02} shows the dominant azimuthal modes
(beyond $m=1$) at $\Gamma=1.825$ ($m=6$ and $m=7$) and at
$\Gamma=2.200$ ($m=4$ and $m=5$).  In both cases, the axial behavior
of the modes shows clear indications of axial wave numbers $k = 1$
and/or $k=2$ and we will show below that these modes indeed fulfill
all conditions of equation~(\ref{eq::triades}) and thus
constitute triadic resonances. In contrast to the behavior at
$\Gamma=2$, the geometric structure of the dominant modes exhibits
only minor temporal variations except strong oscillations of the
amplitude (see animation at {\tt{https://www.hzdr.de/db/VideoDl?pOid=45105}}).

\subsection{Radial dependence}

\begin{figure}[t!]
  \begin{center}
    \includegraphics[width=15.5cm]{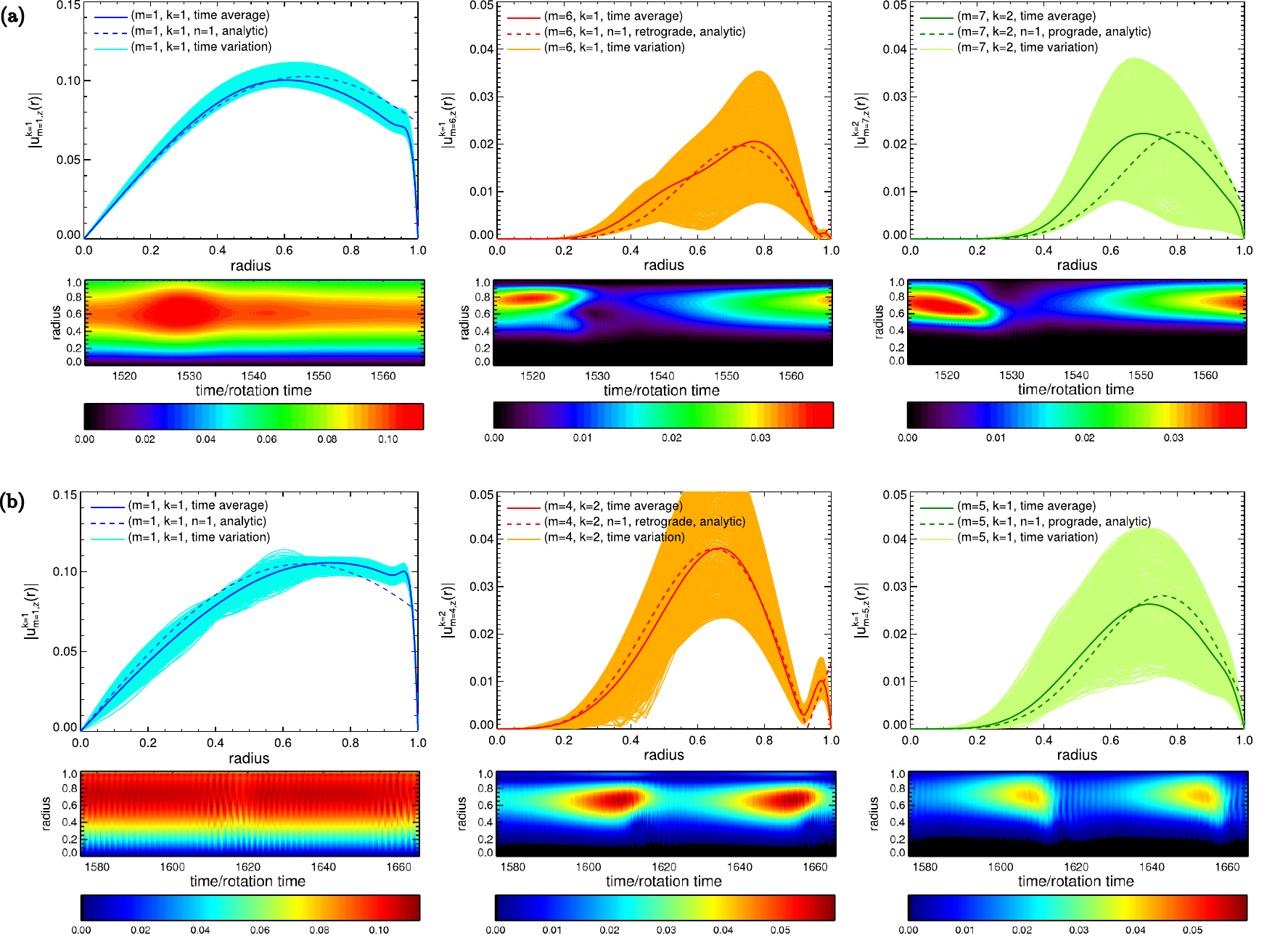}
    \caption{Radial dependence of the axial velocity decomposed into the
      dominant Fourier- and axial modes according to
      equation~(\ref{eq::axial_decomposition}). 
      (a) $\Gamma=1.825$, from left to right: ($m=1, k=1$), ($m=6, k=1$) and
      ($m=7, k=2$), 
      (b) $\Gamma=2.200$, from left to right: ($m=1, k=1$), ($m=4, k=2$) and
      ($m=5, k=1$).
      The solid thick curves show the temporal average and the weakly
      saturated colored curves show radial profiles at different time
      steps representing the temporal variations.
      The thick dashed curves denote the theoretical profiles from the
      linear inviscid solutions ($\propto J_m(\lambda_j r)$, see
      equation~(\ref{eq::kelvinmodes})).    
      The colored contour plots below each profile show 
      the corresponding variation of the radial profiles in time. 
      \label{fig::axialdecomp}}
  \end{center}
\end{figure}
\begin{figure}[t!]
\begin{center}
\includegraphics[width=14.5cm]{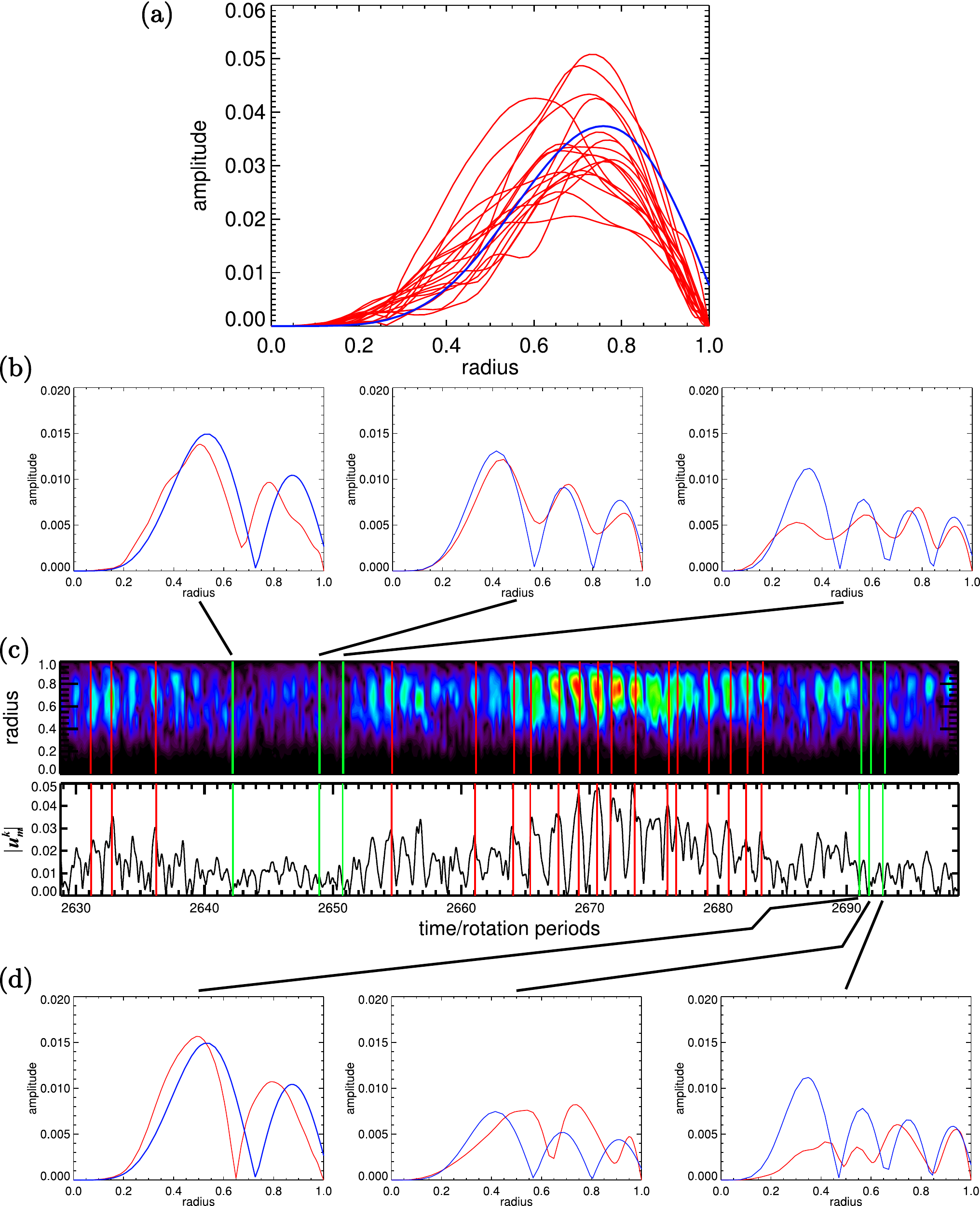}
\caption{Snapshots of radial profiles  $\widehat{{u}}_{m,z}^k(r)$
  with $m=5$ and $k=1$ at $\Gamma=2$. 
  (a) snapshots taken at local maxima of the ($m=5, k=1$) mode
  dominated by $n=1$ (marked by the the vertical red lines in panel (c)). 
  The blue curve represents the analytical behavior $\propto
  J_5(\lambda_{j}r)$ with $\lambda_j$ the first root of the dispersion
  relation.
  (b), (d): snapshots of the radial profiles that match to a higher radial
  wave number. The red curves denote the numerical data and the blue curves
  denote a fit $\propto J_5(\lambda_j r)$ with $\lambda_j$ corresponding to
  the second, third and fourth root of the dispersion 
  relation~(\ref{eq::dispersion}). The corresponding time steps are marked by
  the green lines in panel c. 
  (c) temporal variations of the radial profiles and of the amplitude
  $|\widehat{u}^1_{5,z}|$ at $r=0.67$. 
\label{fig::radprofiles_gam2p000}}
\end{center}
\end{figure}
In order to quantify the visual observations made in the previous
section and to estimate amplitude and frequency of individual Kelvin
modes we use a discrete sine-transformation (DST) applied to each
Fourier mode
\begin{equation}
\vec{\widehat{u}}^k_m(r)=\sum\limits_{q=1}^{N_z}\vec{\widetilde{u}}_m(r,z_n)\sin 
\left(\frac{q k\pi}{N_z+1}\right)\label{eq::axial_decomposition} 
\end{equation}
with $N_z$ the number of points in axial
direction. Figure~\ref{fig::axialdecomp} shows radial profiles for the
dominant contributions to the axial velocity at $\Gamma=1.825$ (a) and
at $\Gamma=2.200$ (b) where we only present the three leading modes
for each case. The solid thick curves present the time average and the
bright colors in the background represent the variation of the radial
profiles in time.  For comparison, the thick dashed curves show the
analytical behavior $\propto J_m(\lambda_{j}r)$ predicted by the
linear in-viscid approximation (\ref{eq::kelvinmodes}) with
$\lambda_j$ the first root of the dispersion
relation~(\ref{eq::dispersion}). The differences between the numerical
data and the analytical curves are surprisingly small with deviations
mainly in the boundary layers which can be explained by the different
boundary conditions for the simulations (no slip) and the in-viscid
solution (free slip). Except for (considerable) changes in the
amplitude the pattern of the radial profiles exhibits only minor
temporal changes and shows no significant contributions from higher
radial modes (see color coded contour plots in
figure~\ref{fig::axialdecomp} that show the variations of the radial
profiles in time).

The behavior becomes more complex at $\Gamma=2$ at which we see
many modes with different $m$ and $k$ with nearly the same amplitude
and clear changes in the radial
structure. Figure~\ref{fig::radprofiles_gam2p000} shows characteristic
radial profiles for the mode ($m = 5, k = 1$) and their temporal
behavior as a typical example. The amplitude exhibits considerable
fluctuations and, unlike in the previous cases, we see additional
changes in the radial structure. When the amplitude is around a
(local) maximum, the radial velocity profile is in accordance with a
wavenumber that corresponds to the first root of the dispersion
relation and the radial profiles can nicely be described by a function
$\propto J_5(\lambda_1 r)$ (see red curves in
figure~\ref{fig::radprofiles_gam2p000}a). However, when the amplitude
is weak we see clear signatures of higher radial wave numbers (see
figure~\ref{fig::radprofiles_gam2p000} b and d). Even though the
higher radial modes remain weak, they might represent a channel for
the transfer of energy from the (periodically occurring) maxima of a
mode into smaller scales.

\subsection{Amplitudes and frequencies}
\subsubsection{Height equal diameter $(\Gamma=2)$}\label{sec::451}

\newcommand{\figwidth}{4.6cm}
\newcommand{\backjump}{-0.45cm}
\begin{figure}[h!]
\begin{center}
\includegraphics[width=\figwidth]{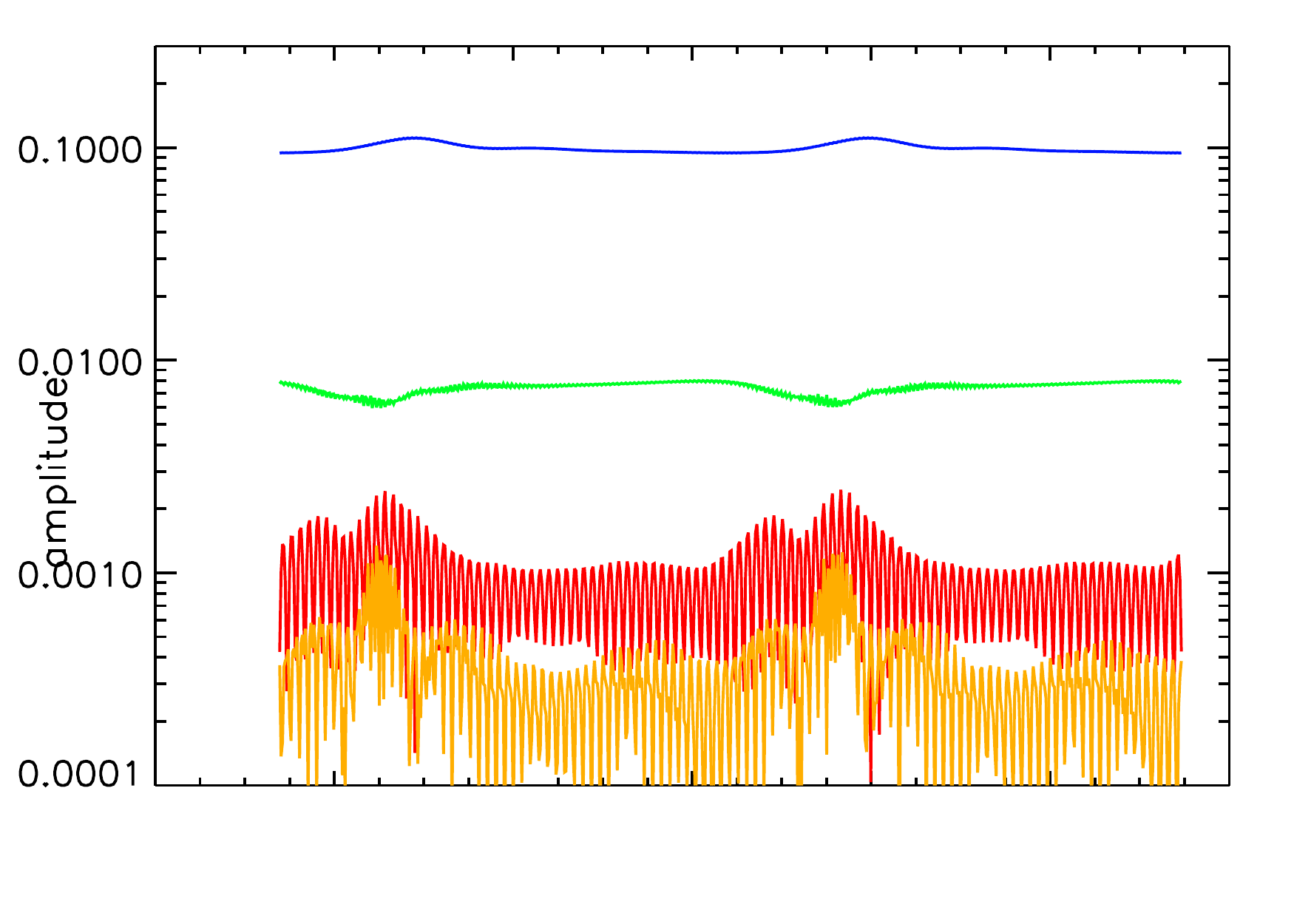}
\includegraphics[width=\figwidth]{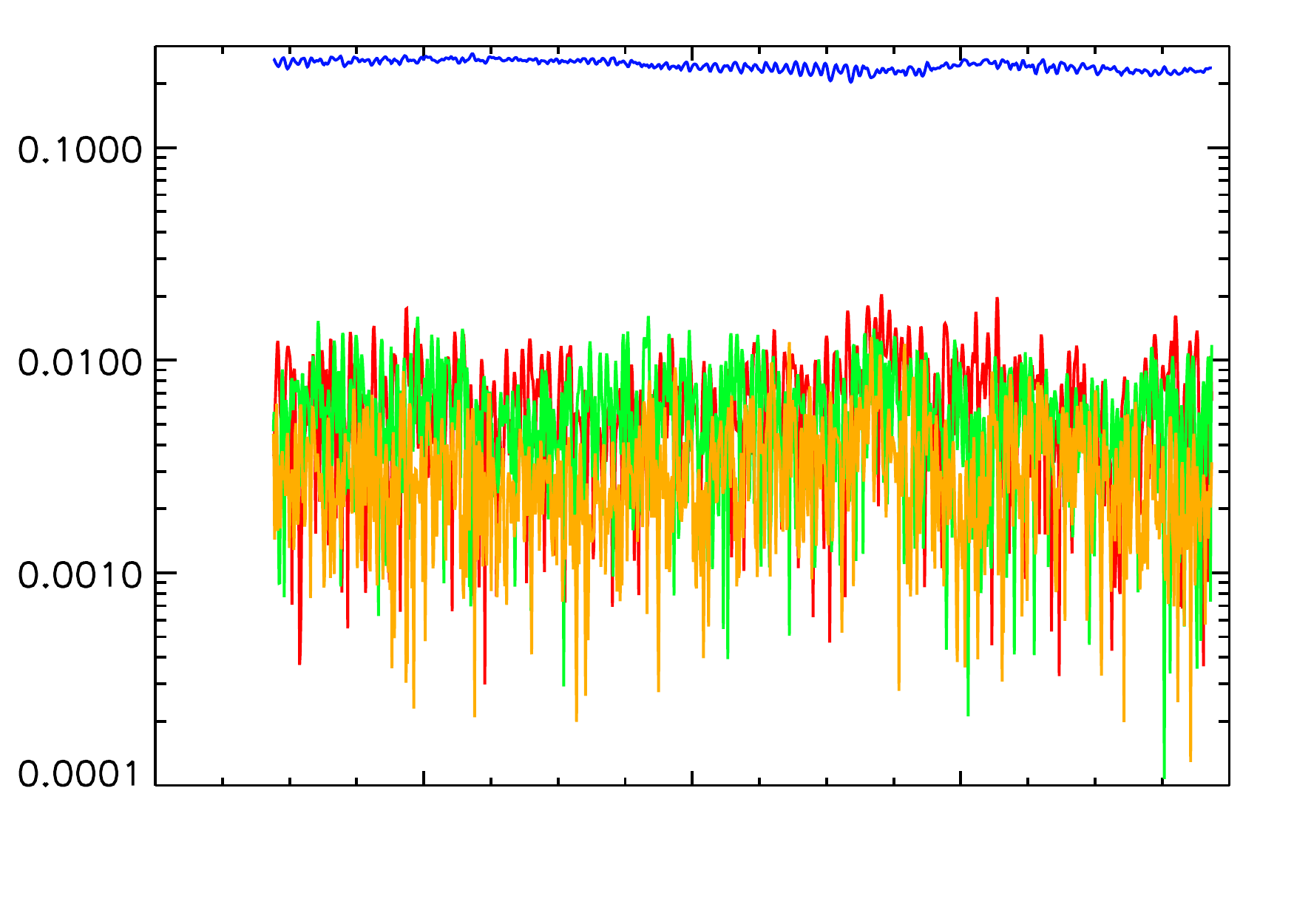}
\includegraphics[width=\figwidth]{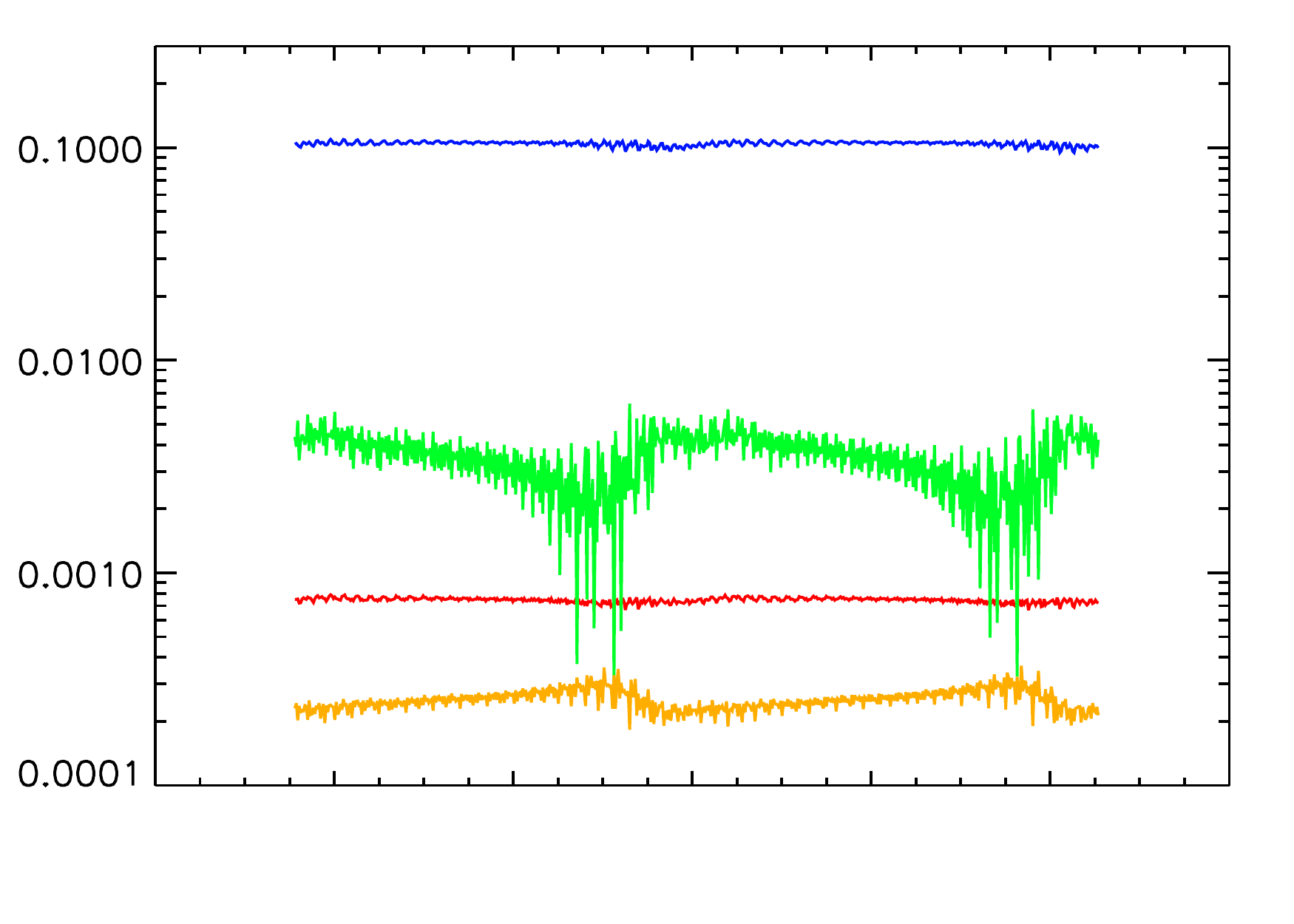}
\\[\backjump]
\includegraphics[width=\figwidth]{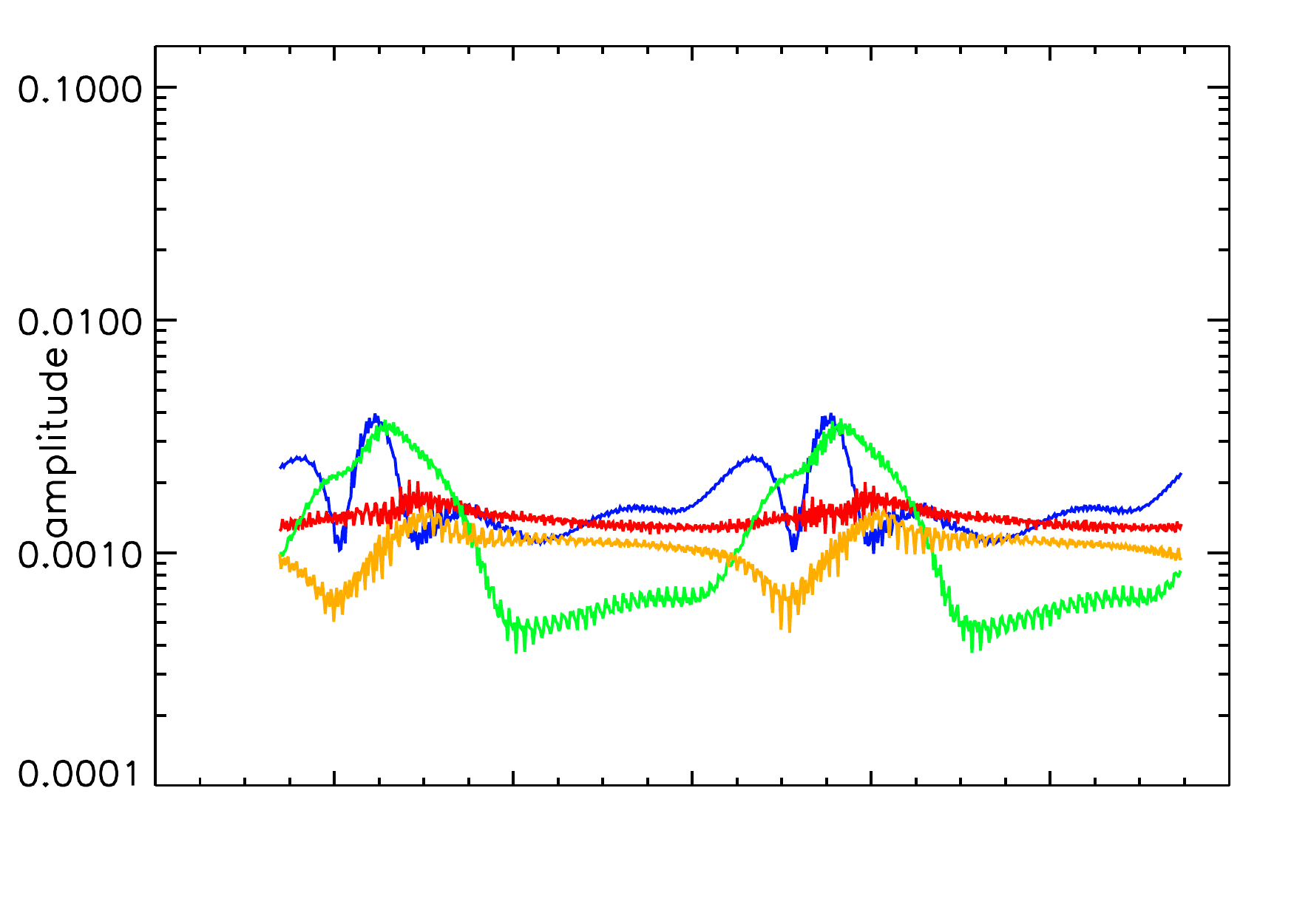}
\includegraphics[width=\figwidth]{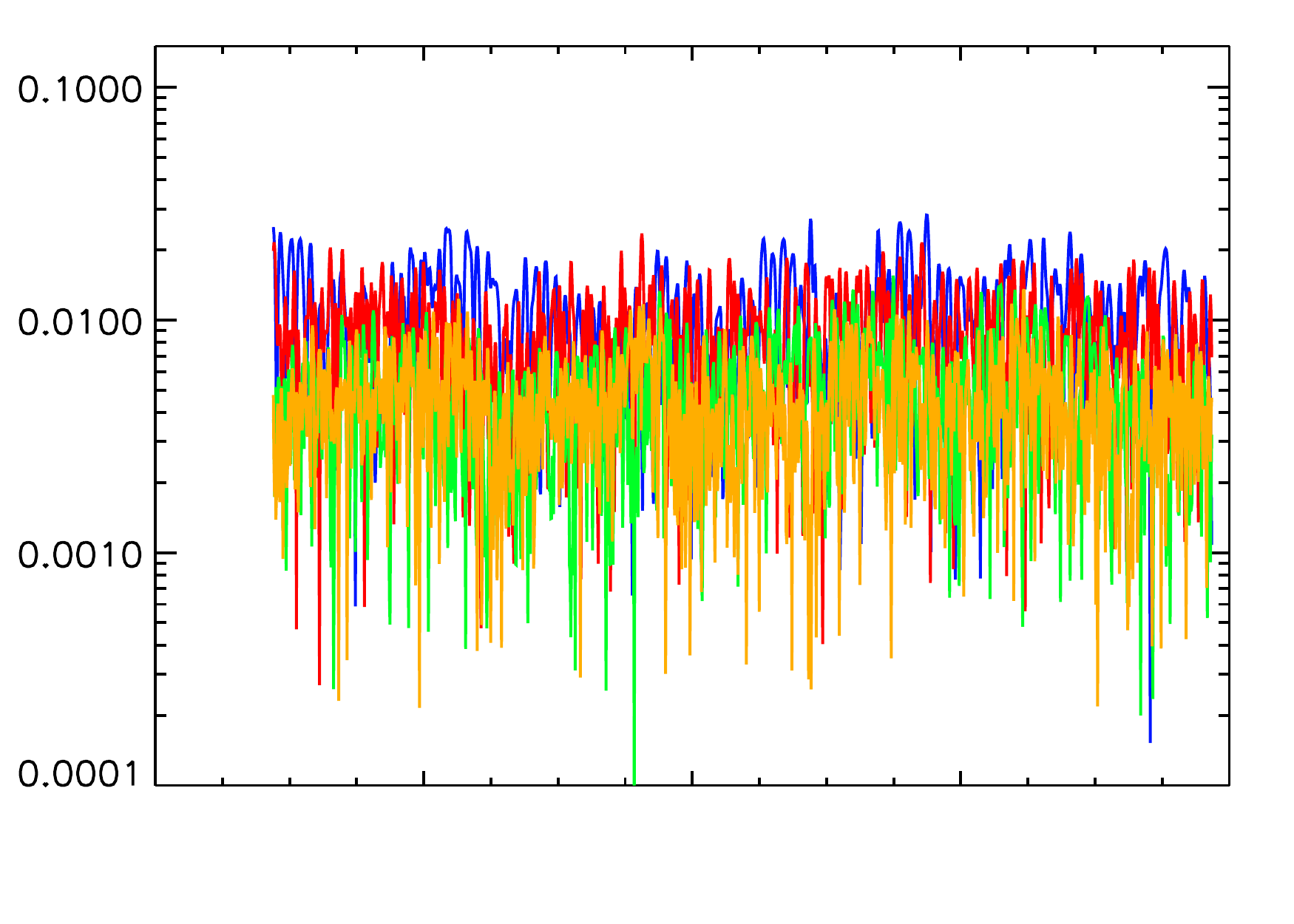}
\includegraphics[width=\figwidth]{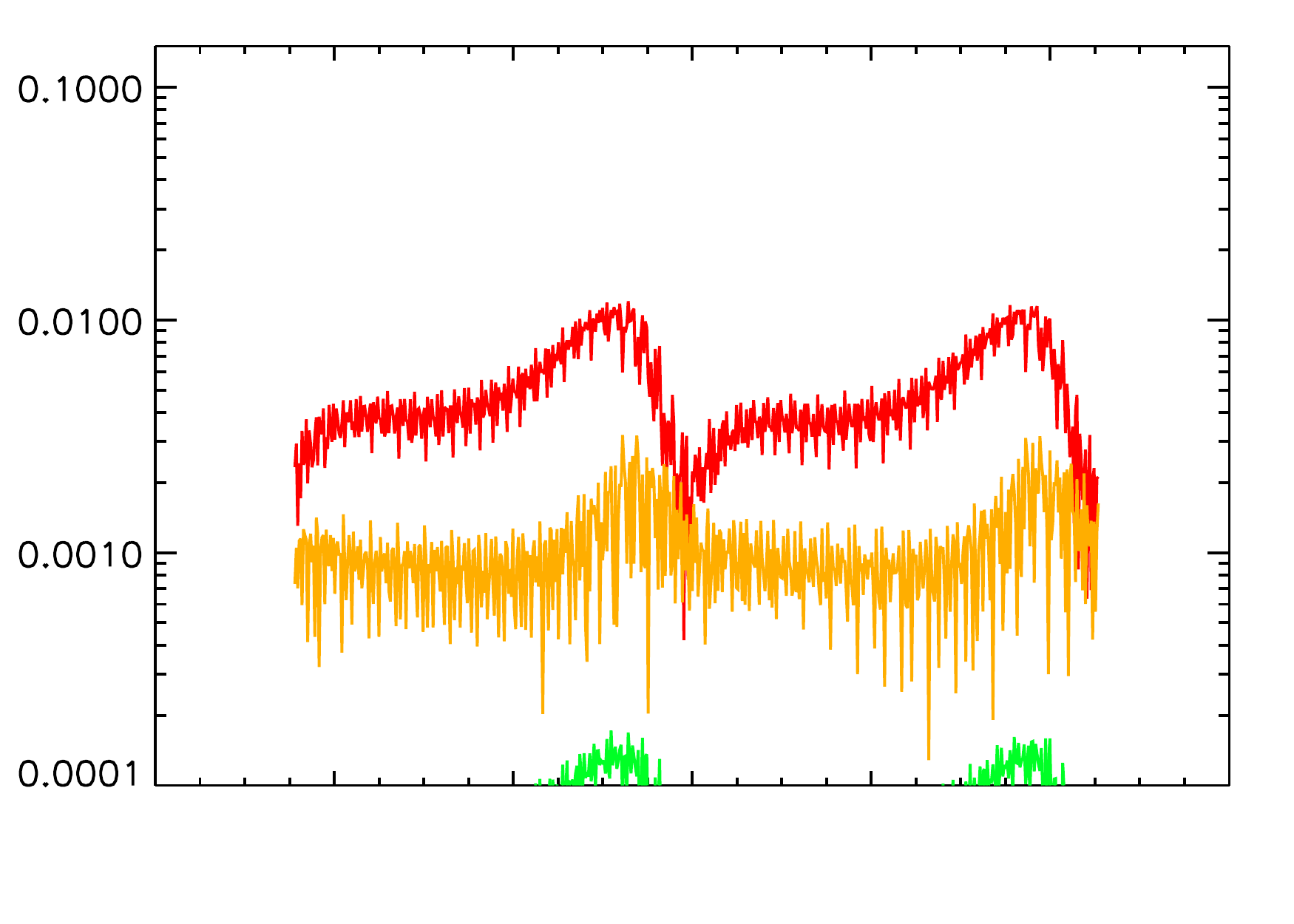}
\\[\backjump]
\includegraphics[width=\figwidth]{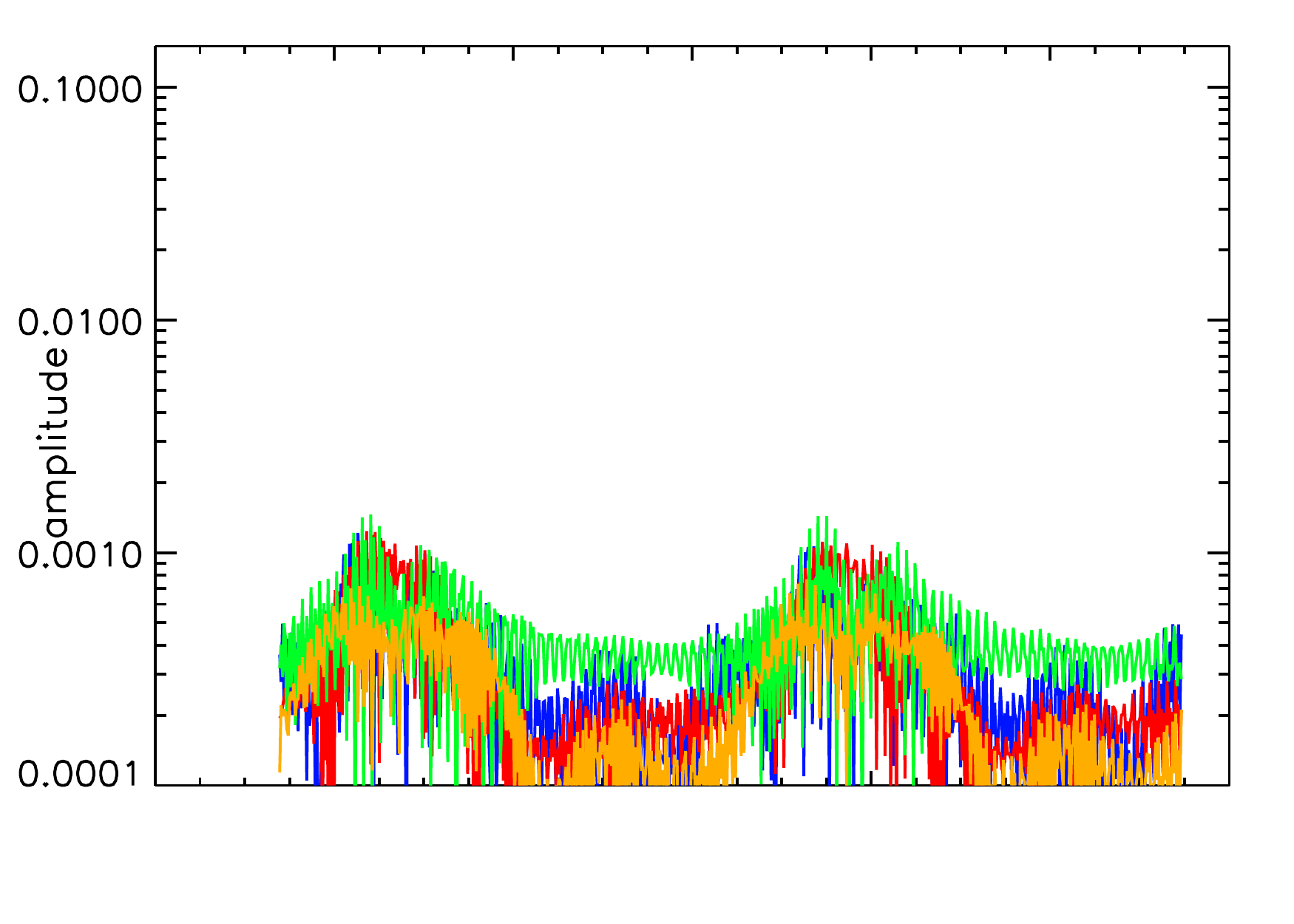}
\includegraphics[width=\figwidth]{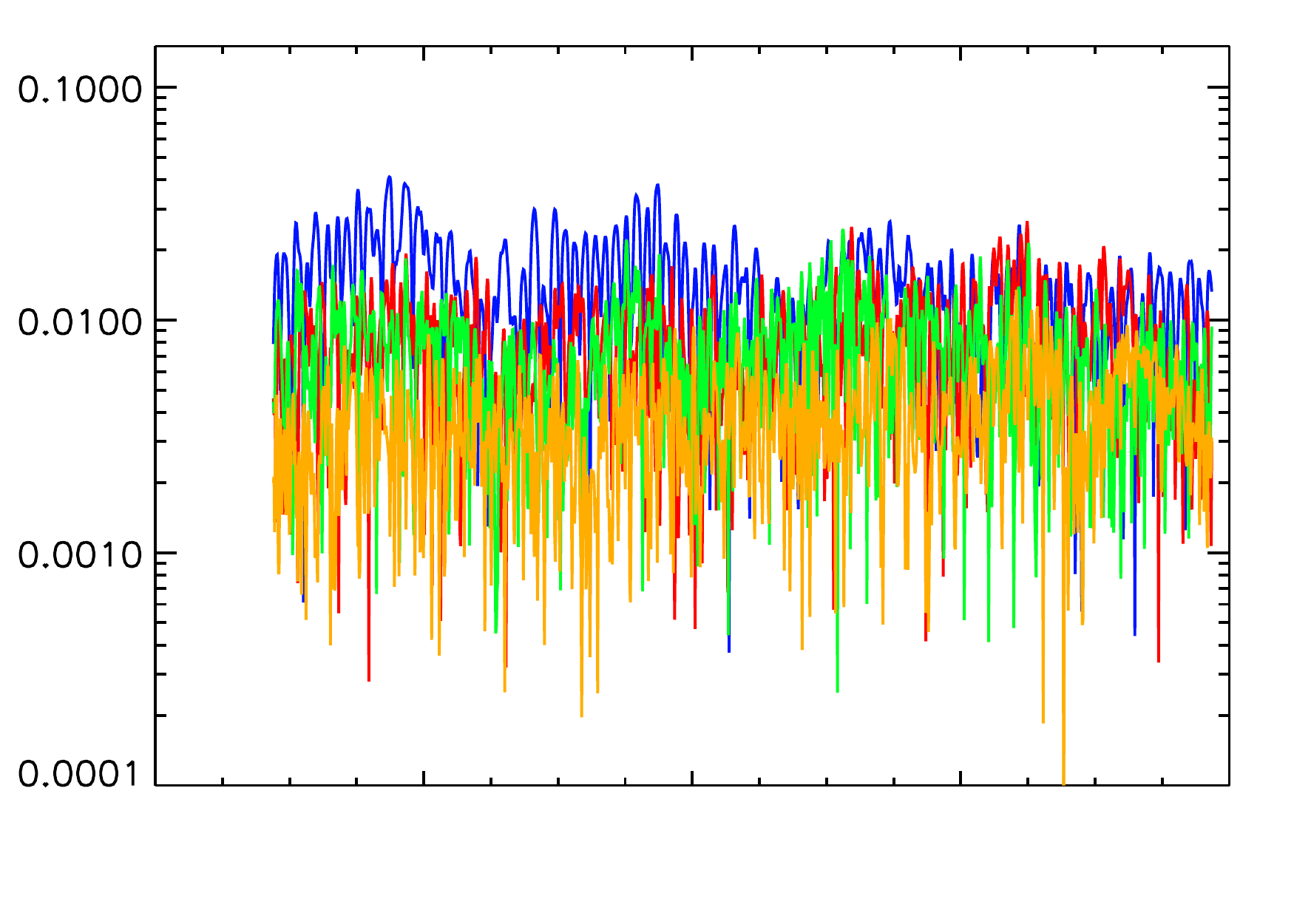}
\includegraphics[width=\figwidth]{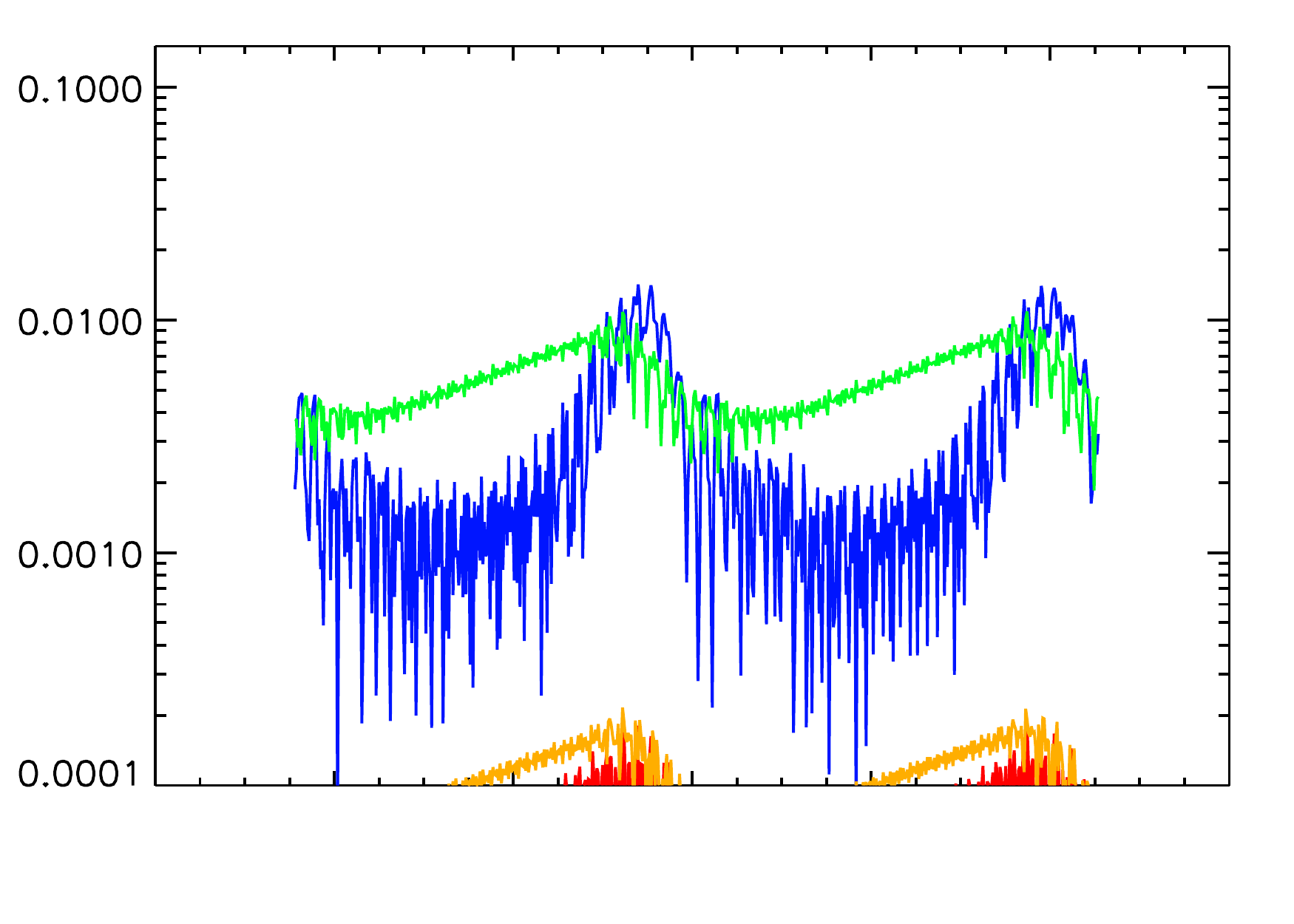}
\\[\backjump]
\includegraphics[width=\figwidth]{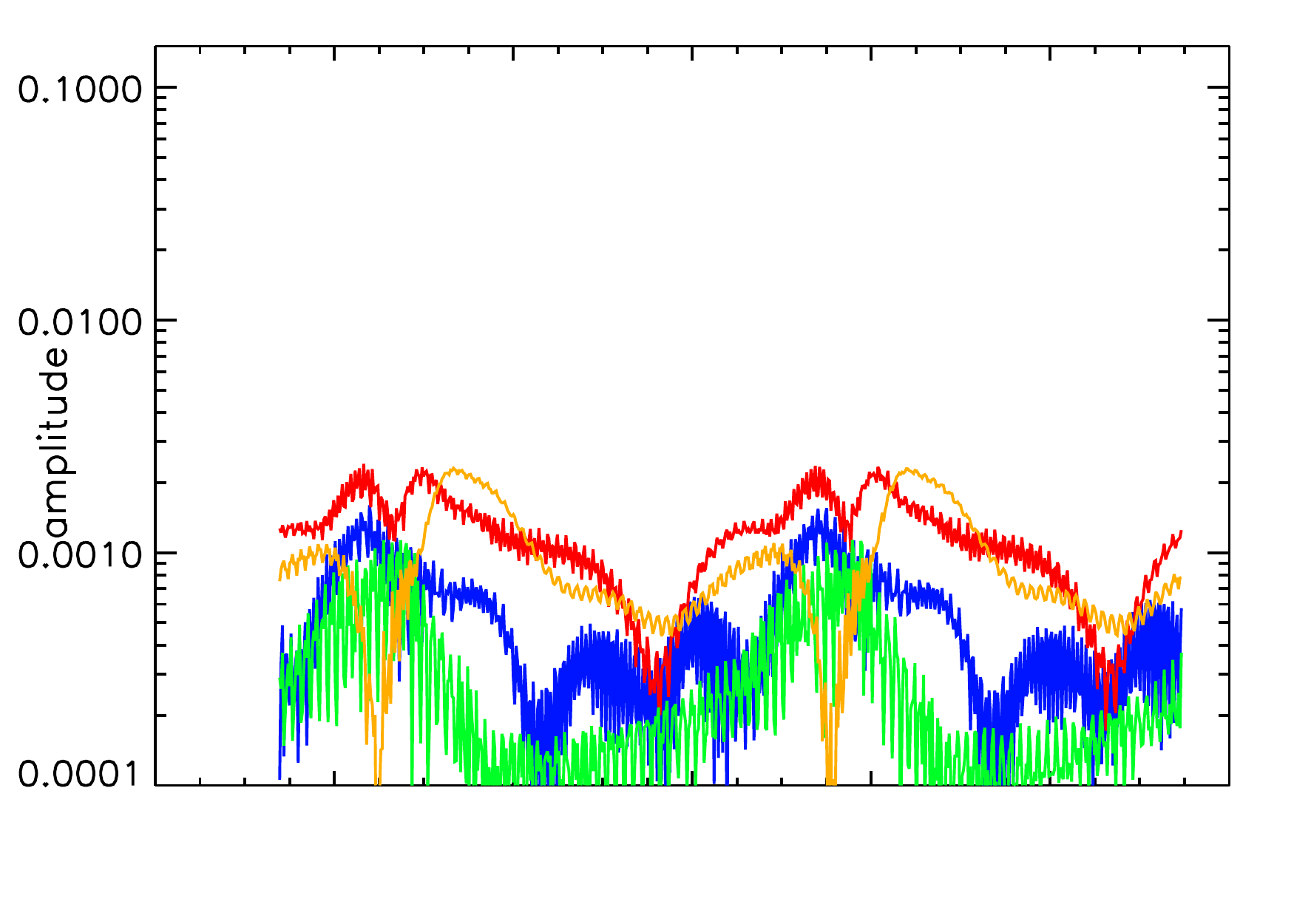}
\includegraphics[width=\figwidth]{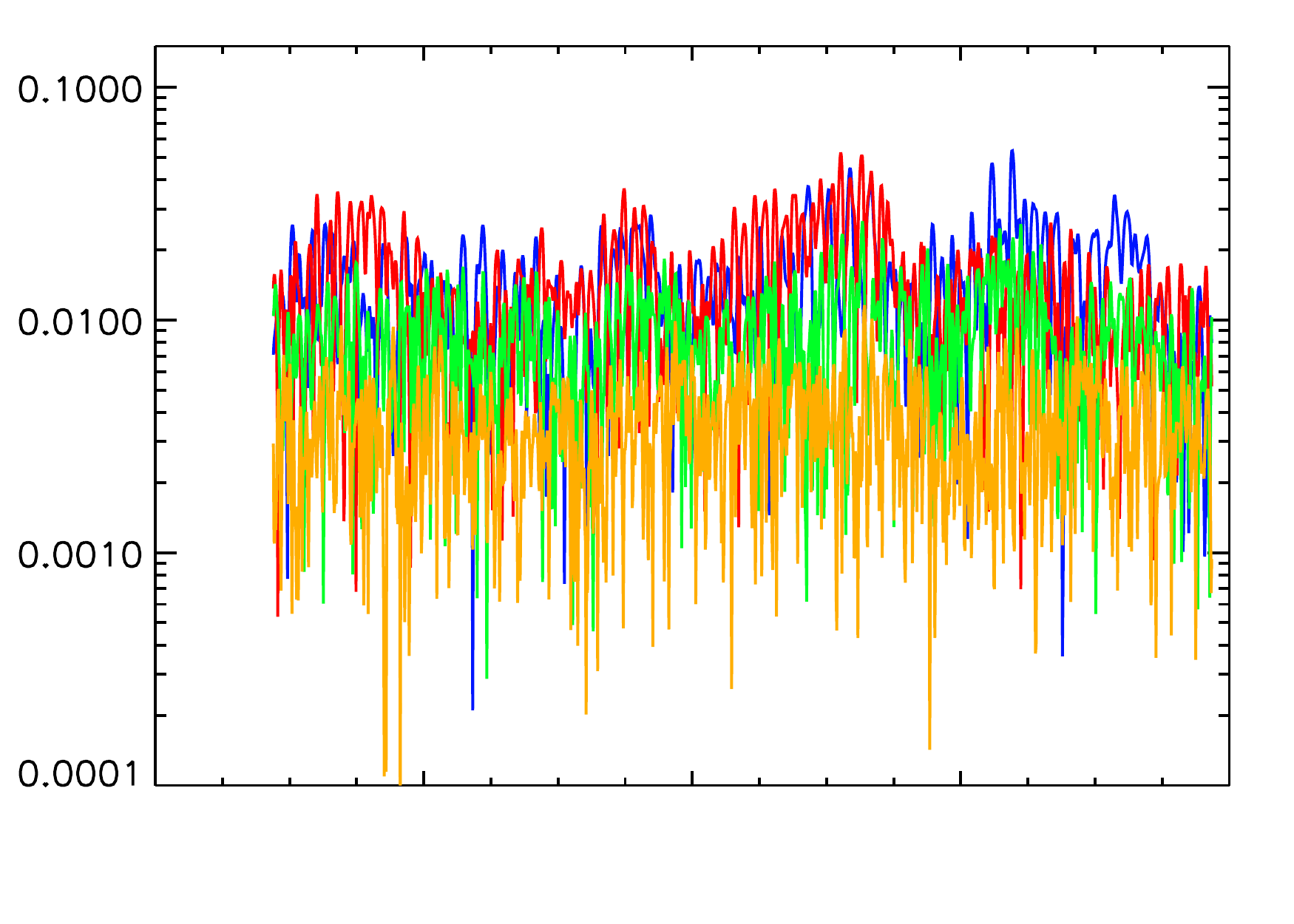}
\includegraphics[width=\figwidth]{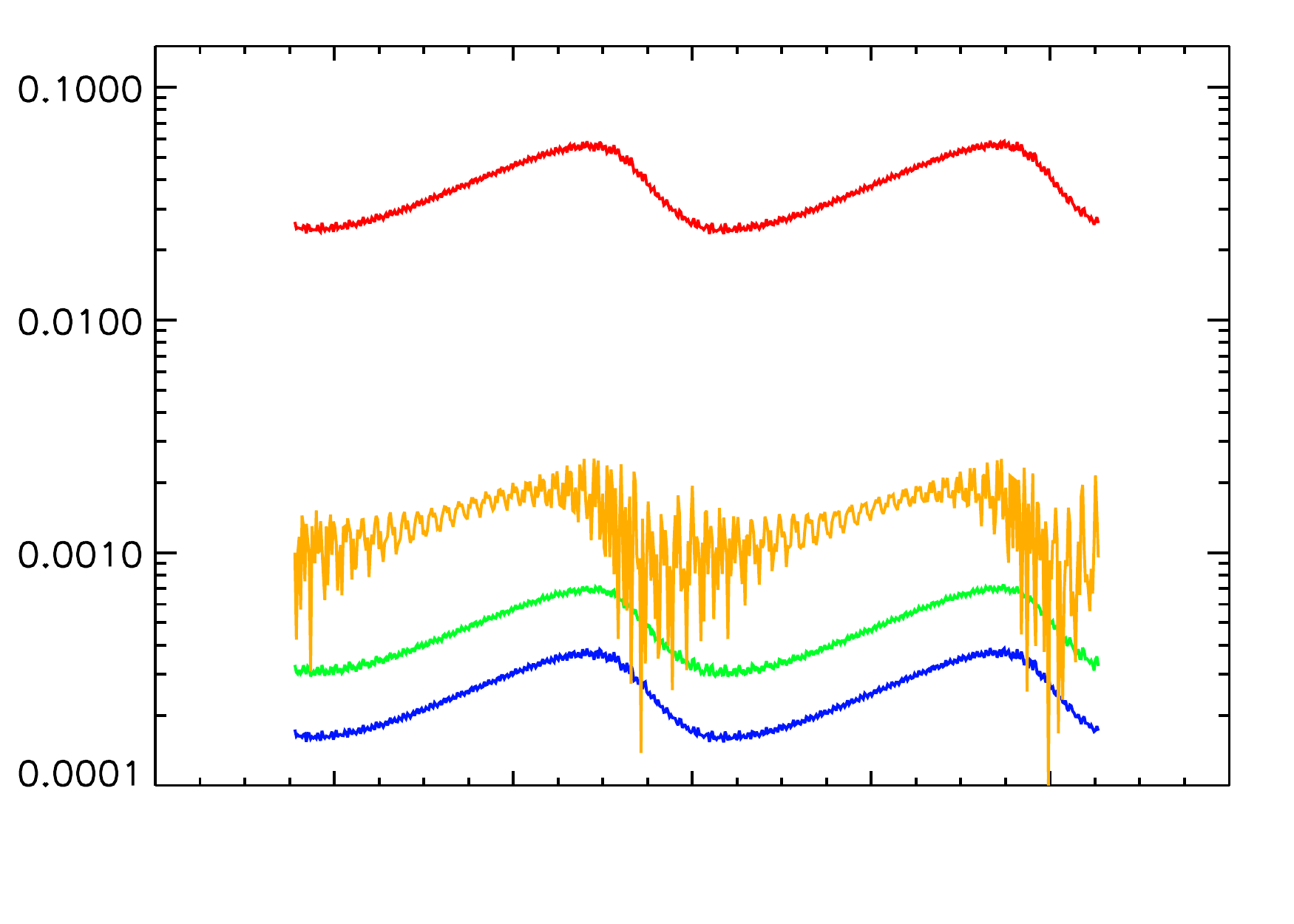}
\\[\backjump]
\includegraphics[width=\figwidth]{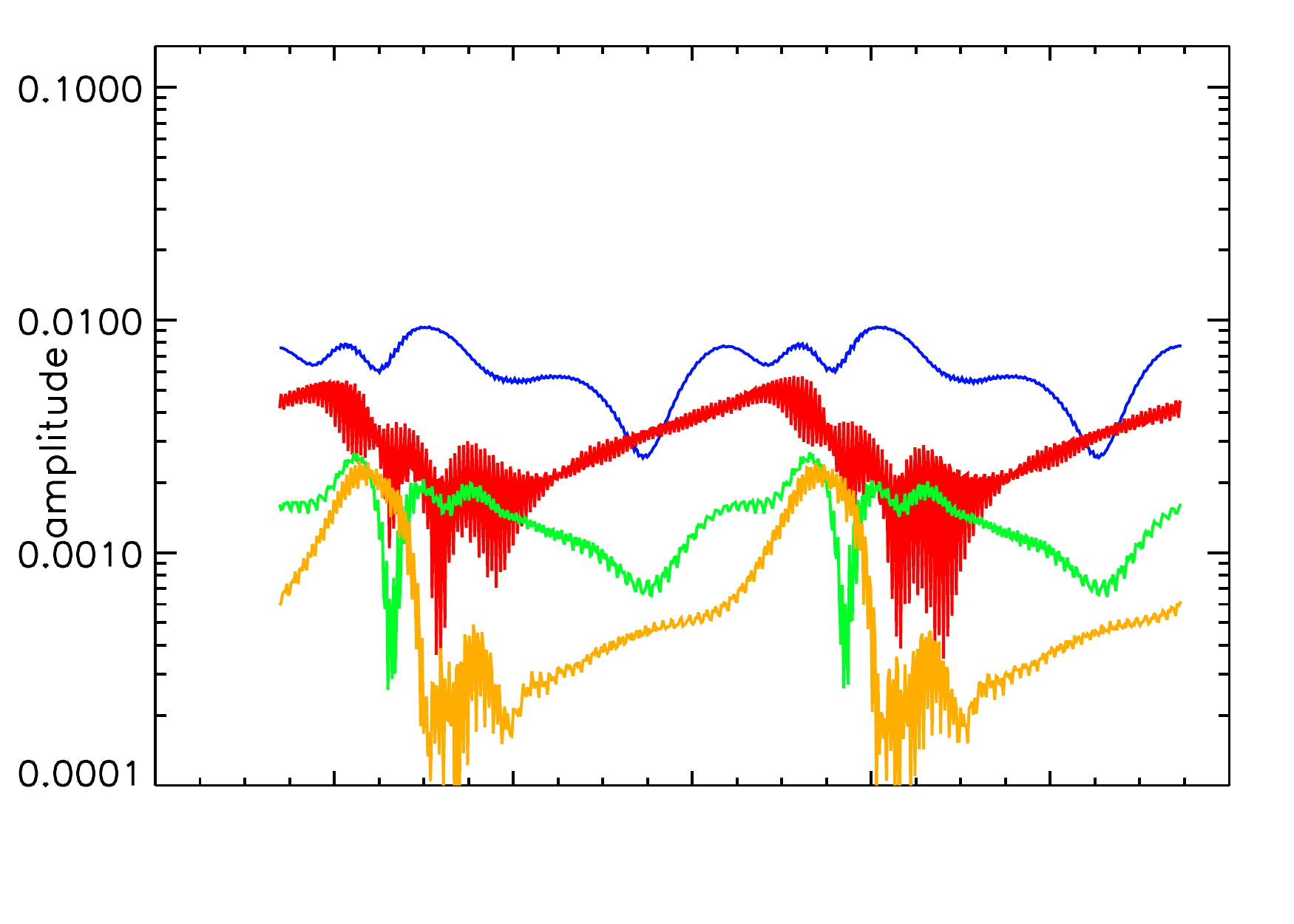}
\includegraphics[width=\figwidth]{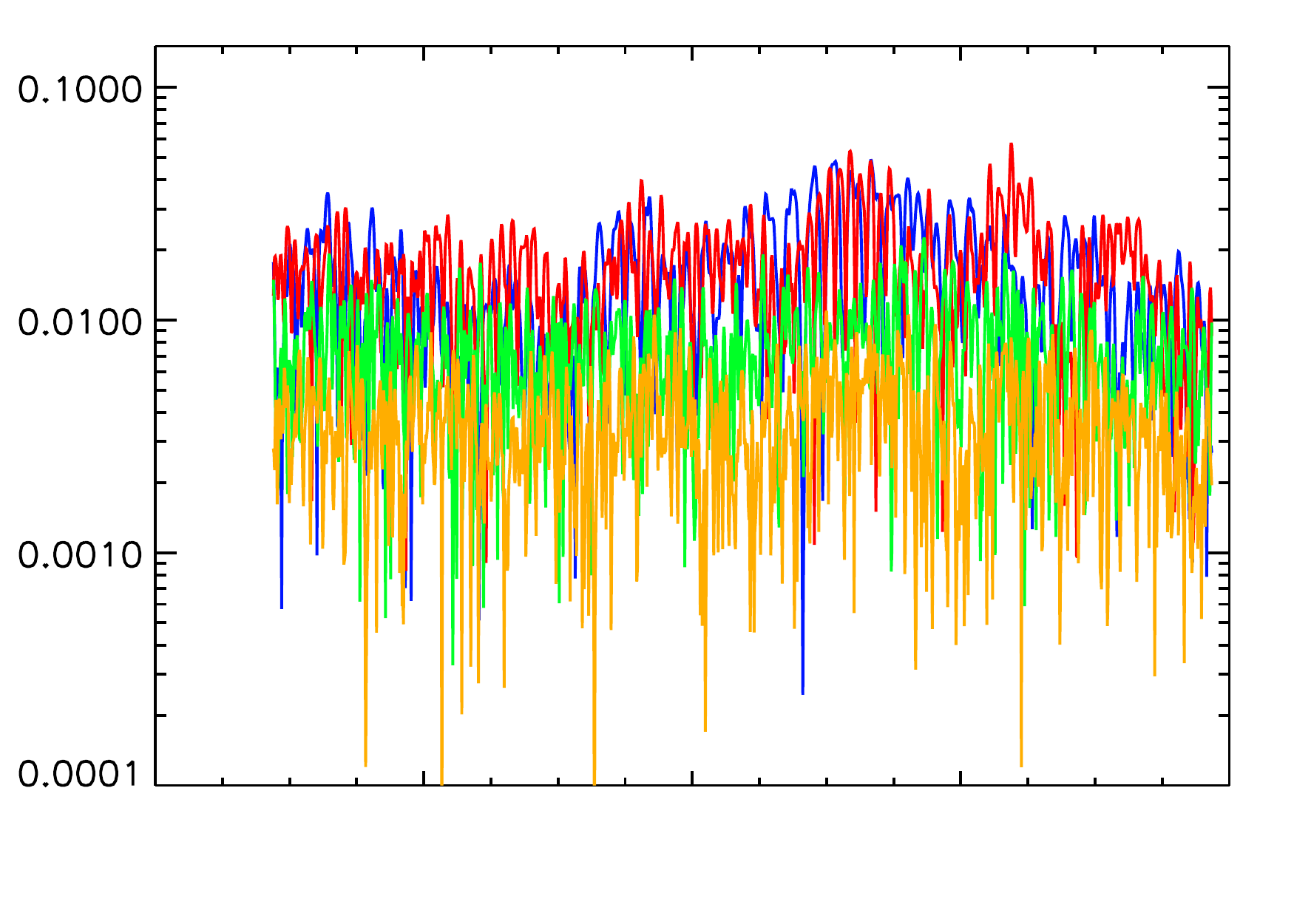}
\includegraphics[width=\figwidth]{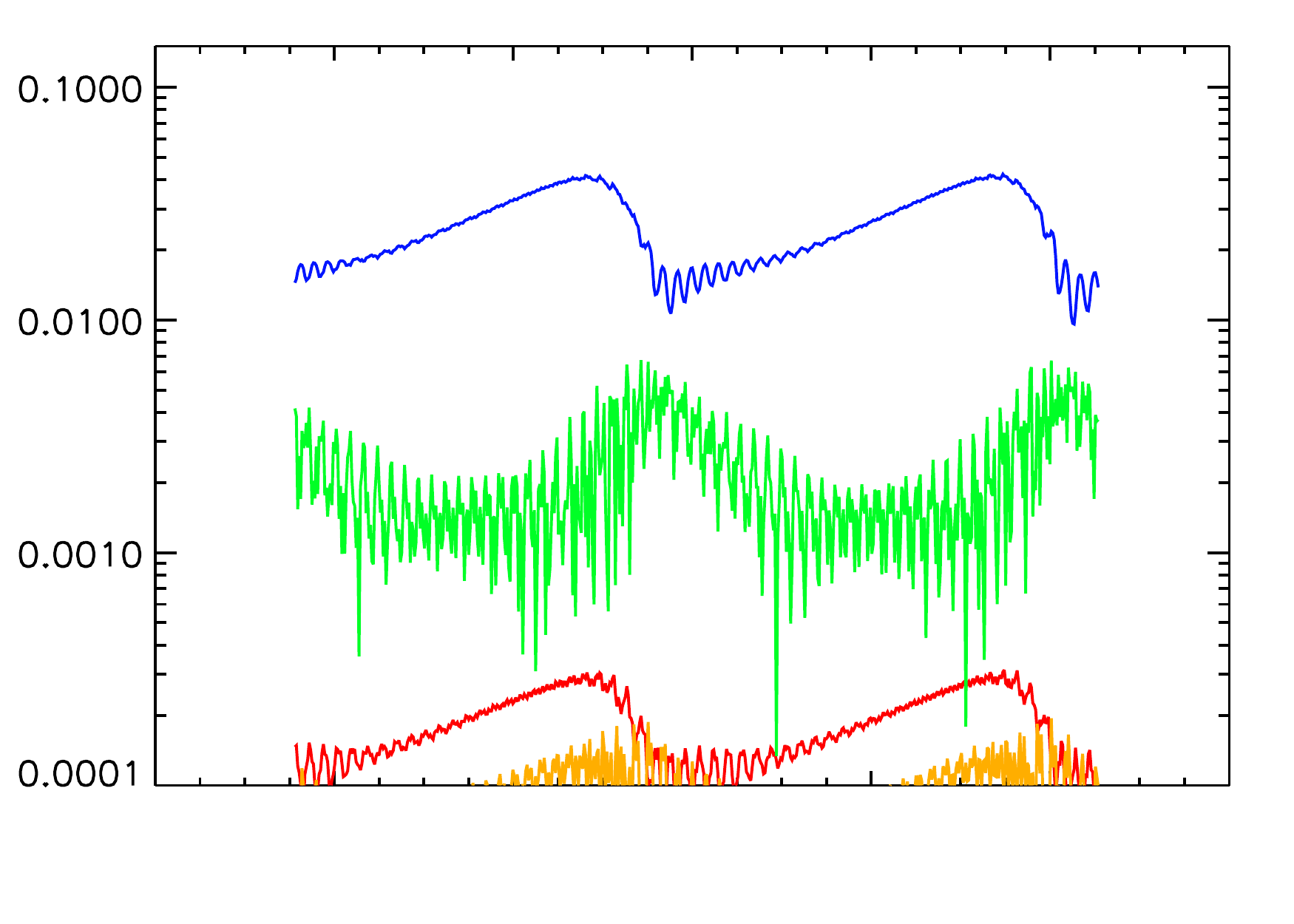}
\\[\backjump]
\includegraphics[width=\figwidth]{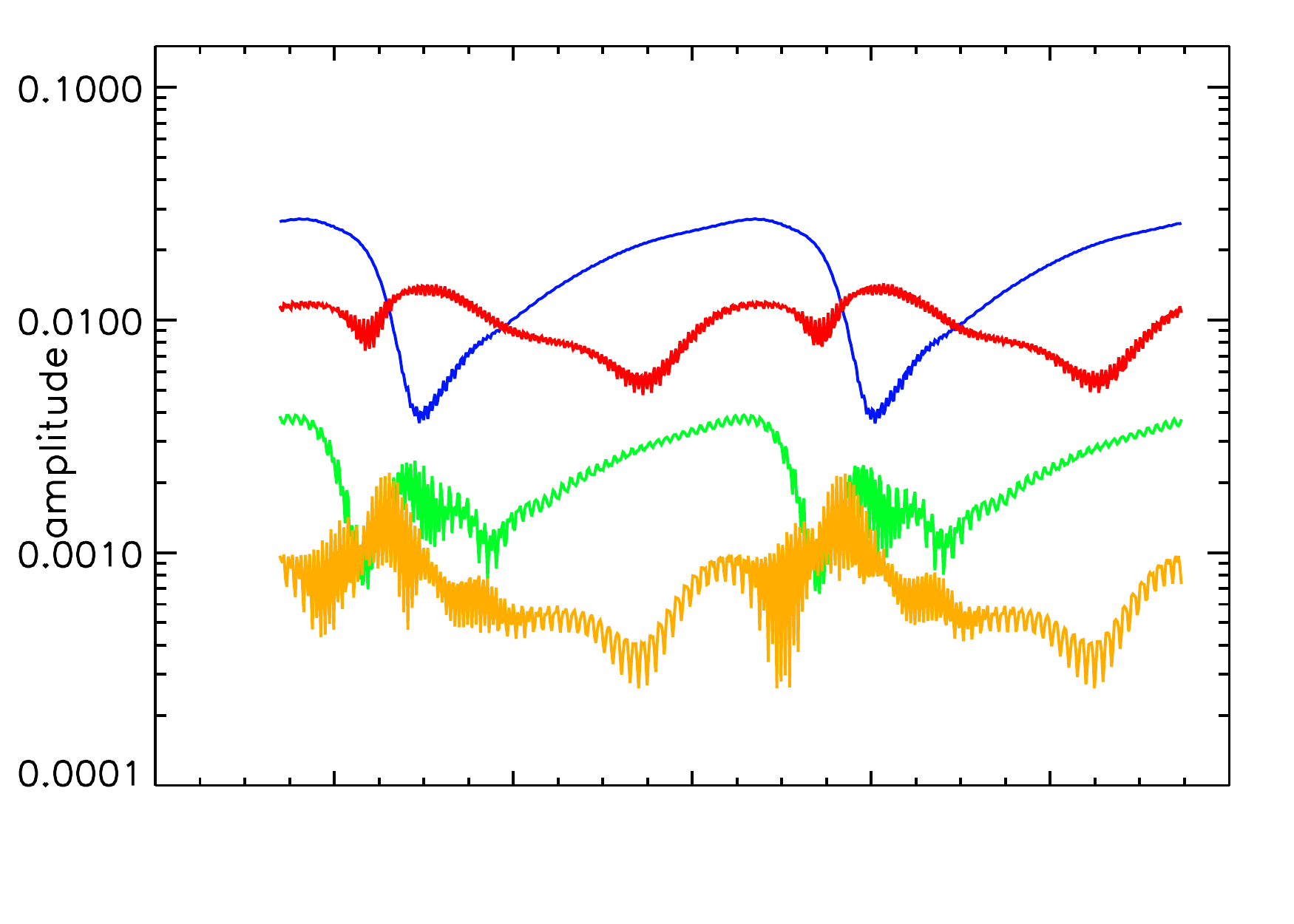}
\includegraphics[width=\figwidth]{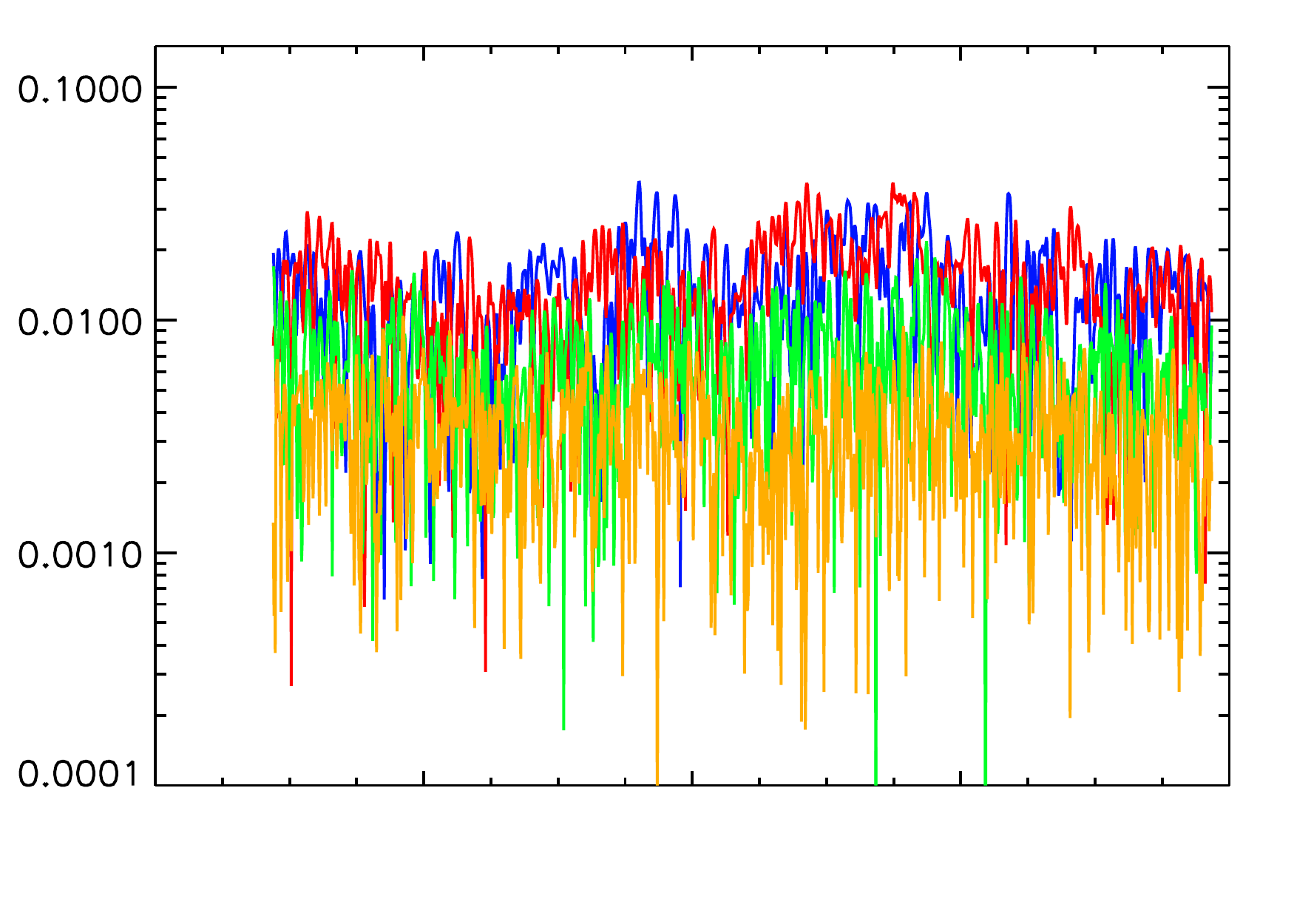}
\includegraphics[width=\figwidth]{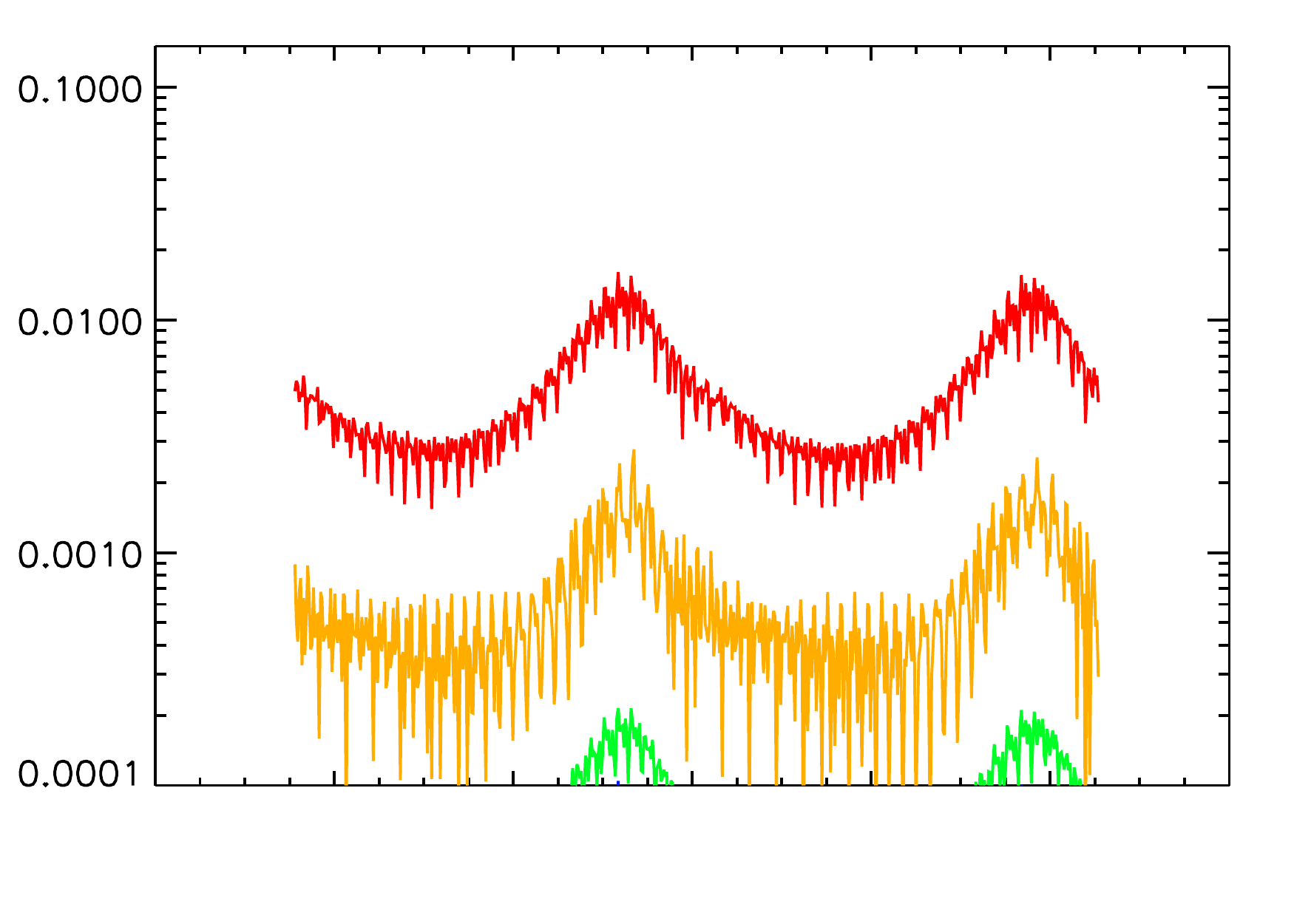}
\\[\backjump]
\includegraphics[width=\figwidth]{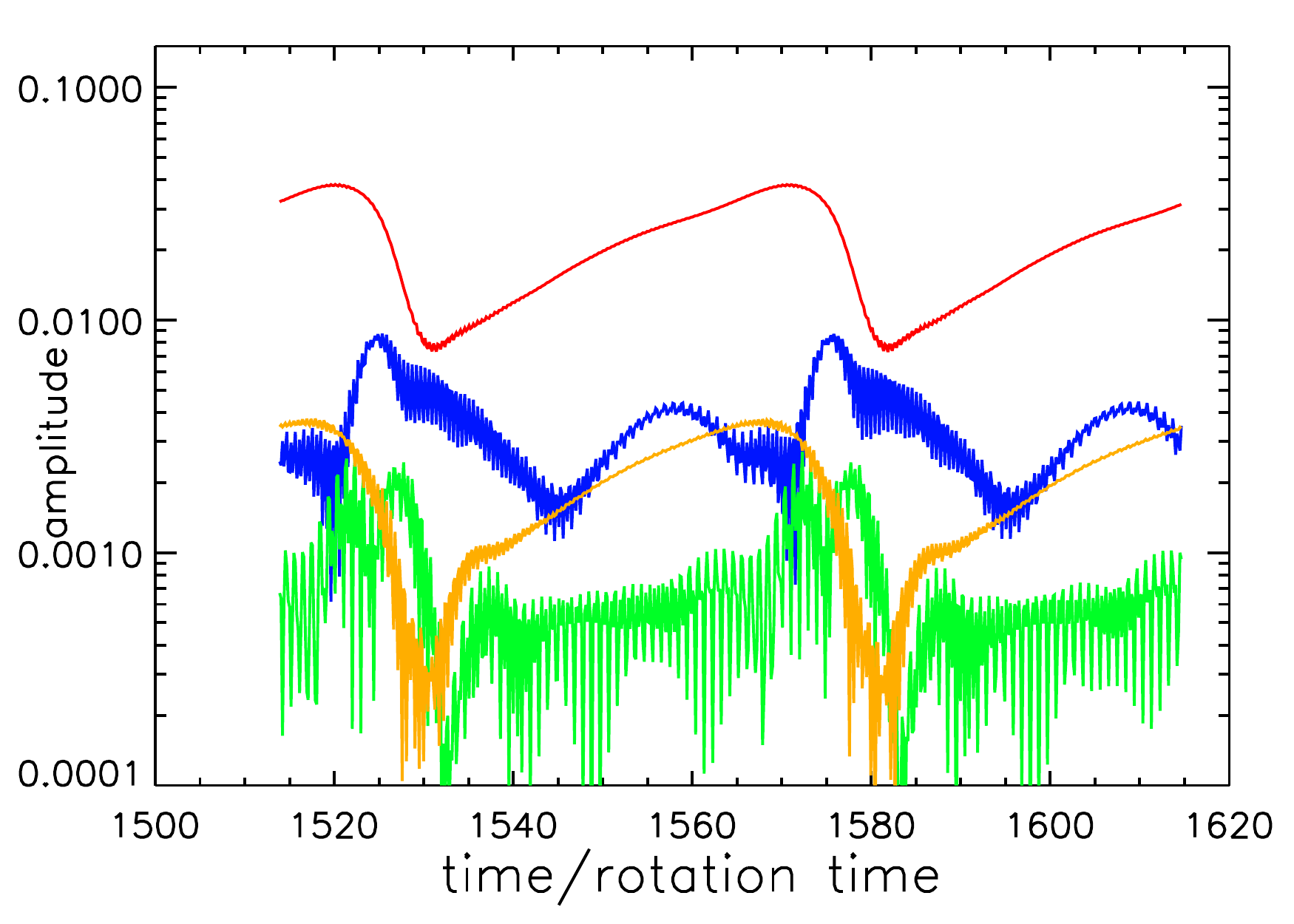}
\includegraphics[width=\figwidth]{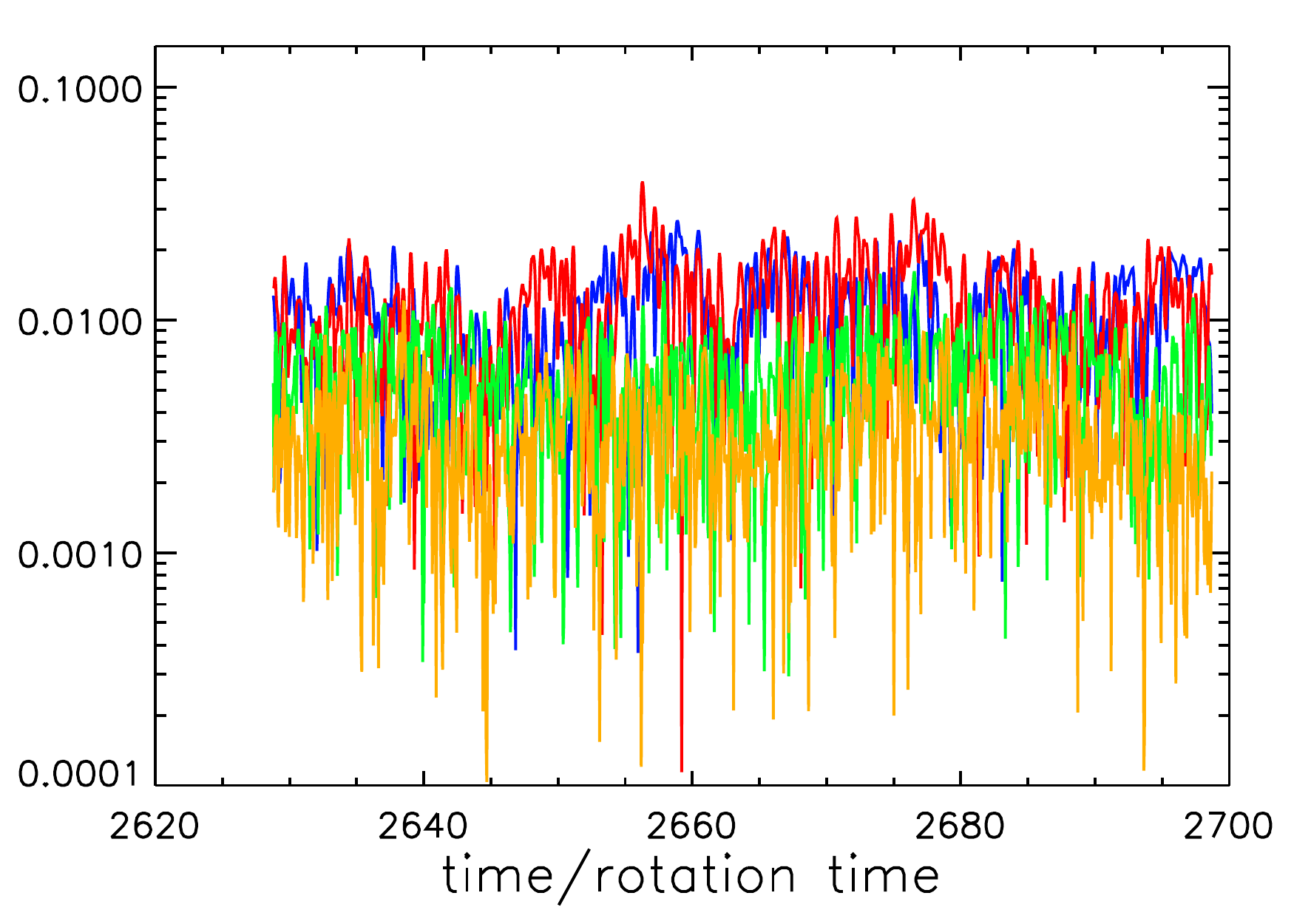}
\includegraphics[width=\figwidth]{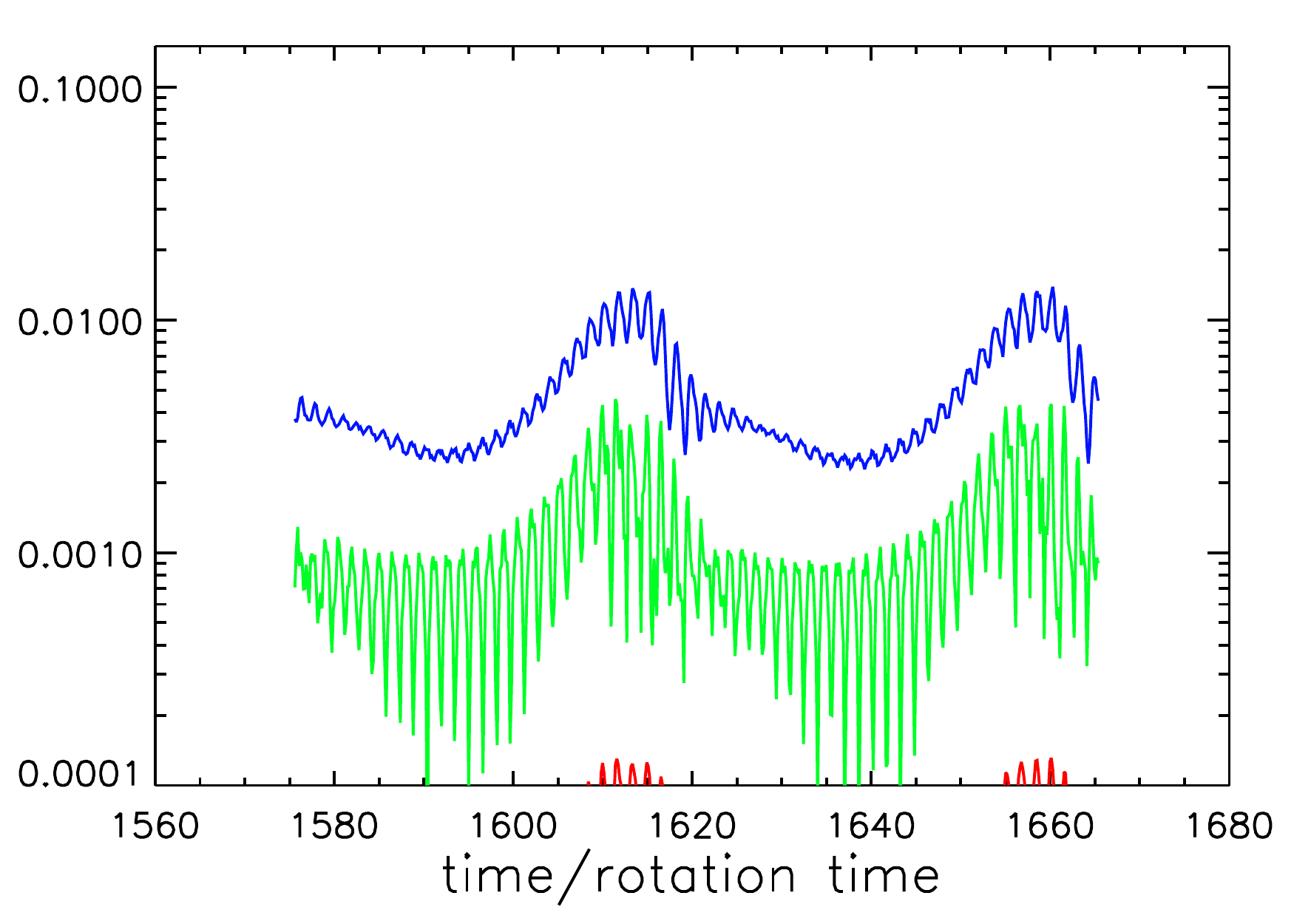}
\caption{Temporal behavior of the amplitude of the flow
  $\widehat{{u}}^k_{m,z}$ at $r=0.67$ decomposed into Fourier- and 
  axial modes with $m=1...7$ (from top to bottom) and
  $k=1...4$ (blue, red, green, yellow curve). From left to right:
  $\Gamma=1.825$, $\Gamma=2.0$ and $\Gamma=2.2$.}\label{fig::kelvin_amp_vs_time}
\end{center}
\end{figure}
\begin{table}[b!]
\begin{center}
\begin{tabular}{c|cccc|l}
${\rm{Re}}$ & $m$ & $k$ & $\max |u_m^k|$ & $\Omega_{\rm{num}}$ & $\Omega_{\rm{theo}}$ \\
\hline
\hline
6500 & $1$ & $1$ & 0.2673 &  $-1.0000$  & $-1.0000$ (forced)\\
&       $3$ & $1$ & $0.0412$ & $+0.2911$ & $+0.3342$ (for $n=2$)\\
&       $3$ & $2$ & $0.0266$ & $+0.3154$ & $+0.3333$ (for $n=5$)\\
&       $4$ & $1$ & $0.0536$ & $+0.0357$ & no corresponding mode\\
&       $4$ & $2$ & $0.0525$ & $+0.3593$ & $+0.3647$ (for $n=4$)\\
&       $5$ & $1$ & $0.0491$ & $+0.2278$ & $+0.2030$ (for $n=3$)\\
&       $5$ & $2$ & $0.0578$ & $+0.4692$ & $+0.5157$ (for $n=2$)\\
&       $6$ & $1$ & $0.0392$ & $+0.2006$ & $+0.1868$ (for $n=3$)\\
&       $6$ & $2$ & $0.0391$ & $+0.6239$ & $+0.6309$ (for $n=1$)\\
\hline
\hline
10000& $1$ & $1$ & 0.2775 & $ -1.0000$ & $-1.0000$ (forced)\\
      &  $3$ & $1$ & 0.0329 & $+0.3090$ & $+0.3342$ (for $n=2$)\\
      &  $3$ & $2$ & 0.0302 & $+0.3013$ & $+0.3333$ (for $n=5$)\\
      &  $4$ & $1$ & 0.0562 & $+0.1687$ & $+0.1532$ (for $n=5$)\\
      &  $4$ & $2$ & 0.0420 & $+0.3511$ & $+0.3647$ (for $n=4$)\\
      &  $5$ & $1$ & 0.0570 & $+0.2001$ & $+0.2030$ (for $n=3$)\\
      &  $5$ & $2$ & 0.0574 & $+0.5168$ & $+0.5157$ (for $n=2$)\\
      &  $6$ & $1$ & 0.0417 & $+0.3452$ & $+0.3204$ (for $n=1$)\\
      &  $6$ & $2$ & 0.0437 & $+0.6410$ & $+0.6309$ (for $n=1$)\\
\hline
\hline
\end{tabular}
\caption{Amplitudes and frequencies from the simulations at
  $\Gamma=2$.  Only modes with well-defined drift behavior are
  listed.  The columns denote (from left to right): Reynolds number,
  azimuthal wave number, axial wave number, maximum of the amplitude,
  azimuthal drift frequency from the numerical data, and theoretical
  frequency from the dispersion relation~(\ref{eq::dispersion}) for
  larger radial wave numbers $n$ that fit best to the observed
  frequencies.  }\label{tab::amp_freq_onres}
\end{center}
\end{table}
Figure~\ref{fig::kelvin_amp_vs_time} shows the temporal behavior of
$|{\widehat{u}}_{m,z}^k|$ at $r=0.67$ for $m=1$ to $m=7$ (from top to
bottom) and for $k=1$ to $k=4$ (blue, red, green, yellow curve). The
peak values of the amplitudes are additionally listed in
table~\ref{tab::amp_freq_onres} ($\Gamma=2$) and in
table~\ref{tab::amp_freq_offres} ($\Gamma=1.825$ and $\Gamma=2.2$)
together with the frequencies obtained from the time derivative of the
azimuthal phase of each mode. Note that we only list modes with a
regular behavior of the frequencies, i.e., contributions which exhibit
a steady and unique azimuthal drift that allows a conclusive
computation of the time derivative of the azimuthal phase. The results
quantitatively complement the observations made in the previous
paragraph. For $\Gamma=2$ (central column and
table~\ref{tab::amp_freq_onres}) we see a dominant forced mode with
$m=1$ and $k=1$ and the contributions with $k > 1$ are negligible
(smaller by a factor of 20). The higher azimuthal modes ($m=2\dots 7$)
reach approximately $10\%$ of the amplitude of the $m=1$ mode with
$k=1$ and $k=2$ slightly prevailing over $k=3$ and $k=4$.  A striking
property of the flow at $\Gamma=2$ is the common orientation of the
azimuthal drift motion of all modes (see
table~\ref{tab::amp_freq_onres}). We only find modes with a retrograde
azimuthal drift so no combination is possible that fulfills the
requirements for a triadic resonance (in all cases $\delta\Omega \neq
1$, see table~\ref{tab::amp_freq_onres}). The presence of distinct
frequency signals that fit to higher radial modes (which however are
hardly evident in the radial profiles for the ($m=5, k=1$) mode, see
figure~\ref{fig::radprofiles_gam2p000}) shows that there must be
further contributions with chaotic behavior which are probably
dominant and having no regular azimuthal drift.

We also performed simulations at a slightly larger Reynolds number
${\rm{Re}}=10000$ (at $\Gamma=2$) and found only minor changes with
respect to the run at ${\rm{Re}}=6500$ except for the modes $(m=4,
k=1)$ and $(m=6,k=1)$ which have lower frequencies at
${\rm{Re}}=6500$. However, so far our data is not sufficient to
explain the impact of ${\rm{Re}}$ on the drift frequencies or to
establish a conclusive scaling towards more realistic ${\rm{Re}}$ that
will be reached in the experiment. Hence, we refrain from any further
discussion of the properties of the flow at larger ${\rm{Re}}$.

\subsubsection{Triadic resonances at $\Gamma=1.825$ and $\Gamma=2.200$}

The behavior of the amplitude confirms the modified characteristics of
the flow when the forced mode is off-resonance ($\Gamma=1.825$, left
column in figure~\ref{fig::kelvin_amp_vs_time}, or $\Gamma=2.2$, right
column in figure~\ref{fig::kelvin_amp_vs_time}; see also
table~\ref{tab::amp_freq_offres}). For $\Gamma=2.200$ we find a
well-defined single triadic resonance with unique wave numbers ($m=4,
k=2$ and $m=5, k=1$) and frequencies ($\Omega_a=-0.6368$ and
$\Omega_b=0.3632$).  These values fulfill the conditions for a triadic
resonance and are quite close to the theoretical values obtained from
the dispersion relation~(\ref{eq::dispersion}) (see
table~\ref{tab::amp_freq_offres}).
\begin{table}[b!]
\begin{center}
\begin{tabular}{c|cc|cc|l}
$\Gamma$ & $m$ & $k$ & $\max |u_m^k|$ & $\Omega_{\rm{num}}$ & $\Omega_{\rm{theo}}$ \\
\hline
\hline
 1.825 & $1$ & $1$ & $0.1113$ & $-1.0000$ & $-1.0000$ (forced mode)\\
       & $6$ & $1$ & $0.0272$ & $-0.2979$ & $-0.3045$ (at $\Gamma=1.80074$)\\
       & $7$ & $2$ & $0.0382$ & $+0.7010$ & $+0.6956$ (at $\Gamma=1.80074$)\\
       & $5$ & $1$ & $0.0094$ & $-0.2596$ & $-0.3511$ (at $\Gamma=1.94082$)\\
       & $6$ & $2$ & $0.0144$ & $+0.7405$ & $+0.6489$ (at $\Gamma=1.94082$)\\
\hline
\hline
2.200 & $1$ & $1$ & $0.1095$ & $-1.0000$ &  $-1.0000$ (forced mode)\\
       & $4$ & $2$ & $0.0596$ & $-0.6368$ & $-0.6650$ (at $\Gamma=2.17832$)\\
       & $5$ & $1$ & $0.0423$ & $+0.3632$ & $+0.3350$ (at $\Gamma=2.17832$)\\
\hline
\hline
\end{tabular}
\caption{Amplitudes and frequencies from the simulations for the
  dominant modes at $\Gamma=1.825$ and $\Gamma=2.200$.  The columns
  denote (from left to right): aspect ratio, azimuthal wave number,
  axial wave number, maximum of the amplitude, azimuthal drift
  frequency from numerical data, and theoretical frequency from the
  dispersion relation~(\ref{eq::dispersion}) at the aspect ratio at
  which the linear inviscid approximation predicts a triadic resonance
  (always supposing a radial wave number corresponding to the first
  root of the dispersion relation which is confirmed in
  figure~\ref{fig::axialdecomp}).  }\label{tab::amp_freq_offres}
\end{center}
\end{table}
The emergence of a triadic resonance is less explicit at
$\Gamma=1.825$. However, at this aspect ratio we find even two triadic
resonances (see figure~\ref{fig::kelvin_amp_vs_time}) with
$(m,k)=(6,1)$ and $(7,2)$ and a second resonance with $(m,k)=(5,1)$
and $(6,2)$ (with much weaker amplitude, see
table~\ref{tab::amp_freq_offres}). In all cases the amplitude of the
free modes that constitute a triadic resonance strongly oscillates
with a maximum of up to 50\% of the forced mode ($\Gamma=2.2, m=4,
k=2$) which corresponds to roughly 6\% of the angular velocity of the
cylindrical container.

\subsection{Azimuthal shear flow}
\begin{figure}[h!]
\begin{center}
\includegraphics[width=6.6cm]{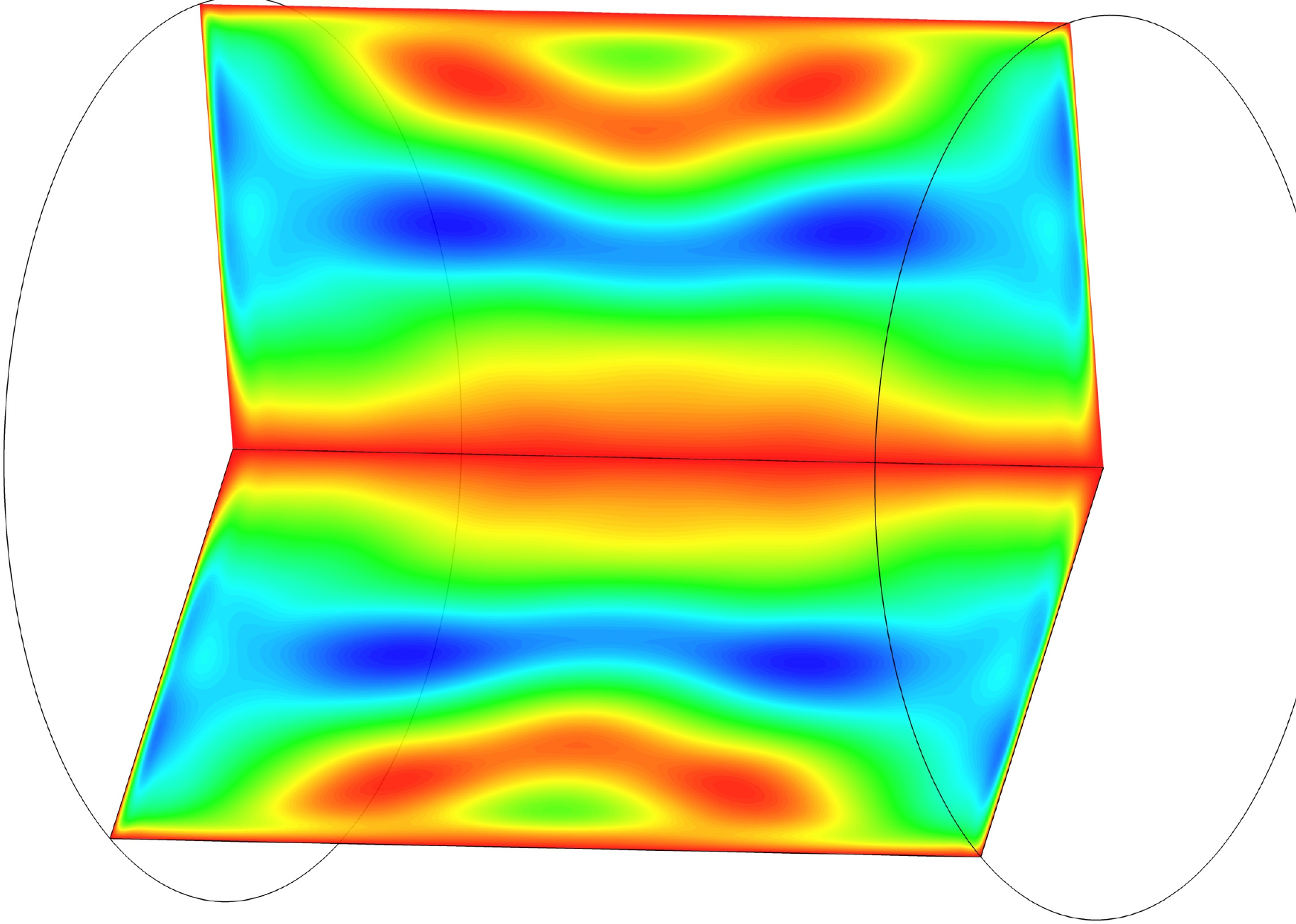}
\includegraphics[width=7.6cm]{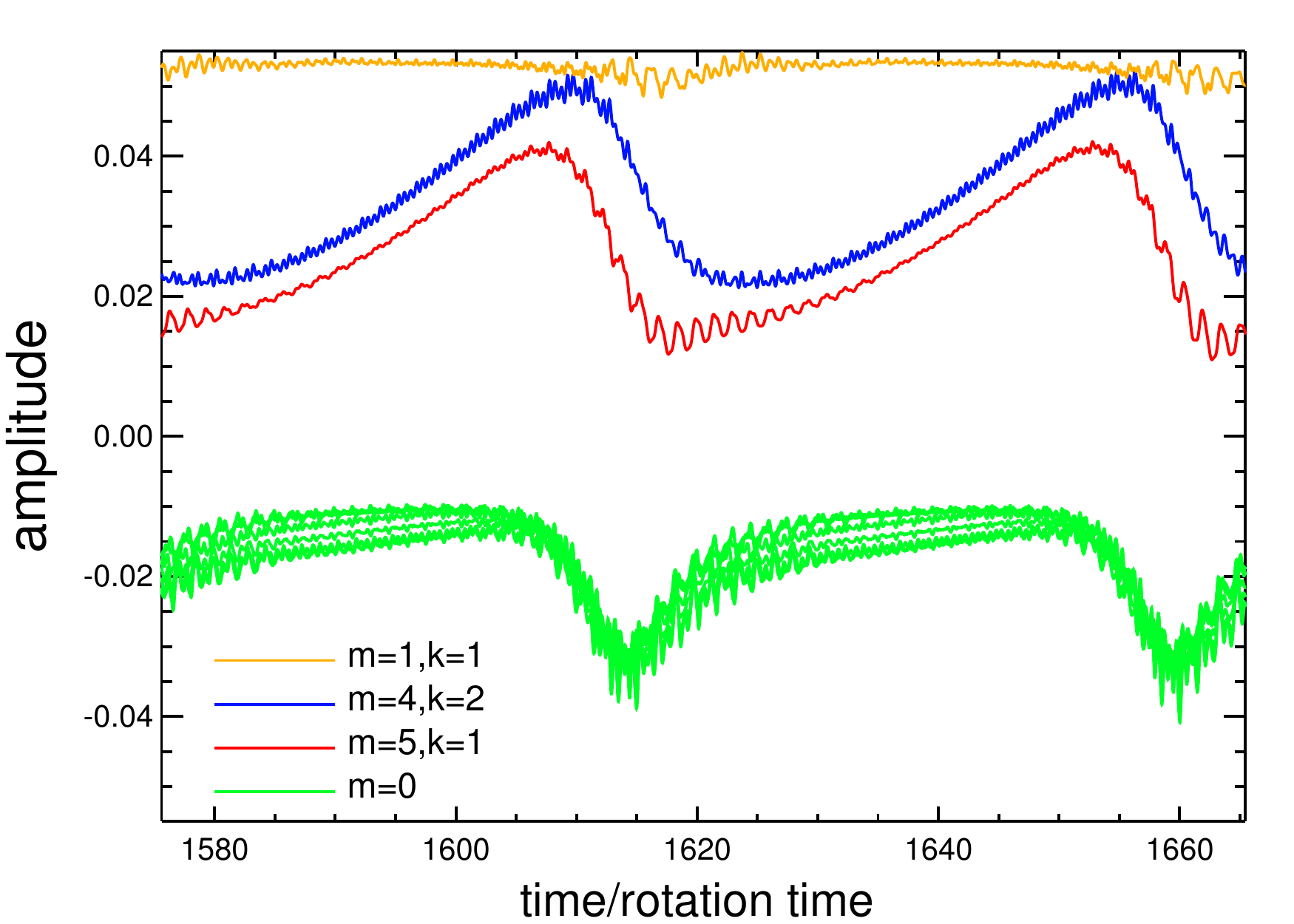}
\caption{Left: Pattern of the induced azimuthal shear flow at
  $\Gamma=2.2$. Right: Temporal development of the amplitude at
  $r=0.67$ of the azimuthal shear flow (green curve), the free Kelvin
  modes (red and green curve) and the forced mode (yellow
  curve).}\label{fig::shearflow}
\end{center}
\end{figure}

The growth of the triads goes along with a growth of an axisymmetric
azimuthal shear flow that is mostly geostrophic (see left panel in
figure~\ref{fig::shearflow}). The axisymmetric mode emerges with a
slight delay with respect to the free Kelvin modes which indicates a
saturation process by a detuning of the resonance frequencies (see
right panel in figure~\ref{fig::shearflow}). The induced
axialsymmetric flow component is negative on average and, hence,
causes a breaking of the solid body rotation which increases with
increasing precession ratio. In the extreme case the breaking of the
induced axisymmetric flow entirely cancels the solid body rotation,
giving the impression that in the laboratory frame the cylindrical
container is rotating around a standing fluid \cite{johann}. This
effect is closely connected to the abrupt transition to a chaotic flow
which takes place at a critical precession ratio
\cite{johann,2014PhFl...26e1703K}.

\section{Conclusions}

We performed numerical simulations of a precession driven flow in
cylindrical geometry. The simulations confirm that the energy that can
be injected via precessional forcing is very sensitive to the aspect
ratio of the cylindrical container. Significant contributions of
higher non-axisymmetric components appear only in a limited range of
aspect ratios (for $\Gamma\in[1.81;2.24]$) and it turns out that in
that regime the behavior of the forced mode (with $m=1$) cannot be
described by a linear theory. Nevertheless, in all cases the forced
Kelvin mode ($m=1$) dominates and free Kelvin modes with higher
azimuthal wave number appear as distortions of the forced mode. The
maximum response of the flow takes place at $\Gamma_{\rm{max}}\approx
1.871$ which is remarkable far away from the resonance predicted by
the linear in-viscid approximation (at $\Gamma=1.98982$). At the
resonance maximum our results yield an amplitude for the $z$-component
of the flow of $|{{u}}^{\rm{max}}_{z}| \approx 0.27$, which --
assuming isotropy -- roughly agrees with the estimations of
\citeasnoun{leorat2}. The amplitude of the forced mode decreases with
increasing distance from the resonance maximum, and in the regimes
with triadic resonances, we still observe values of
$|{{u}}^{\rm{max}}_{z}| \approx 0.10$ (in terms of the azimuthal
velocity of the container). In real units (assuming
$\Omega_{\rm{c}}=2\pi \cdot 10\mbox{Hz}$ and $R=1\,{\rm{m}}$) this
would correspond to fluid velocities of $15\,\mbox{m/s}$ (at
resonance) and $6\,\mbox{m/s}$ (off resonance) which is of the same
order as the typical flow speed in the Riga dynamo
experiment. Regarding the higher Kelvin modes,
the saturated amplitudes of the strongest free Kelvin modes are mostly
independent of the forced mode and achieve values of about $0.05$ to
$0.06$ in terms of the angular velocity of the container corresponding
to $\sim 3.5\,{\rm{m}}/{\rm{s}}$ in the experiment.

At the aspect ratio envisaged for the experiment ($\Gamma=2$ with
$H=2\,{\rm{m}}$ and $R=1\,{\rm{m}}$) we may expect an energy of the
fluid flow at roughly two thirds of the maximum value at the optimum
aspect ratio ($\Gamma=1.871$). However, this optimum aspect ratio
probably depends on the Reynolds number and on the forcing, so that we
believe that the chosen geometry for the dynamo experiment is a good
compromise to ensure an efficiently driven fluid flow in a container
with a ratio of diameter to height that at least approximately
reflects the geometry of planetary bodies.

Although at present we do not know whether the flow fields that emerge
in our simulations will be able to drive a dynamo, we may try to
estimate typical magnetic field strengths that can be expected in the
planned dynamo experiment.  We assume a saturation of the magnetic
energy at roughly $5\%$ of the kinetic energy of the hydrodynamic flow
which is substantiated by the saturation behavior of the Riga dynamo
\cite{2008CRPhy...9..721G}. We further presuppose that the internal
velocity caused by the precessional forcing linearly scales with the
rotation of the container (and hence with ${\rm{Re}}$) and assume an
angular velocity of $\Omega_{\rm{c}}=2\pi\cdot 10\,\mbox{Hz}$, a
radius $R=1\,\mbox{m}$, a height $H=2\,\mbox{m}$ and a density of
liquid sodium of $\rho\approx 930\,\mbox{kg/m}^3$ (at $\sim\! 400\,{\rm{K}}$). After the change over to
real units the kinetic energy obtained in our simulations at
$\Gamma=2$ ($E_{\rm{kin}}\approx 0.006$) corresponds approximately to
a magnetic field strength $B \approx\sqrt{0.05\mu_0\rho
  E_{\rm{kin}}\Omega_{\rm{c}}^2}\approx 40\,\mbox{mT}$. This is in the
range of the Riga dynamo which is not particularly surprising since
the typical velocities obtained in our simulations indeed match the
velocity arising in the Riga dynamo. However, given that the
corresponding $m=1$ mode has produced no dynamo in the kinematic
simulations it might be more realistic to refer to the free Kelvin
modes with higher $m$ which have less kinetic energy. In that case the
typical magnetic field strength amounts to only $B\approx
5\,\mbox{mT}$ which, however, is still in the range of the values
obtained at the Von-K{\'a}rm{\'a}n-Sodium dynamo
\cite{2007PhRvL..98d4502M}. We expect a stronger response and hence a
larger saturation field strength for increasing ${\rm{Po}}$, at least
as long as we remain below the critical Poincare number for the sudden
transition to a chaotic state
\citeaffixed{johann}{${\rm{Po}}^{\rm{crit}}\approx 0.0725$ at
  ${\rm{Re}}=5.65\times 10^5$, see}. However, unless we know whether
or not the precession driven flow will have the right structure and
sufficient magnitude to excite dynamo action at all, these estimations
are only useful to demonstrate that the reference values assumed for
the planned dynamo experiment will allow magnetic fields with a
reasonable field strength.

Regarding the structure of the flow, we find various manifestations of
non-axisymmetric contributions in terms of free Kelvin modes with
higher azimuthal wave numbers $m$. We find free Kelvin modes in
resonance with the forced mode only when the forced mode is
sufficiently far away from its primary resonance, i.e., when the
amplitude of the forced mode is not too strong. Triadic resonances can
be clearly identified using a combined Fourier-Discrete Sine
transformation in the azimuth and along the axis with structure and
frequencies close to predictions from linear theory. Actually, we were
also expecting triads around $\Gamma\approx 2$ with $m=5$ and
$m=6$. Indeed, we do find patterns of free Kelvin modes in that
regime, but no combination of them satisfies the triadic resonance
conditions. Instead, we see free Kelvin modes solely with retrograde
drift motion and for each azimuthal wavenumber $m$ the contributions
with $k=1$ and $k=2$ are approximately of equal strength so that in
sum we see a clear breaking of the equatorial symmetry of the
flow. This symmetry breaking has been essential for the functioning of
a dynamo in simulations of \citeasnoun{2005PhFl...17c4104T} in a
precessing sphere whereas it seemed less important in the direct
numerical simulations of \citeasnoun{2011PhRvE..84a6317N}. However,
preliminary kinematic dynamo simulations we conducted recently with an
analytic flow of Kelvin modes suggest that the joint appearance of
contributions with $k = 1$ and $k = 2$ considerably facilitates the
occurrence of dynamo action.

The triadic instability requires a rather long time to emerge and
saturates to a final state with slow periodic growth and decay. This
period is likely related to the precession time scale but a conclusive
correlation requires further simulations at different ${\rm{Po}}$. The
periodic growth followed by a fast decay of the free Kelvin modes in
the regime with triadic resonances reminds of the so called resonant
collapse \cite{1970JFM....40..603M,1992JFM...243..261M}. However, in
our simulations the energy is essentially distributed among very few,
large scale modes. These modes are the fundamental forced mode and two
free Kelvin modes that establish the triad and an axisymmetric,
essentially geostrophic mode. We do not spot any turbulence-like
behavior during the decay of the free Kelvin modes, i.e. there is no
transition from the dominance of the large scale mode into small scale
turbulent flow which would be characteristic for the collapse
phenomenon. This behavior might change when the Reynolds number
approaches larger (i.e. more realistic) values. This is substantiated
by the simulations at $\Gamma=2$.  Here we see no triadic resonance,
but rather a quasi-periodic occurrence of free Kelvin modes that
exhibit a trend towards larger radial wave numbers during their decay
which may represent (part of) a cascade to small scale structures.

In principle the occurrence of very few dominant modes calls for a
low-dimensional model. However, the ansatz~(\ref{eq::triadenansatz})
used in this study to explain the occurrence of triadic resonances is
too simplistic to allow for a reasonable description of the non-linear
behavior and the saturation obtained in the simulations presented
above. A more sophisticated approach is presented
by~\citeasnoun{FLM:7951619} who use a low-dimensional model with four
coupled ordinary differential equations that describe the amplitude of
the forced mode, the free Kelvin modes and an axisymmetric geostrophic
mode. This model allows for saturation via the axisymmetric mode and
for a reinforcement of the forced mode by the free Kelvin modes and
qualitatively reproduces the results obtained in our
simulations. However, in order to yield a quantitative agreement,
advanced models are still required.

\ack{The authors acknowledge support from the Helmholtz-Allianz
  LIMTECH. This research was supported in part by the National Science
  Foundation under Grant No. NSF PHY11-25915. AG acknowledges the
  experiences he could make in participating in the program on
  {\it{Wave-Flow Interaction in Geophysics, Climate, Astrophysics, and
      Plasmas}} at the Kavli Institute for Theoretical Physics, UC
  Santa Barbara. The authors kindly acknowledge the discussions with
  Andreas Tilgner and Oliver G{\"o}pfert from the Institute of
  Geophysics at the University of G{\"o}ttingen.}

\section*{References}
\bibliographystyle{jphysicsB}

\end{document}